\documentclass[amsmath,amssymb,showpacs,showkeywords,twocolumn]{revtex4}
\usepackage{graphicx}
\usepackage{times}
\usepackage{mathrsfs}
\usepackage{amsmath}
\usepackage{graphicx}
\usepackage{epsfig}
\usepackage{dcolumn}
\usepackage{bm}
\usepackage{multirow}
\usepackage{pdfpages}
\usepackage[utf8x]{inputenc}
\DeclareUnicodeCharacter{2212}{-}
\begin{document}
\title{Interface controlled spin filtering and nonreciprocal transport in altermagnet/Ising - superconductor junctions}
\author{Arindam Boruah\footnote{arindamboruah@dibru.ac.in}, Saumen Acharjee\footnote{saumenacharjee@dibru.ac.in} and Prasanta Kumar Saikia\footnote{saikiapk@dibru.ac.in}}
\affiliation{Department of Physics, Dibrugarh University, Dibrugarh 786 004, 
Assam, India}

\begin{abstract}
We investigate theoretically spin-resolved transport, spin filtering, and nonreciprocal effects in an altermagnet/Ising superconductor (AM/ISC) junction with a spin-active interface. Using a modified Bogoliubov–de Gennes  framework within the scattering formalism, we demonstrate that the interplay among intrinsic spin-orbit coupling (ISOC), anisotropic AM spin texture and spin-dependent interfacial scattering gives rise to strongly anisotropic charge and spin conductance. In the weak spin-mixing regime, transport remains predominantly helicity conserving and exhibits pronounced angular dependence governed by the relative orientation between the AM spin texture and interface magnetization. Increasing ISOC enhances spin conductance and leads to spin-selective Andreev reflection resulting in finite spin filtering. In contrast, the strong spin-mixing regime exhibits enhanced angular anisotropy and robust spin-polarized transport over a broad energy range. Conventional Andreev reflection becomes strongly suppressed, accompanied by substantial spectral redistribution. We further show that nonreciprocal transport persists throughout the single-band, intermediate and double-band ISC regime. 
The spin polarization and spin-filter efficiency exhibit nonmonotonic dependence on system parameters, reaching values up to $\sim 86\%$, with characteristic angular modulation determined by the AM spin texture. Finite-energy analysis reveals enhanced spin selectivity at low energies and suppression near the superconducting gap. Furthermore, strong spin mixing at the AM/ISC junction produces asymmetric conductance patterns, indicating nonreciprocal transport. Our results establish AM/ISC junctions as a versatile platform for tunable superconducting spintronics and directional spin transport.
\end{abstract}

\pacs{74.45.+c, 85.75.-d, 74.90.+n, 75.76.+j}
\maketitle

\section{Introduction}
The generation, manipulation, and detection of spin-polarized transport in superconducting hybrid systems have emerged as a central theme in modern superconducting spintronics~\cite{Zutic2004,Aoki2001,Bergeret2018,Pfleiderer2001,Linder2015,Saxena2000,Buzdin2005,Bergeret2005}. In particular, the interfaces where superconducting correlations coexist with spin-dependent interactions
provide a unique platform to realize unconventional transport phenomena such as dissipationless spin currents~\cite{Jacobsen2016,Han2020,Samokhvalov2021}, spin filtering~\cite{Tokuyasu1988,Shevtsov2014,Eschrig2015} and nonreciprocal quantum transport~\cite{Nadeem2023,Sun2025,Zhang2024a}. Beyond their fundamental significance, they hold the potential for next generation low power quantum and spintronic devices. 
A key ingredient underlying these functionalities is spin-dependent scattering at interfaces, commonly described by a spin-active interface, which incorporate spin-mixing and spin-flip processes~\cite{Linder2010,Lu2025,Panda2026,Li2026,Amin2016,Ouassou2016,Triola2014,Hirai2003}. 
In conventional superconductor (SC)/ferromagnet (FM) junctions, exchange splitting of the FM lifts spin degeneracy through explicit breaking of time-reversal symmetry~\cite{Buzdin2005}. However, such approaches are inherently constrained by the presence of net magnetization, stray fields, and their limited compatibility with singlet superconducting correlations~\cite{Acharjee2016}. 

From a symmetry perspective, this motivates the exploration of new materials that lift spin degeneracy without invoking macroscopic time-reversal symmetry breaking. In this context, altermagnets (AMs) have emerged as a distinct class of magnetic materials that offer  unconventional spin-dependent transport properties~\cite{Boruah2025,Smejkal2022b,Occhialini2022,Smejkal2022a,Smejkal2020,Feng2022,Rial2024,Morano2025,Cheong2024}. Experimental realizations in systems such as RuO$_2$, MnF$_2$, MnRe, Mn$_5$Si$_3$ and La$_2$CuO$_4$~\cite{Feng2022,Occhialini2022,Rial2024,Morano2025,Cheong2024} underscore their experimental relevance. Unlike conventional magnets, AMs exhibit vanishing net magnetization while hosting momentum-dependent spin splitting dictated by crystalline symmetries. Thus, they provide a symmetry-driven route for spin selectivity without breaking global time-reversal symmetry~\cite{Beenakker2023,Ouassou2023,Sun2023,Papaj2023,Cheng2024}. 

In contrast, Ising superconductors (ISCs) provide a natural and complementary platform for the symmetry-driven spin structure of AMs. In particular, monolayer transition-metal dichalcogenides such as NbSe$_2$, MoS$_2$, WSe$_2$ and TaS$_2$ lack in-plane inversion symmetry and thereby host strong intrinsic spin orbit coupling (ISOC), which locks electron spins out of the plane and stabilizes Cooper pairing against in-plane magnetic perturbations~\cite{acharjee1021,barrera,idzuchi3,tang2,lu2,saito2,xi2,dvir2,costanzo2, li2}. From a symmetry perspective, the combination of broken inversion symmetry and strong ISOC generates an effective valley-dependent Zeeman field at the $(K,-K)$ points, leading to robust superconductivity beyond the conventional Pauli limit~\cite{dvir2,costanzo2,sohn2,li2,acharjee1021}. This intrinsic spin-valley locking and enhanced resilience of the superconducting state make ISCs an ideal counterpart to AMs. Moreover, when interfaced with spin-textured or magnetic materials, the strong ISOC of the ISCs significantly modifies quasiparticle dynamics and Andreev reflection processes  leading to unconventional transport mechanism~\cite{acharjee1021}. Consequently, ISC/AM junctions provide a versatile platform to explore the interplay between symmetry-protected spin splitting and superconducting correlations~\cite{Boruah2025}. Thus they offer direct implications for tunable spin filtering and nonreciprocal transport~\cite{Zhang2024,Giil2024,Yi2026}.

Previous works on FM/SC and AFM/SC junctions have demonstrated key functionalities such as spin-polarized Andreev reflection~\cite{Sun2023,Papaj2023,Boruah2025,Cheng2024}, generation of long-range triplet correlations~\cite{meng,trifunovic,annunziata}, and interface driven spin filtering~\cite{Ge2025}  underscoring the central role of spin-dependent scattering at interfaces in controlling the quasiparticle dynamics. Furthermore, the emergence of nonreciprocal transport phenomena including superconducting diode effects~\cite{Boruah2025, Cheng2024} and magnetochiral anisotropy highlight the direction dependent dissipationless transport in the presence of broken spatial and time-reversal symmetries~\cite{Sun2023,Ouassou2023,Beenakker2023}. The discovery of AMs and ISCs have opened new opportunities for device applications by enabling control of spin without macroscopic magnetization and enhancing superconducting robustness under strong SOC~\cite{mishra2021,martinez2020,daldin2024}. These properties make them promising candidate for low-dissipation spintronic architectures and nonreciprocal superconducting devices~\cite{Cheong2024}. However, despite their individual advantages, their combined potential in hybrid junctions remains largely unexplored, especially in realistic settings where interface effects dominate transport. In particular, the role of spin-active interfaces as a tunable control parameter for spin filtering and nonreciprocal response has not been systematically investigated. In this work, we investigate quantum transport in an AM/ISC junction with a spin-active interface. We demonstrate that the interplay between momentum-dependent spin splitting in the AM, strong ISOC of the ISC, and spin-mixing and spin-flip processes at the interface enables controllable spin filtering and nonreciprocal charge transport.

The paper is structured as follows: In Sec. II, we define Bogoliubov–de Gennes (BdG) Hamiltonian and propose theoretical framework for the AM/ISC hybrid separated by a spin-active interface. In Sec. III, we discuss the charge and spin conductance spectra and their implications. We perform the symmetry analysis and the discuss origin of non reciprocity in the conductance spectra in Sec. IV. In Sec. V  we discuss the  spin polarization and spin filter efficiency for the proposed junction. Finally, Sec. VI presents our conclusions.

\section{Formulation}
\begin{figure}[t]
\centerline
\centerline{ 
\includegraphics[scale=0.27]{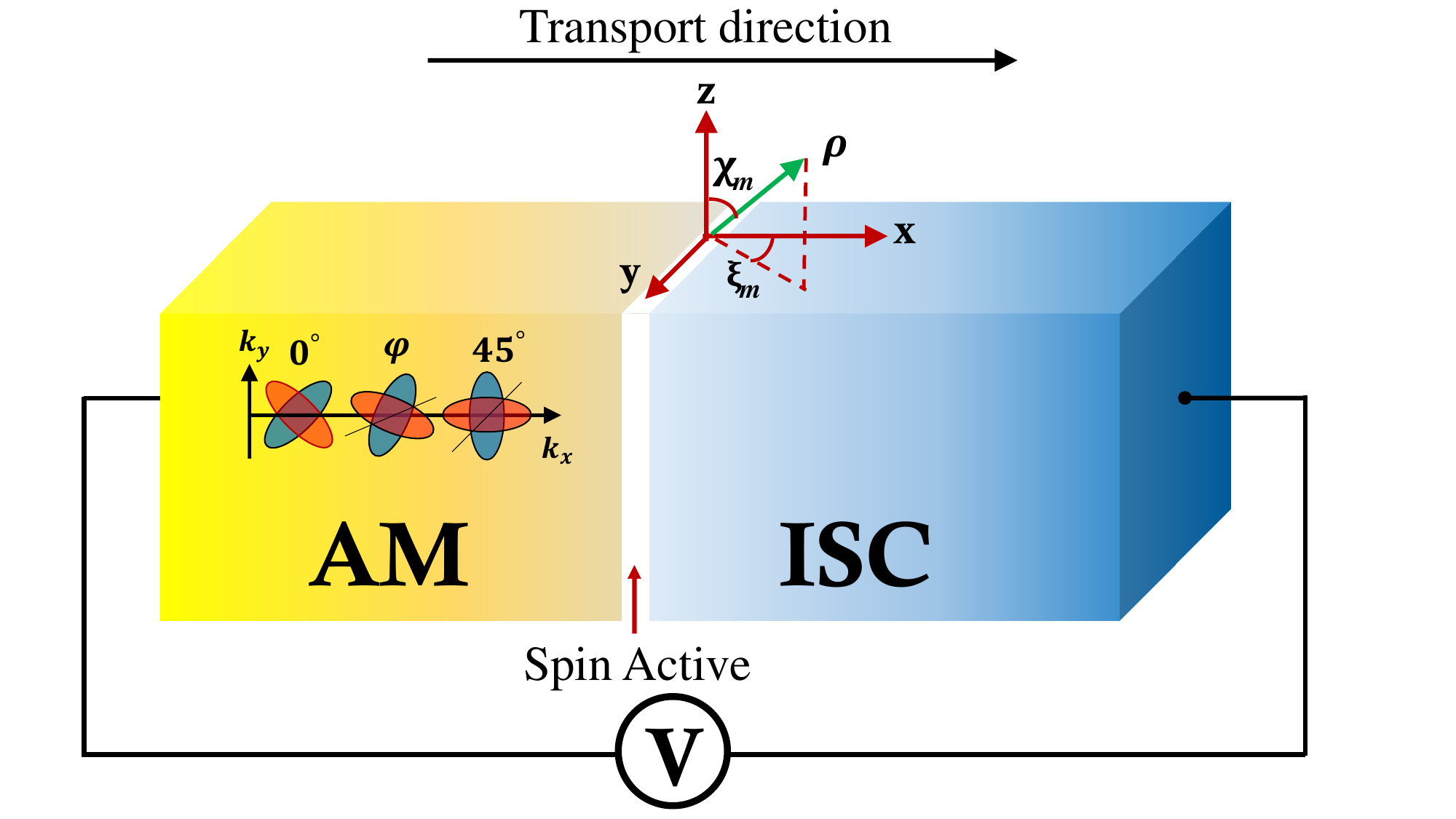}
\vspace{0.6cm}
}
\centerline{ 
\includegraphics[scale=0.27]{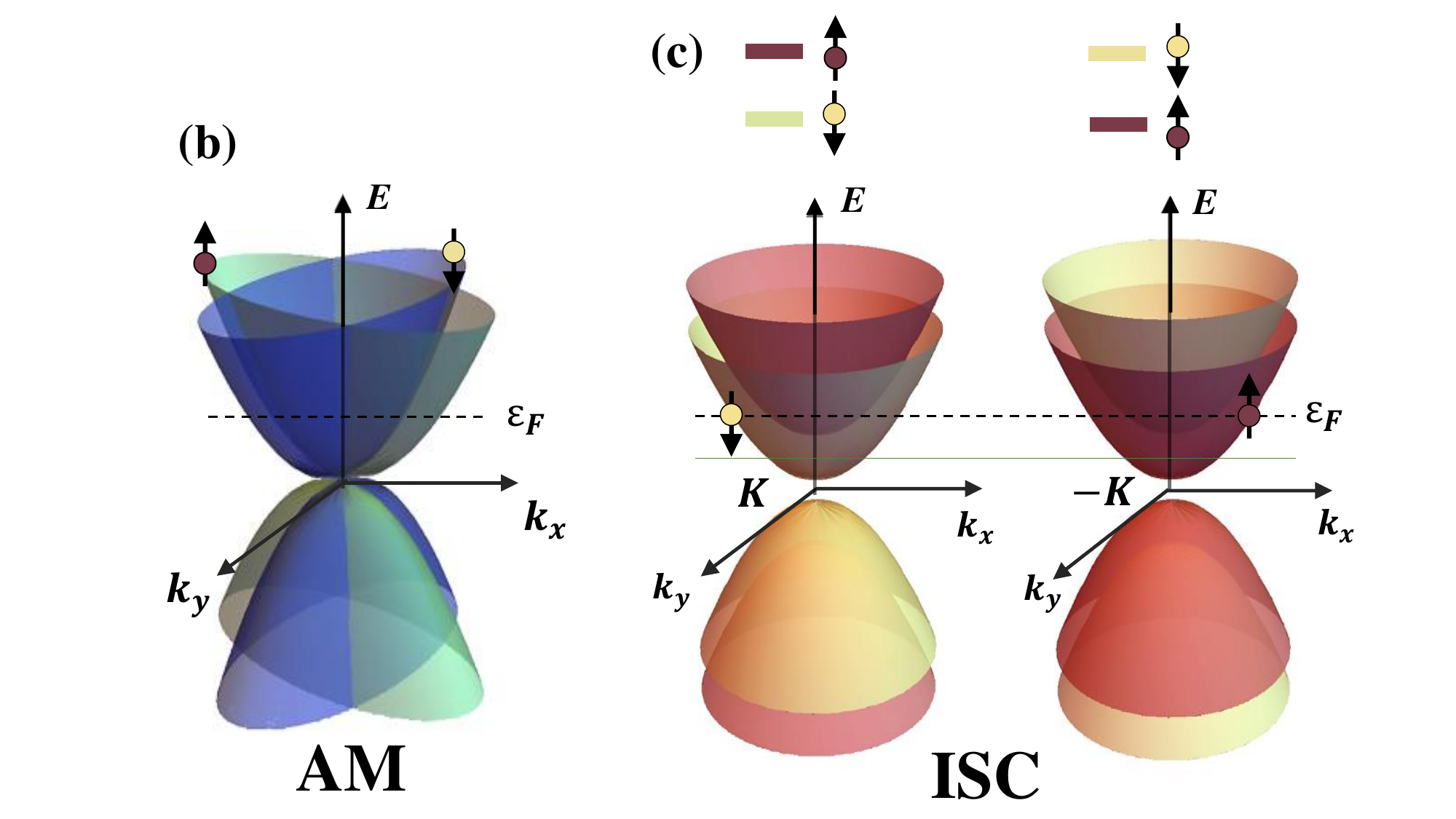}
\vspace{0.6cm}
}
\caption{(a) Schematic illustration of a heterostructure consisting of an Altermagnet (AM) and an Ising Superconductor (ISC), separated by a spin active interface at $x = 0$. The interface is characterized by a magnetic barrier strength ($\rho$) and orientation specified by the polar angle ($\chi_m$) and azimuthal angle ($\xi_m$) respectively. The parameter $\varphi$ denotes the crystallographic orientation of the AM, which effectively rotates the spin-split Fermi surfaces corresponding to majority (blue) and minority (red) spin channels. (b) Schematic energy band structure of AM illustrating momentum-dependent spin splitting. The black dashed line corresponds to Fermi energy of the AM. (c) Schematic energy band structure of an ISC near the $\pm K$ valleys, showing spin-valley locking induced by the ISOC. The red (yellow) solid bands corresponds to the spin up (spin down) configuration.  Both double-band (left) and single-band (right) ISC regimes are depicted, with their Fermi energies $\varepsilon_\text{F}$ are indicated by black dotted and solid green lines respectively.}
\label{fig1}
\end{figure}

We consider a heterostructure consisting of an arbitrary angle oriented Altermagnet (AM) ($x < 0 $) and an Ising Superconductor (ISC) ($x > 0$) separated by a spin active interface at $x = 0$ as shown in Fig. \ref{fig1}(a). The spin active barrier is characterized by the spin dependent potential $\Phi_m = \rho\Phi_0(\sin{\chi_m}\cos{\xi_m}, \sin{\chi_m}\sin{\xi_m}, \cos{\chi_m})$, where, $\chi_m$ and $\xi_m$ are the polar and azimuthal angles of the interfacial magnetic moment with respect to the $z$ axis. The parameter $\Phi_0$ represents the intrinsic spin independent barrier strength and $\rho \equiv |\Phi_m|/\Phi_0$, defines the relative strength of the spin dependent scattering compared to the scaler barrier. The transverse components of $\Phi_m$ ($\propto \sin\chi_m$) give rise to spin-flip scattering processes, while the longitudinal component ($\propto \cos\chi_m$) leads to spin-dependent phase shifts, commonly referred to as spin mixing. Consequently, the parameters $\rho$ and $\chi_m$ provide independent control over spin mixing and spin-flip processes at the interface.  We consider transport along the $x$-direction, while translational invariance along the transverse direction ensures conservation of the momentum component $k_y$.

The quasiparticle excitations are described  by the Bogoliubov-de Gennes (BdG) equation
$\check{\mathcal{H}}_\text{BdG}\Psi = \varepsilon \Psi$, where $\Psi = (\psi_\uparrow, \psi_\downarrow, \psi_\uparrow^\dagger, \psi_\downarrow^\dagger)^\text{T}$ denotes the Nambu spinors. The system is described by the BdG Hamiltonian $\check{\mathcal{H}}_\text{BdG}(x) =
\check{\mathcal{H}}_\text{AM}\,\Theta(-x)
+
\check{\mathcal{H}}_\text{ISC}\,\Theta(x)$. For an arbitrary angle oriented AM, the Hamiltonian can be  defined as \cite{Papaj2023}
\begin{equation}
\label{eq1}
\check{\mathcal{H}}_\text{AM} (\mathbf{k}) = \left(
\begin{array}{cc}
\hat{\mathcal{H}}_0(\mathbf{k})   & 0 \\
 0 & -\sigma_y\hat{\mathcal{H}}^\ast_0(\mathbf{-k})\sigma_y 
 \\
\end{array}\right)
\end{equation} 
\noindent where, the single-particle Hamiltonian of the AM region is given by
\begin{equation}
\label{eq2}
\hat{\mathcal{H}}_0(\mathbf{k}) = \xi_{\mathbf{k}} \hat{\mathrm{I}} + \lambda\, \mathbf{g}(\mathbf{k},\delta)\cdot \hat{\boldsymbol{\sigma}},
\end{equation}
where $\xi_{\mathbf{k}} = \frac{\hbar^2 k^2}{2m} - \mu_{\mathrm{AM}}$ denotes the kinetic energy measured from the chemical potential $\mu_{\mathrm{AM}}$. $\lambda$ characterizes the strength of the AM exchange field. The vector $\hat{\boldsymbol{\sigma}} = (\sigma_x, \sigma_y, \sigma_z)$ represents the Pauli matrices in spin space. The exchange field in AM is momentum dependent, resulting in spin splitting without net magnetization. Although global time-reversal symmetry is preserved, the anisotropic structure of $\mathbf{g}(\mathbf{k},\delta)$ lifts spin degeneracy locally in momentum space. As a result, the spin quantization axis becomes momentum dependent and aligns with $\mathbf{g}(\mathbf{k},\delta)$. Thus, spin is not a globally conserved quantum number. This feature plays a central role in determining interfacial transport, where scattering processes become intrinsically anisotropic and helicity dependent.

The momentum dependent exchange field is chosen to capture the characteristic $d$-wave AM symmetry and is expressed as~\cite{Boruah2025, Sun2023}
\begin{equation}
\label{eq3}
\mathbf{g}(\mathbf{k},\delta) =
\left(
k_x k_y\sin{2\delta},\;
(k_x^2 - k_y^2)\cos{2\delta},\;
0
\right).
\end{equation}
This form of $\mathbf{g}(\mathbf{k})$ follows from the symmetry of AM systems preserving combined $\mathcal{PT}$ symmetry while breaking idividual $\mathcal{P}$ and $\mathcal{T}$ symmetry. The lowest-order spin splitting allowed by crystal symmetry transforms according to the $B_{1g}$ irreducible representation, leading to a $d$-wave structure of the form $(k_x k_y, k_x^2 - k_y^2, 0)$~\cite{Smejkal2020,Smejkal2022a}. This structure has been shown to describe AM spin splitting in materials such as RuO$_2$ and MnTe. The parameter $\delta = \frac{1}{2}\tan^{-1}(\alpha_1/\alpha_2)$ parametrizes the orientation of the AM lobes relative to the transport direction, while the overall strength of the exchange interaction is quantified by $\bar{\alpha} = \sqrt{\alpha_1^2 + \alpha_2^2}$. To ensure a physically relevant regime with a closed Fermi surface, the parameters must satisfy the constraint $\bar{\alpha} < \alpha_c \equiv \hbar^2/m$, such that the resulting Fermi surface remains elliptic rather than hyperbolic. The corresponding energy band structure of the AM is depicted in Fig. \ref{fig1}(b).  The orientation angle $\delta$ controls the momentum-space structure of the AM spin-splitting field $\mathbf{g}(\mathbf{k},\delta)$ and thereby modifies the spin-dependent scattering processes at the interface. Different values of $\delta$ rotate the anisotropic spin texture in momentum space, leading to distinct helicity distributions and consequently different spin-resolved transport characteristics.

The Hamiltonian for the ISC region is given by~\cite{acharjee1021,Cheng2024}
\begin{equation}
\label{eq4}
\check{\mathcal{H}}_\text{ISC} (\mathbf{k}) = \left(
\begin{array}{cc}
 \hat{\mathcal{H}}_\pm(\mathbf{k})   & \hat{\Delta}(\mathbf{k}) \\
 -\hat{\Delta }^{\ast}(\mathbf{k}) & -\hat{\mathcal{H}}^{\ast}_\pm(\mathbf{-k})
 \\
\end{array}\right)
\end{equation} 
\noindent where, $\hat{\mathcal{H}}_\pm(\mathbf{k})$ denotes the single-particle Hamiltonian in the presence of the two inequivalent valleys $\mathbf{K}$ and $-\mathbf{K}$ of the ISC. It can be defined as~\cite{tang2}.

\begin{equation}
\label{eq5}
\hat{\mathcal{H}}_\pm(\mathbf{k}) = \left(\frac{\hbar^2k^2}{2m}-\mu_\text{S}\right)\hat{\text{I}} +\epsilon\beta\hat{\sigma}_z 
\end{equation}
\noindent where, $\epsilon = \pm 1$ labels the valley index corresponding to the 
$\pm \mathbf{K}$ points, and 
$\mu_\text{S}$ denotes the chemical potential in the ISC region. The parameter 
$\beta$ characterizes the strength of the Ising spin–orbit coupling (ISOC), while 
$\hat{\sigma}_z$ is the Pauli matrix acting in spin space. The presence of ISOC lifts the spin degeneracy, leading to spin-split subbands with opposite spin polarization in the two valleys as seen from Fig.~\ref{fig1}(c). In particular, at the $\mathbf{K}$ valley the spin-up band resides at higher energy than the spin-down band, whereas the ordering is reversed at the $-\mathbf{K}$ valley. Despite this valley dependent spin splitting, the normal-state Hamiltonian preserves global time-reversal symmetry and retains a residual U$(1)$ spin-rotation symmetry about $z$-axis. Notably due to the large momentum-space separation between the $\mathbf{K}$ and $-\mathbf{K}$ valleys, intervalley scattering induced by disorder is strongly suppressed and can be neglected. As a result, $\epsilon$ remains a good quantum number in the absence of strong perturbations.

The superconducting order parameter for the ISC appears in Eq. (\ref{eq4}) can be defined as \cite{Cheng2024}
\begin{equation}
\label{eq6}
\hat{\Delta}(\mathbf{k}) = \Delta_0 i \hat{\sigma}_y e^{i\phi} \Theta(x) ,
\end{equation}
\noindent where $\Delta_0$ is the magnitude of the superconducting gap and $\phi$ corresponds to the superconducting phase of the ISC. The nature of the superconducting state is governed by the relative magnitude of $\mu_\mathrm{S}$ and $\beta$. For $\mu_\mathrm{S}  <\beta$, both spin-split bands intersect the Fermi level, resulting in a double band superconducting regime. In contrast, for $\mu_\mathrm{S} > \beta$, only a single spin-polarized band contributes, giving rise to an effective single band superconducting state \cite{acharjee1021}. The superconducting pairing occurs between electrons from opposite valleys with opposite momenta and opposite spins, thereby forming Cooper pairs that preserve time-reversal symmetry, as illustrated 
in Fig.~\ref{fig1}(c).

The quasiparticle wave function in the AM region is constructed by considering an incident electron with helicity index $s = \pm$. Due to the momentum-dependent spin texture of the AM, the scattering states are naturally expressed in the helicity basis defined by the eigenstates of the AM Hamiltonian. The wave function in the AM region can be written as
\begin{multline}
\label{eq7}
\Psi_{\mathrm{AM}}^{(s)}(x<0) = \hat{\Omega}^{e}_{s}\, e^{i\kappa^{e}_{s}x} 
+ \sum_{s'=\pm} r_{ss'}\, \hat{\Omega}^{e}_{s'}\, e^{-i\kappa^{e}_{s'}x} \\
+ \sum_{s'=\pm} a_{ss'}\, \hat{\Omega}^{h}_{s'}\, e^{i\kappa^{h}_{s'}x},
\end{multline}

\noindent where $\hat{\Omega}^{e}_{s}$ and $\hat{\Omega}^{h}_{s}$ denote the electron-like and hole-like quasiparticle spinors respectively. The coefficients $r_{ss'}$ and $a_{ss'}$ correspond to the normal and Andreev reflection amplitudes, describing the scattering of an incident electron in helicity channel $s$ into electron-like or hole-like states with helicity $s'$. The helicity spinors are given by
\begin{align}
\nonumber
\hat{\Omega}^{e}_{+} &= \frac{1}{\sqrt{2}}(1, \,\,\, e^{i\phi_{\mathbf{k}}}, \,\,\, 0, \,\,\, 0)^\text{T},\\
\nonumber
\hat{\Omega}^{e}_{-} &= \frac{1}{\sqrt{2}}(1, \,\, -e^{i\phi_{\mathbf{k}}}, \,\, 0, \,\, 0)^\text{T},\\
\nonumber
\hat{\Omega}^{h}_{+} &= \frac{1}{\sqrt{2}}(0, \,\, 0, \,\, 1, \,\, - e^{-i\phi_{\mathbf{k}}})^\text{T},\\
\nonumber
\hat{\Omega}^{h}_{-} & = \frac{1}{\sqrt{2}}(0, \,\, 0, \,\, 1, \,\, e^{-i\phi_{\mathbf{k}}})^\text{T},
\end{align}
\noindent Here, the phase $\phi_{\mathbf{k}} = \arg\!\left[g_x(\mathbf{k},\delta) + i g_y(\mathbf{k},\delta)\right]$ characterizes the momentum-dependent orientation of the AM exchange field in the $xy$-plane. 
\noindent The wave vectors for electron (hole) in the AM region are given by
\begin{equation}
\label{eq8}
\kappa^{e(h)}_{s'} =
\frac{-m\alpha_1 \kappa_y + s' \sqrt{2m \mathcal{Q}_1 \big(\mu_{\mathrm{AM}} \pm E\big) + \kappa_y^2 \mathcal{Q}_2}}{\mathcal{Q}_1},
\end{equation}

\noindent where $s' = \pm$ labels the two helicity branches. Here, $\mathcal{Q}_1 \equiv (\hbar^2 + \tau m \alpha_2)$ and $\mathcal{Q}_2 \equiv m^2 (\alpha_1^2 + \alpha_2^2) - \hbar^4$. The parameter $\mu_{\mathrm{AM}}$ denotes the chemical potential of the AM region, and $\kappa_y$ is the conserved transverse momentum. 

The wave function in the ISC regions can be defined as \cite{Boruah2025, Cheng2024}
\begin{multline}
\label{eq9}
\Psi^{\pm}_{\text{ISC}}(x>0) = 
b^\pm_1 \Phi_1 e^{ik_\pm x}
+b^\pm_2 \Phi_2e^{ik_\mp x}+
b^\pm_3 \Phi_3e^{-ik_\pm x}
\\+b^\pm_4\Phi_4e^{-ik_\mp x}
\end{multline}
\noindent where, $\Phi_1, \, \Phi_2, \, \Phi_3, \, \Phi_4$ are defined as
\begin{align}
    \nonumber
    \Phi_1 &= u e^{i\phi/2}\hat{\chi}_1 + v e^{-i\phi/2}\hat{\chi}_4\\
    \nonumber\Phi_2 &= u e^{i\phi/2}\hat{\chi}_2 - v e^{-i\phi/2}\hat{\chi}_3\\
    \nonumber
    \Phi_3 &= v e^{i\phi/2}\hat{\chi}_1 + u e^{-i\phi/2}\hat{\chi}_4\\
    \nonumber\Phi_4 &= -v e^{i\phi/2}\hat{\chi}_2 +u e^{-i\phi/2}\hat{\chi}_3
\end{align}

\noindent where, we define,  $\hat{\chi}_1 = (1,0,0,0)^\text{T}$, $\hat{\chi}_2 = (0,1,0,0)^\text{T}$, 
$\hat{\chi}_3 = (0,0,1,0)^\text{T}$ and
 $\hat{\chi}_4 = (0,0,0,1)^\text{T}$. Here, $b^\pm_1$ ($b^\pm_2$)  and $b^\pm_3$ ($b^\pm_4$) are the transmission coefficients for up (down) spin electrons and holes respectively.  
 
 The quasi particle momenta of the electron and hole in the ISC region can be expressed as
 \begin{equation}
 \label{eq10}
 k_{e(h),\pm} = \sqrt{\frac{2m}{\hbar^2}(\mu_\text{S} \mp \epsilon \beta +\tau \sqrt{E^2-\Delta^2)}}
\end{equation}   
where, $\tau = + (-)$ for electron (hole) considered for our analysis. Here, $k_{+(-)}$ represents the momenta of the electron (hole) in the ISC. 

The quasiparticle amplitudes $u$ and $v$ appearing in Eq.(\ref{eq9}) are defined as
\begin{align}
\nonumber
 u &=  \frac{1}{\sqrt{2}}\sqrt{1+\sqrt{1-\frac{\Delta^2}{E^2}}},\\
 \label{eq11}
 v &=  \frac{1}{\sqrt{2}}\sqrt{1-\sqrt{1-\frac{\Delta^2}{E^2}}}
\end{align}

The boundary conditions in the presence of a spin-active interface are derived by incorporating a $\delta$-function potential that includes both spin-independent and spin-dependent contributions. The interface is modeled by the potential~\cite{Linder2010}
\begin{equation}
\mathcal{V}_{\text{int}} = \left[ \Phi_0 \hat{I}+ \mathbf{\Phi}_m \cdot \boldsymbol{\sigma} \right]\delta(x),
\label{eq12}
\end{equation}
where, $\hat{I}$ is a $4\times4$ identity matrix, $\Phi_0$ denotes the non magnetic barrier strength and $\mathbf{\Phi}_m$ represents the magnetic component of the interface potential.

By integrating the BdG equation over the interval $[-\epsilon, \epsilon]$, in the limit $\epsilon \rightarrow 0$, one obtains the generalized boundary conditions for the quasiparticle wave function. The wave function remains continuous, while its derivative exhibits a discontinuity proportional to the interface potential. The detailed derivation is provided in Appendix~A. The resulting boundary conditions can be written as
\begin{align}
\Psi_{\text{AM}}^\pm(0^-) &= \Psi_{\text{ISC}}^\pm(0^+), 
\label{eq13}\\
\partial_x\Psi_{\text{ISC}}^\pm(0^+) - \partial_x\Psi_{\text{AM}}^\pm(0^-)
&= \hat{\Lambda}\Psi^\pm_{\text{AM}}(0^-),
\label{eq14}
\end{align}
where $\hat{\Lambda}$ is a $4 \times 4$ matrix describing the effective spin-active interface potential. The interface matrix $\hat{\Lambda}$ includes both scalar and spin-dependent contributions. It can be written as~\cite{Linder2010}
\begin{equation}
\hat{\Lambda} =
\begin{pmatrix}
Z+\Omega_1 & \Omega_2 & 0 & 0 \\
\Omega_2^* & Z^\ast-\Omega_1 & 0 & 0 \\
0 & 0 & Z^\ast+\Omega_1 & \Omega_2^* \\
0 & 0 & \Omega_2 & Z-\Omega_1
\end{pmatrix}
\label{eq15}
\end{equation}
\noindent where, $Z = Z_0+iZ_1$ with the magnitude of the parameters $Z_0= \dfrac{2m\Phi_0}{\hbar^2 k_F \cos\theta}$ and $|Z_1| = \sqrt{\Gamma_1^2 + k_x^2 \Gamma_2^2}$ characterizes the spin-independent barrier and spin dependent barrier strength respectively where $\Gamma_1 \equiv \frac{i\alpha_1 k_y}{\hbar^2}$ and $\Gamma_2 \equiv \frac{\alpha_2 m}{\hbar^2}$. It is to be noted that both $Z_0$ and $Z_1$ are expressed as dimensionless quantities normalized by the Fermi energy $\varepsilon_\text{F}$. The spin dependent barrier terms are given by $\Omega_1 = 2\rho \Phi_0 \cos\chi_m$ and
$\Omega_2 = 2\rho \Phi_0 \sin\chi_m\, e^{-i\xi_m}$. In contrast to the purely spin independent barrier case, the presence of $\hat{\Lambda}$ couples different spin or helicity channels at the interface.  

\begin{figure*}[t]
\centerline{
\includegraphics[scale=0.22]{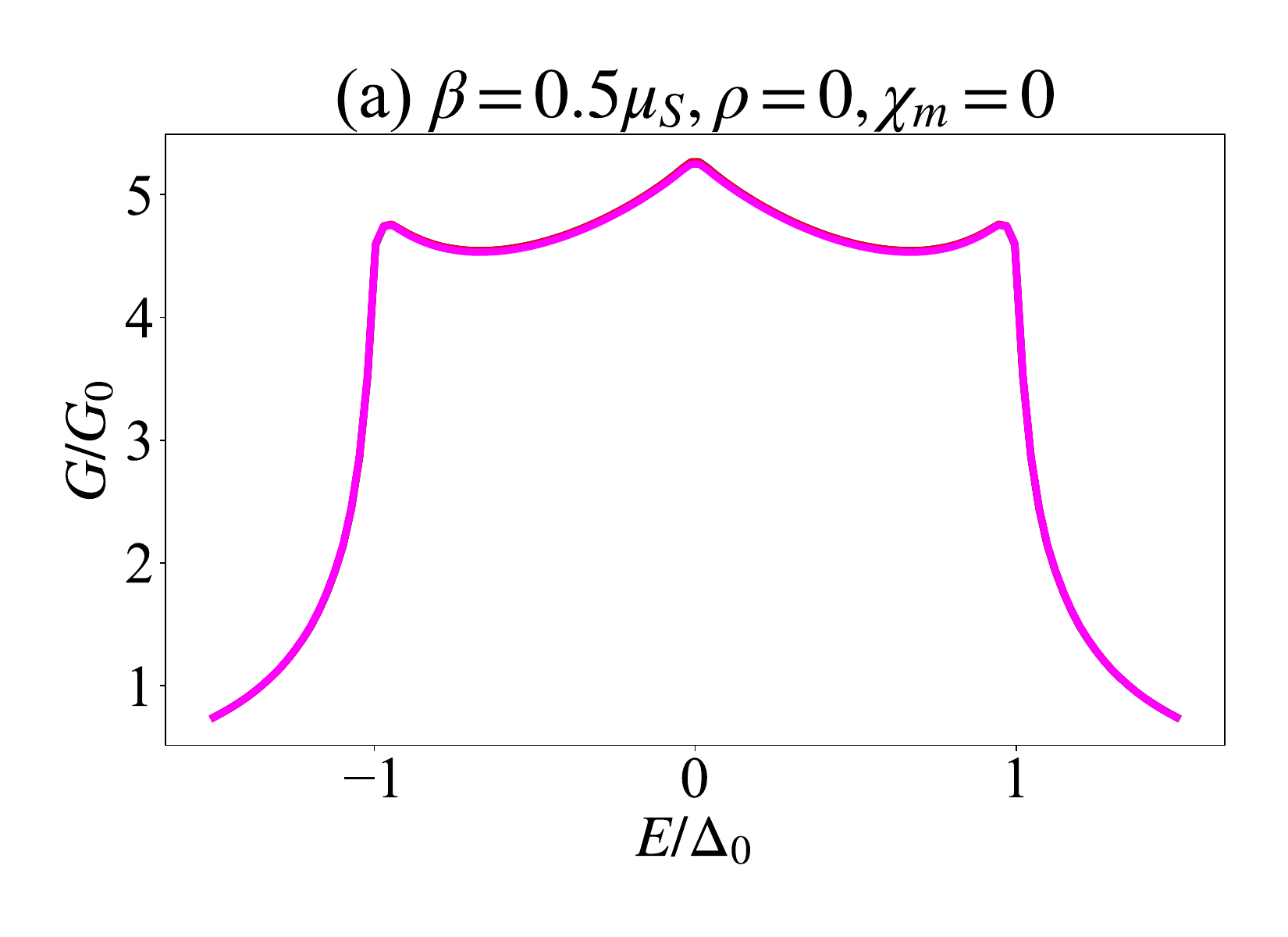}
\hspace{-0.42cm}
\vspace{-0.15cm}
\includegraphics[scale=0.22]{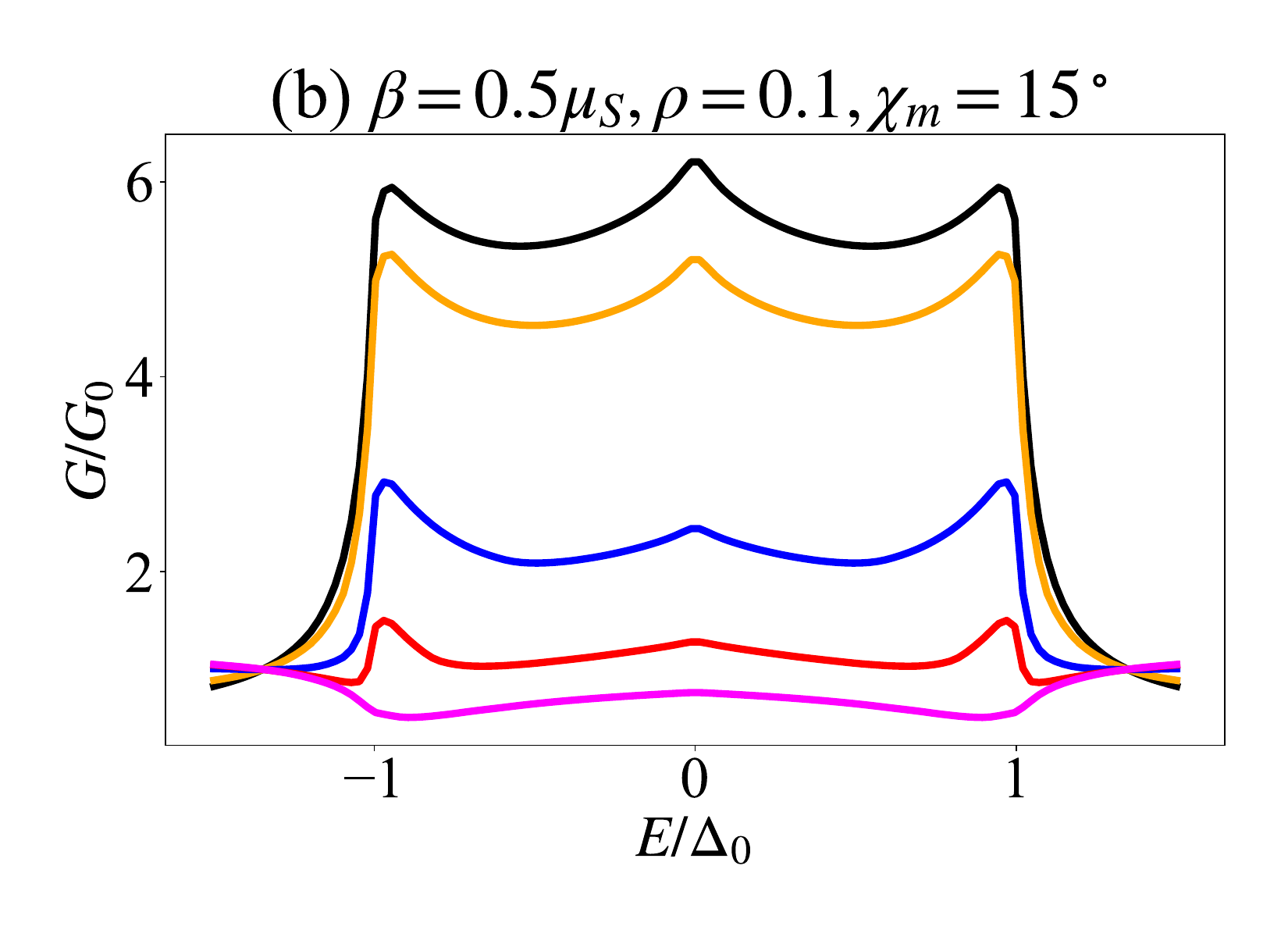}
\hspace{-0.42cm}
\vspace{-0.15cm}
\includegraphics[scale=0.22]{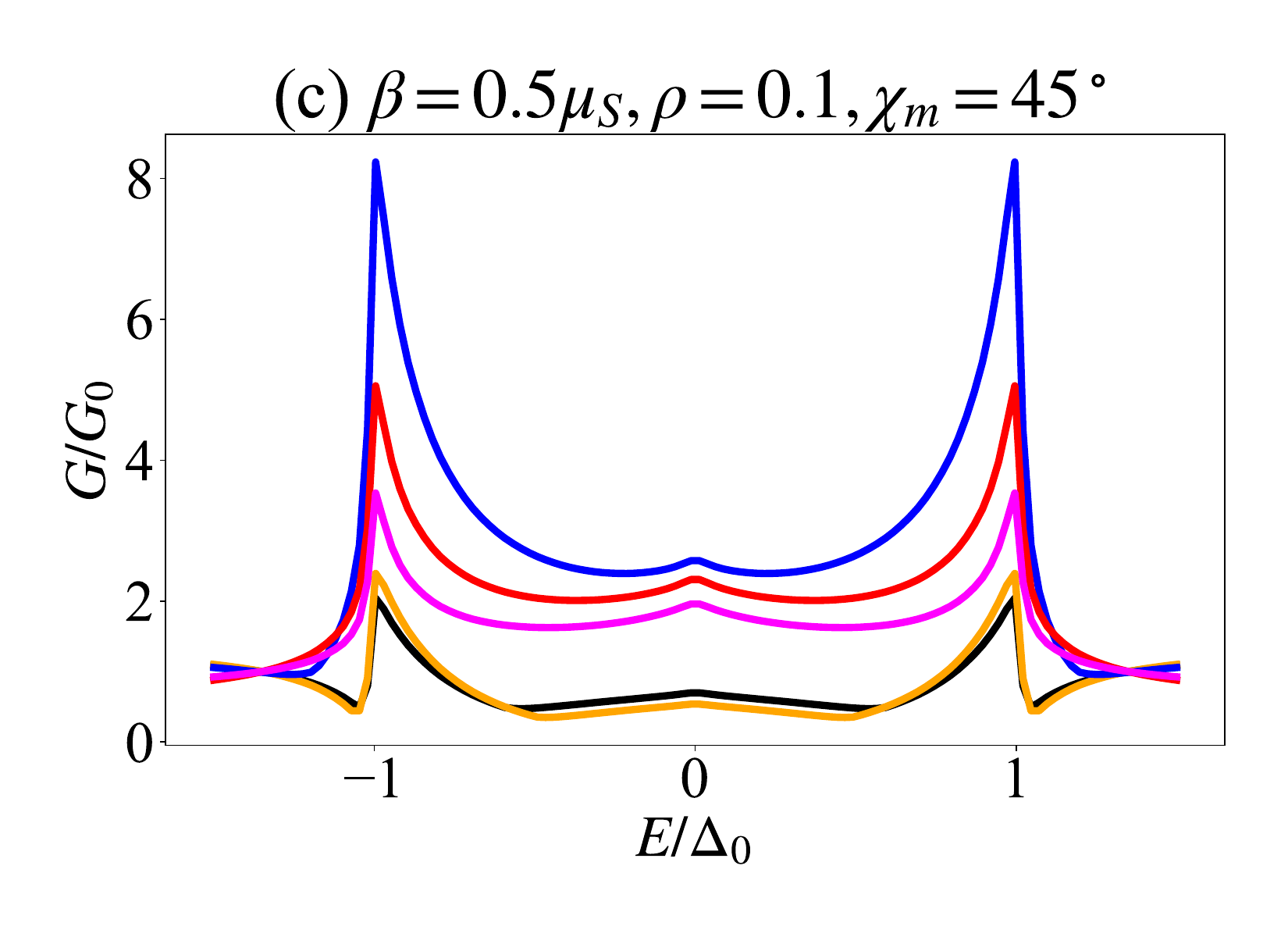}
\hspace{-0.45cm}
\vspace{-0.15cm}}
\centerline{
\includegraphics[scale=0.22]{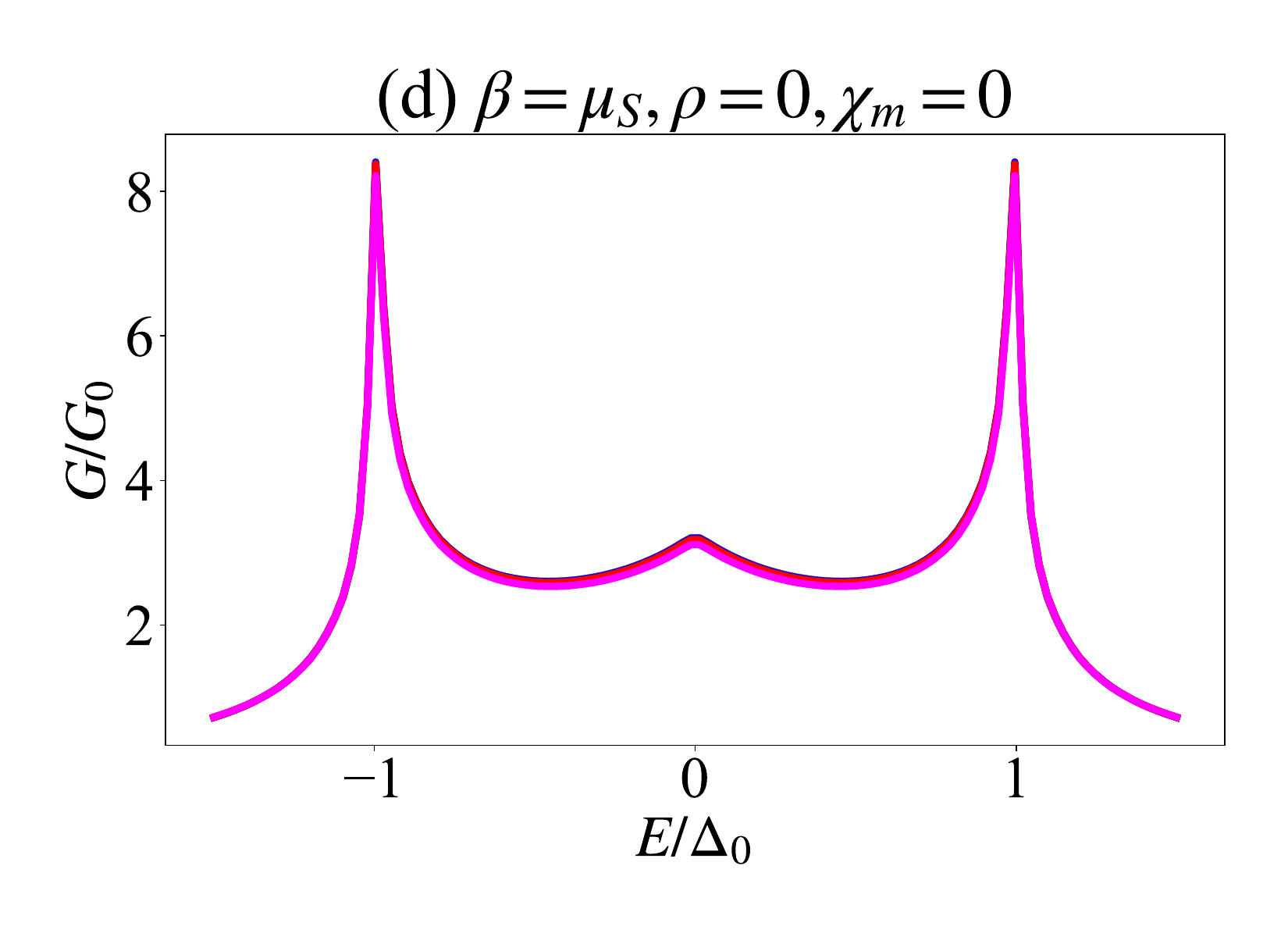}
\hspace{-0.42cm}
\vspace{-0.15cm}
\includegraphics[scale=0.22]{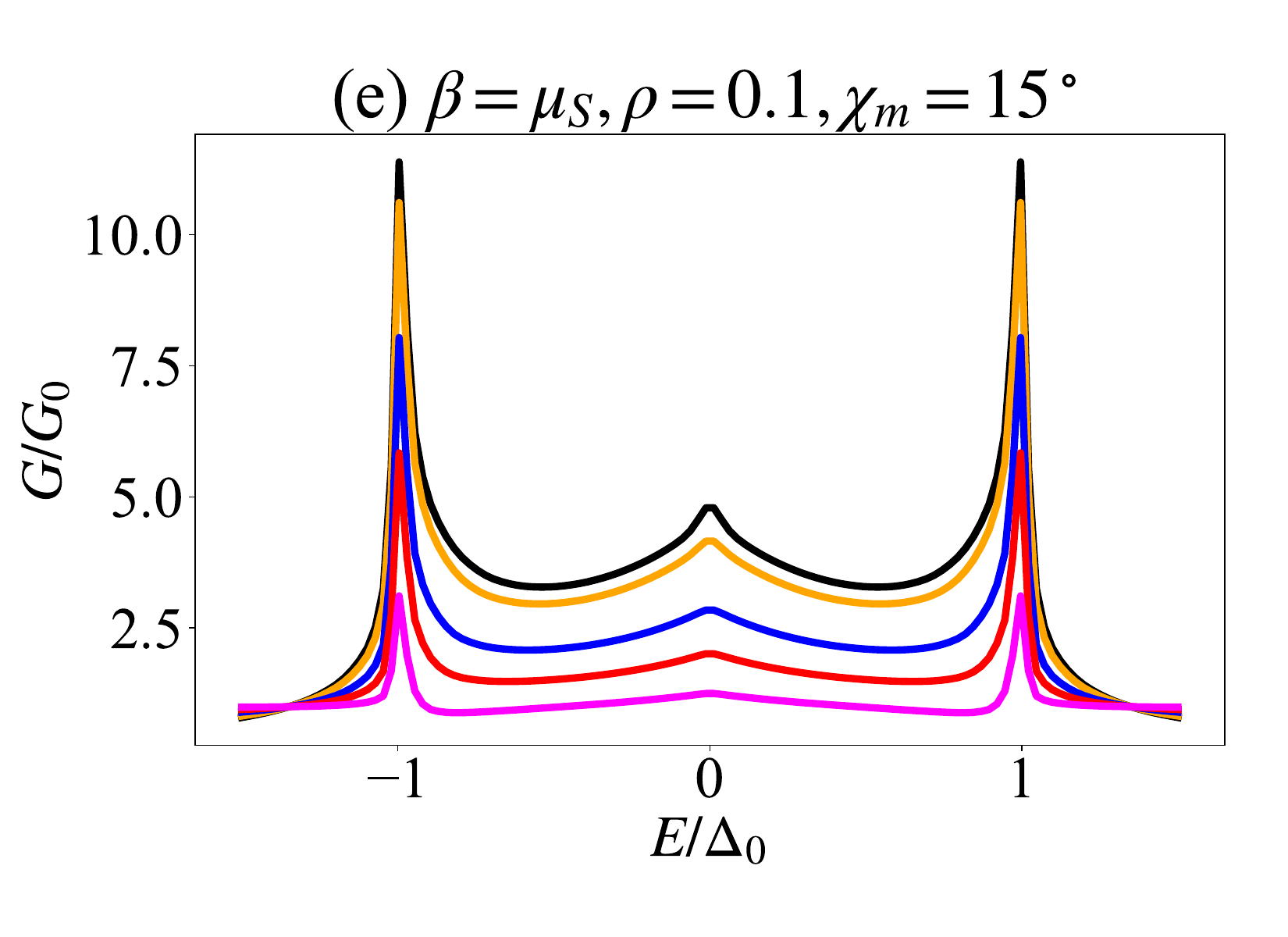}
\hspace{-0.42cm}
\vspace{-0.15cm}
\includegraphics[scale=0.22]{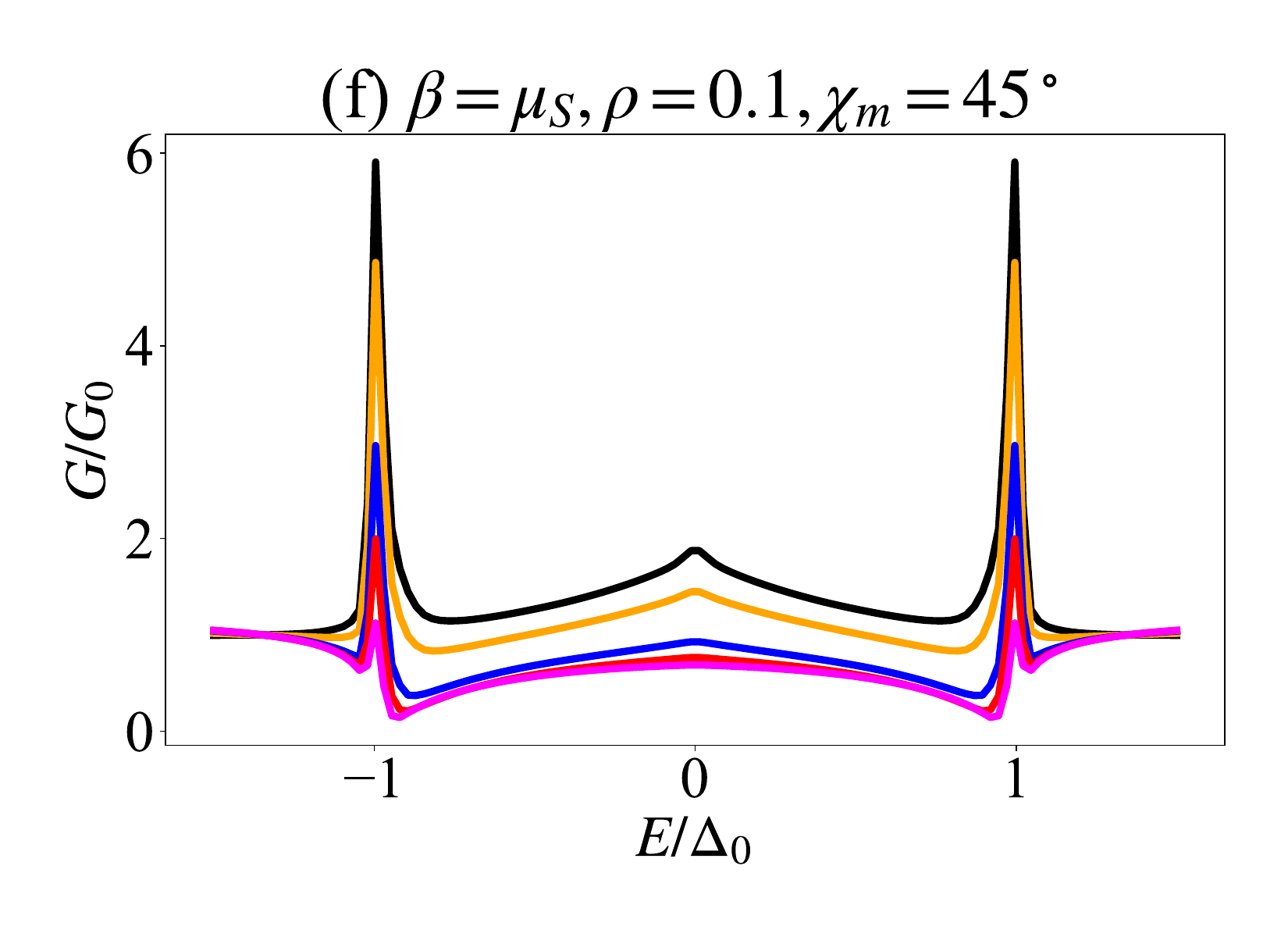}
\hspace{-0.45cm}
\vspace{-0.15cm}}
\centerline{
\includegraphics[scale=0.22]{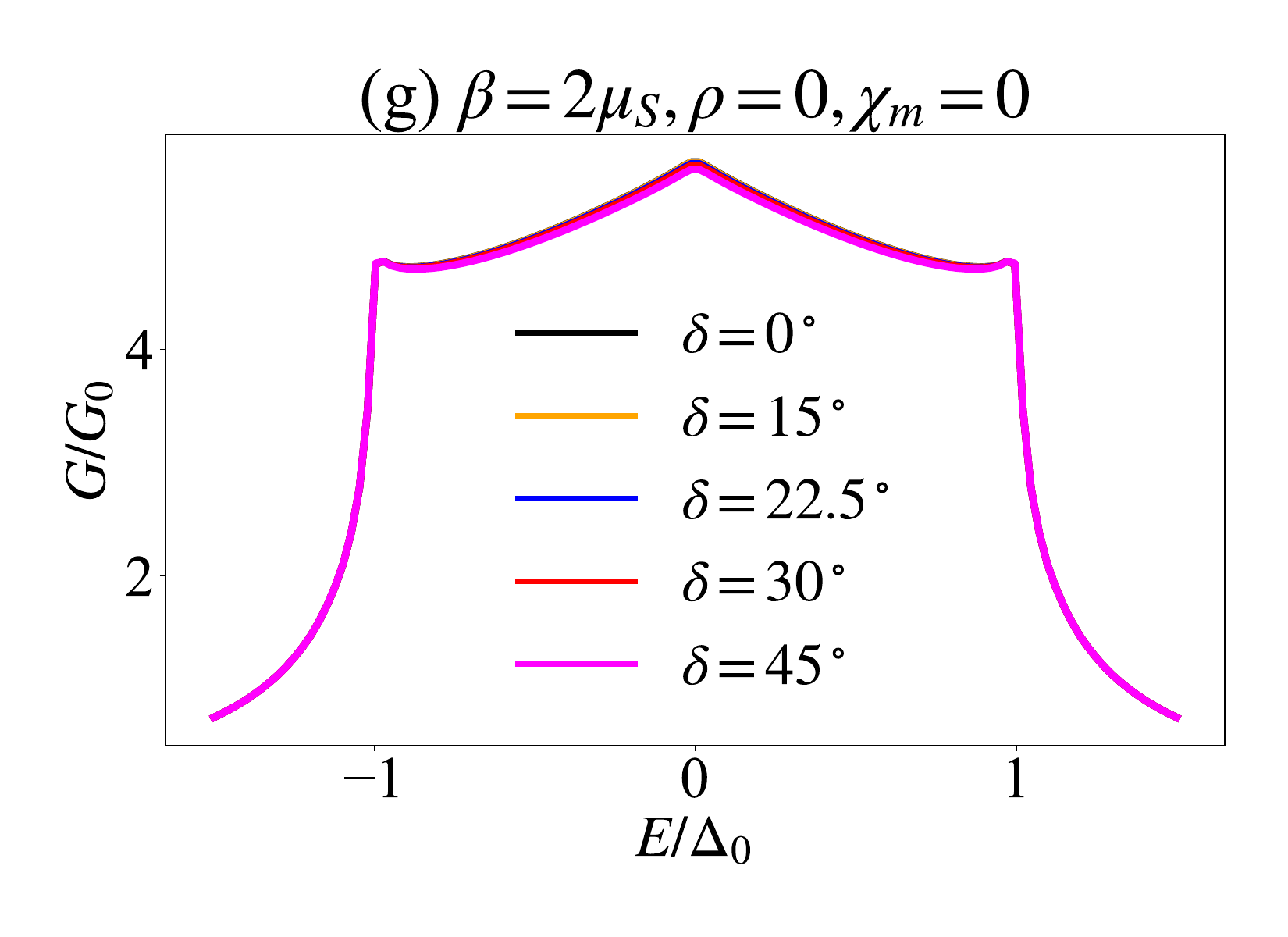}
\hspace{-0.42cm}
\vspace{-0.2cm}
\includegraphics[scale=0.22]{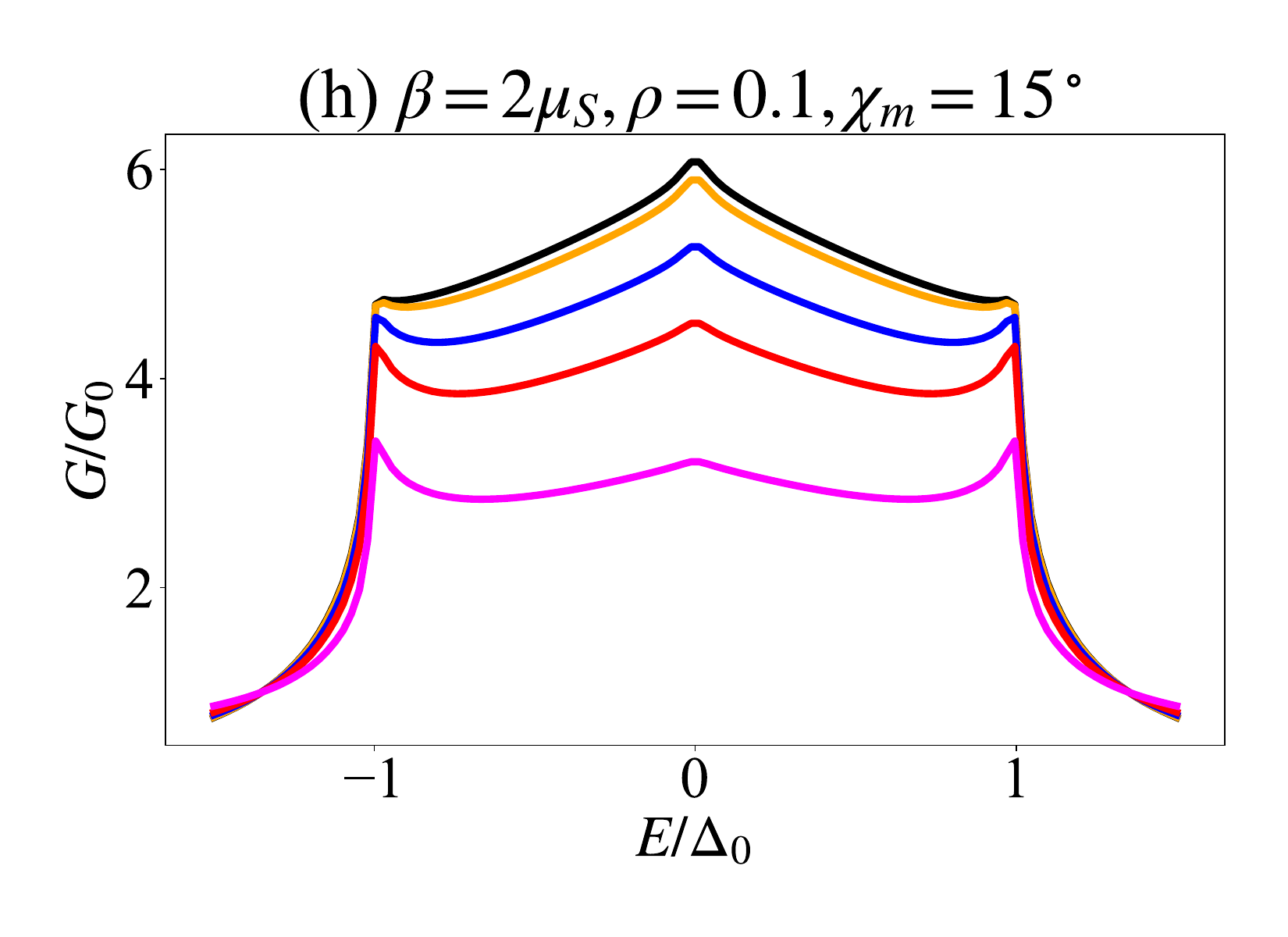}
\hspace{-0.42cm}
\vspace{-0.2cm}
\includegraphics[scale=0.22]{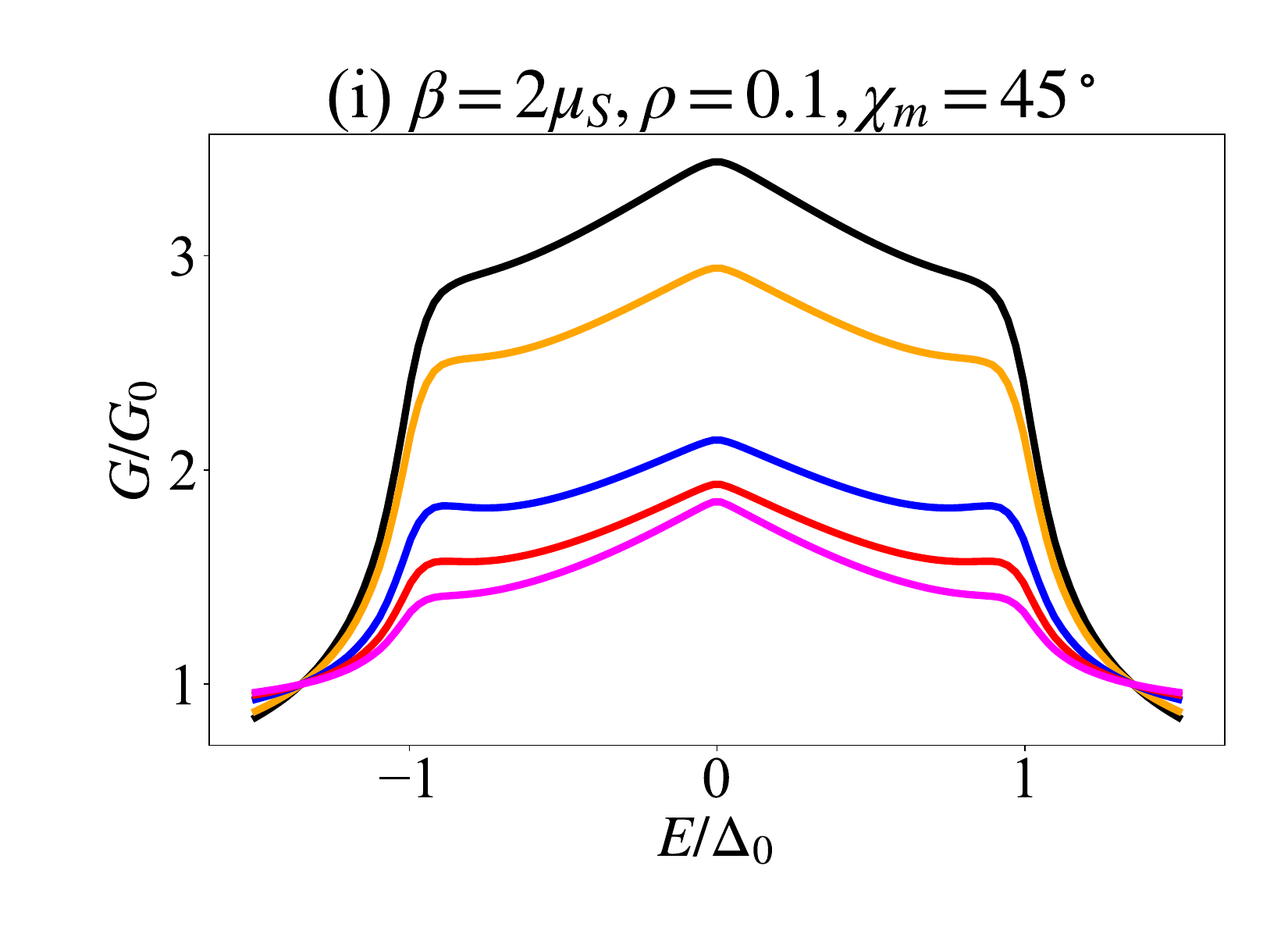}
\hspace{-0.42cm}
\vspace{-0.2cm}}
\caption{Charge conductance spectra for different AM orientation angles ($\delta$) considering $Z_0 = 0.1$ and $\xi_m = 45^\circ$. Panels (a)–(c), (d)–(f), and (g)–(i) correspond to single band ISC ($\beta = 0.5\mu_S$), intermediate ($\beta = \mu_S$) and double band ISC ($\beta = 2\mu_S$) respectively. The interface magnetic moment is $\rho = 0$ and orientation angle $\chi_m = 0^\circ$ in the left column representing non-magnetic barrier while for the middle and right column are for magnetic barrier with $\rho = 0.1$  considering $\chi_m = 15^\circ$ and $45^\circ$ respectively.}
\label{fig2}
\end{figure*} 

\begin{figure*}[t]
\centerline{
\includegraphics[scale=0.22]{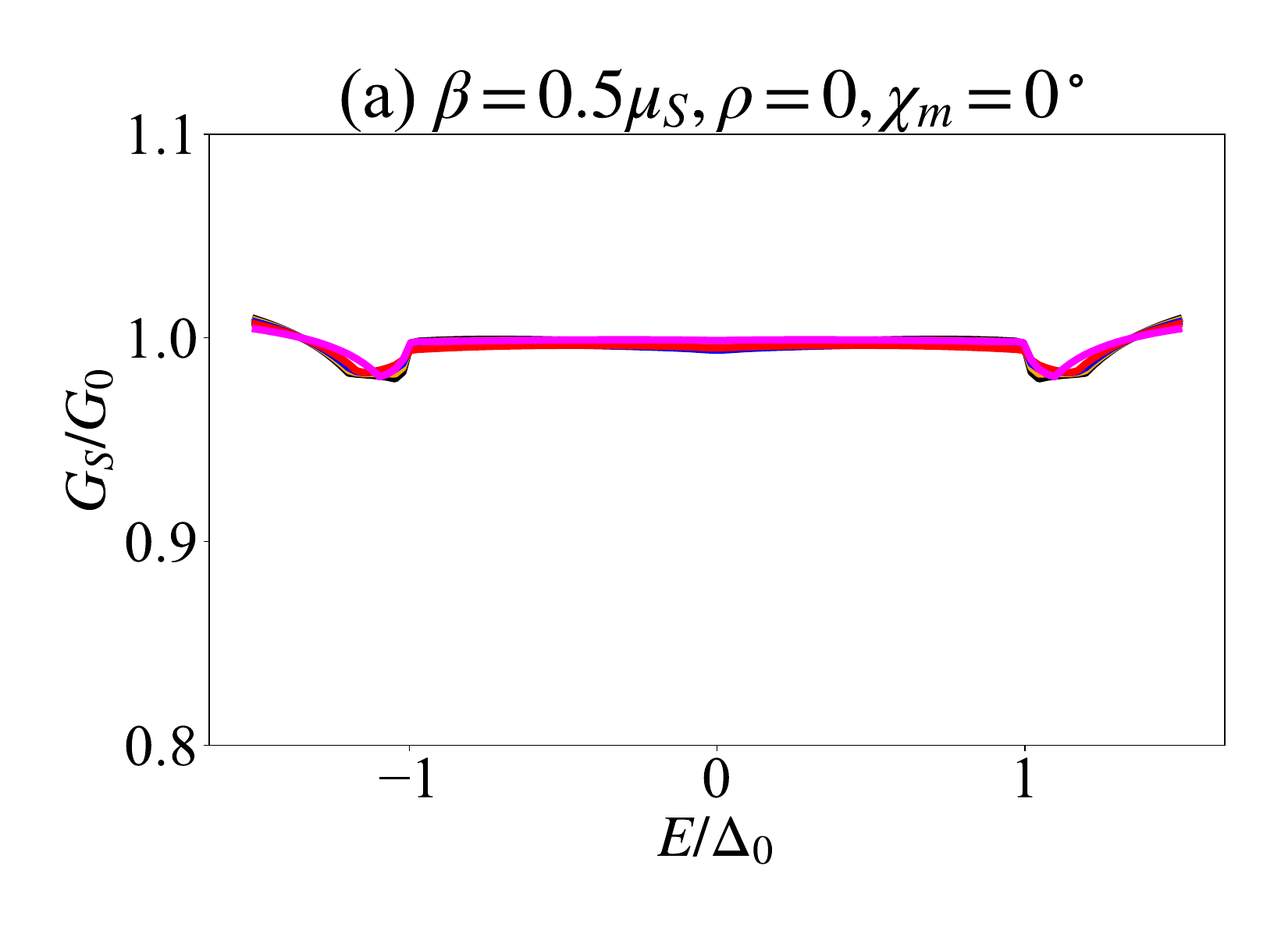}
\hspace{-0.42cm}
\vspace{-0.15cm}
\includegraphics[scale=0.22]{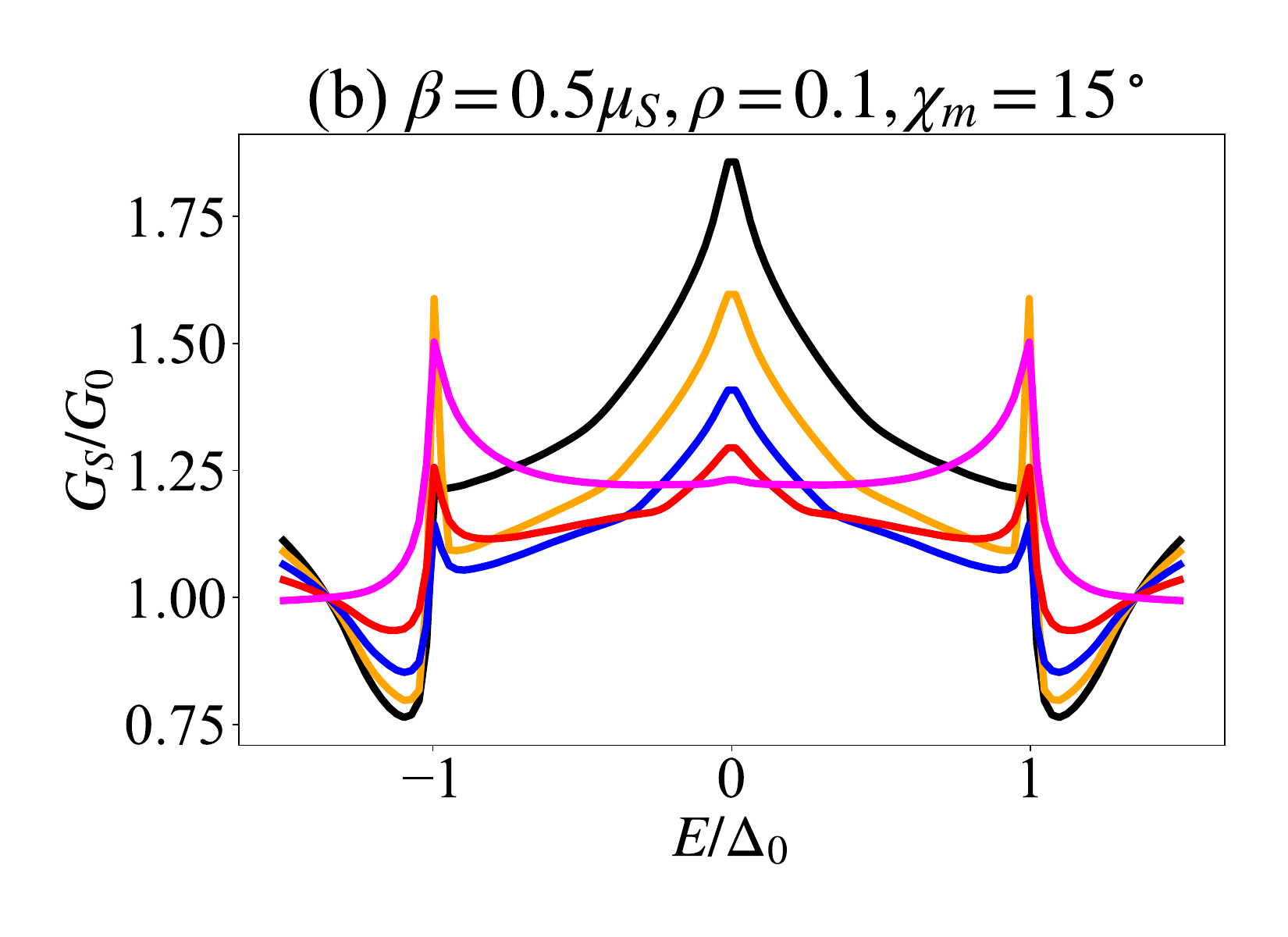}
\hspace{-0.42cm}
\vspace{-0.15cm}
\includegraphics[scale=0.22]{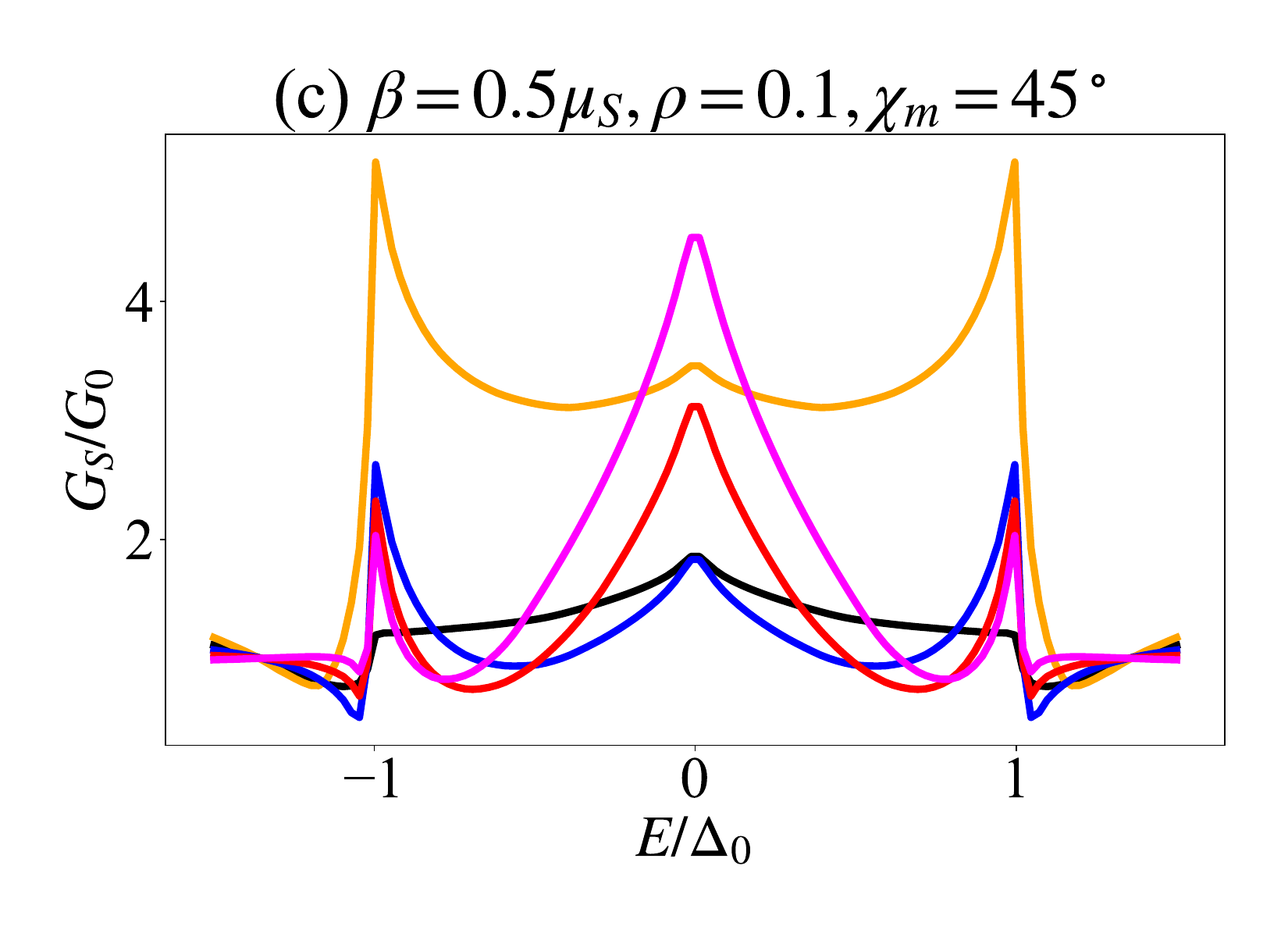}
\hspace{-0.45cm}
\vspace{-0.15cm}}
\centerline{
\includegraphics[scale=0.22]{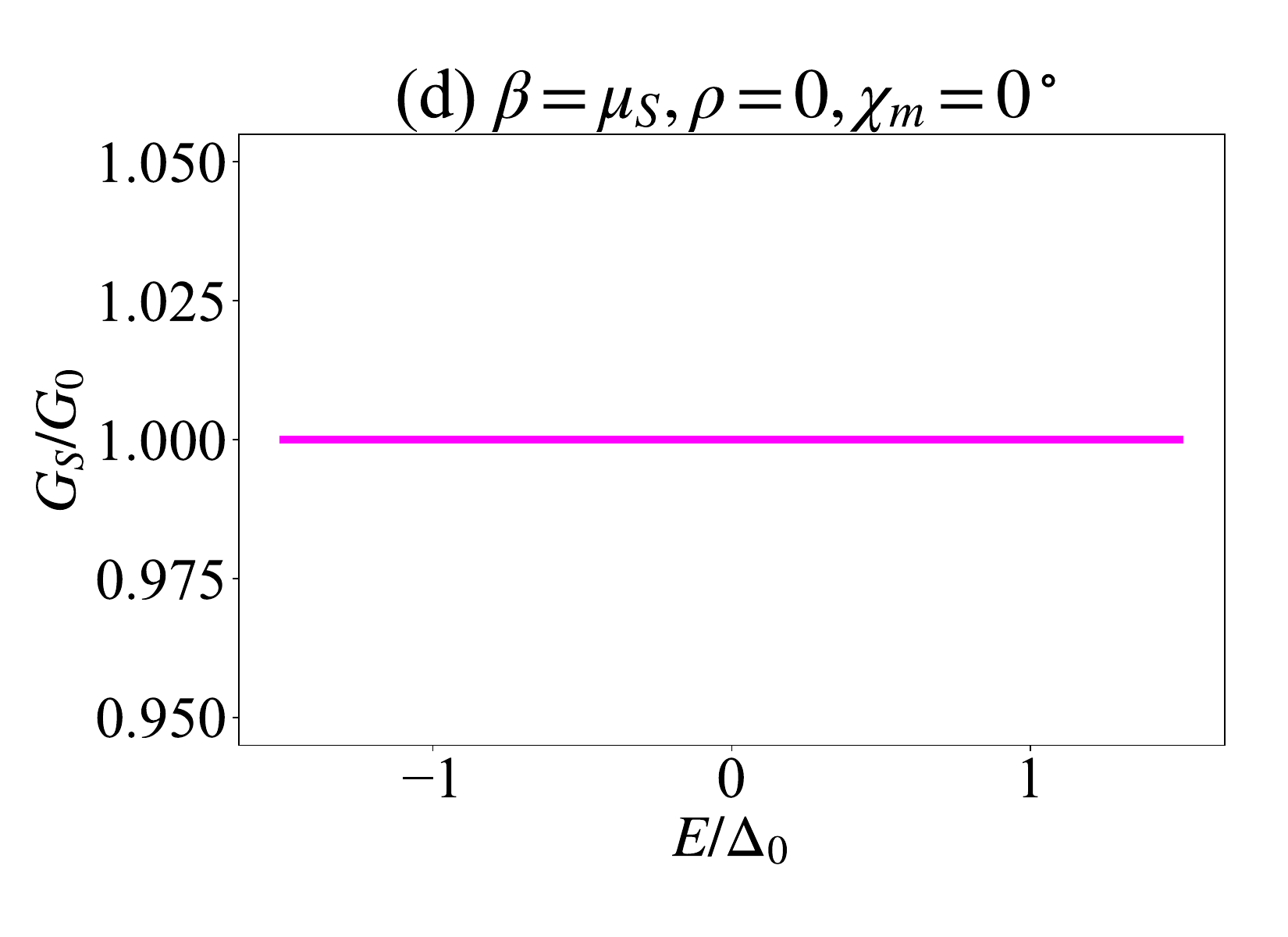}
\hspace{-0.42cm}
\vspace{-0.15cm}
\includegraphics[scale=0.22]{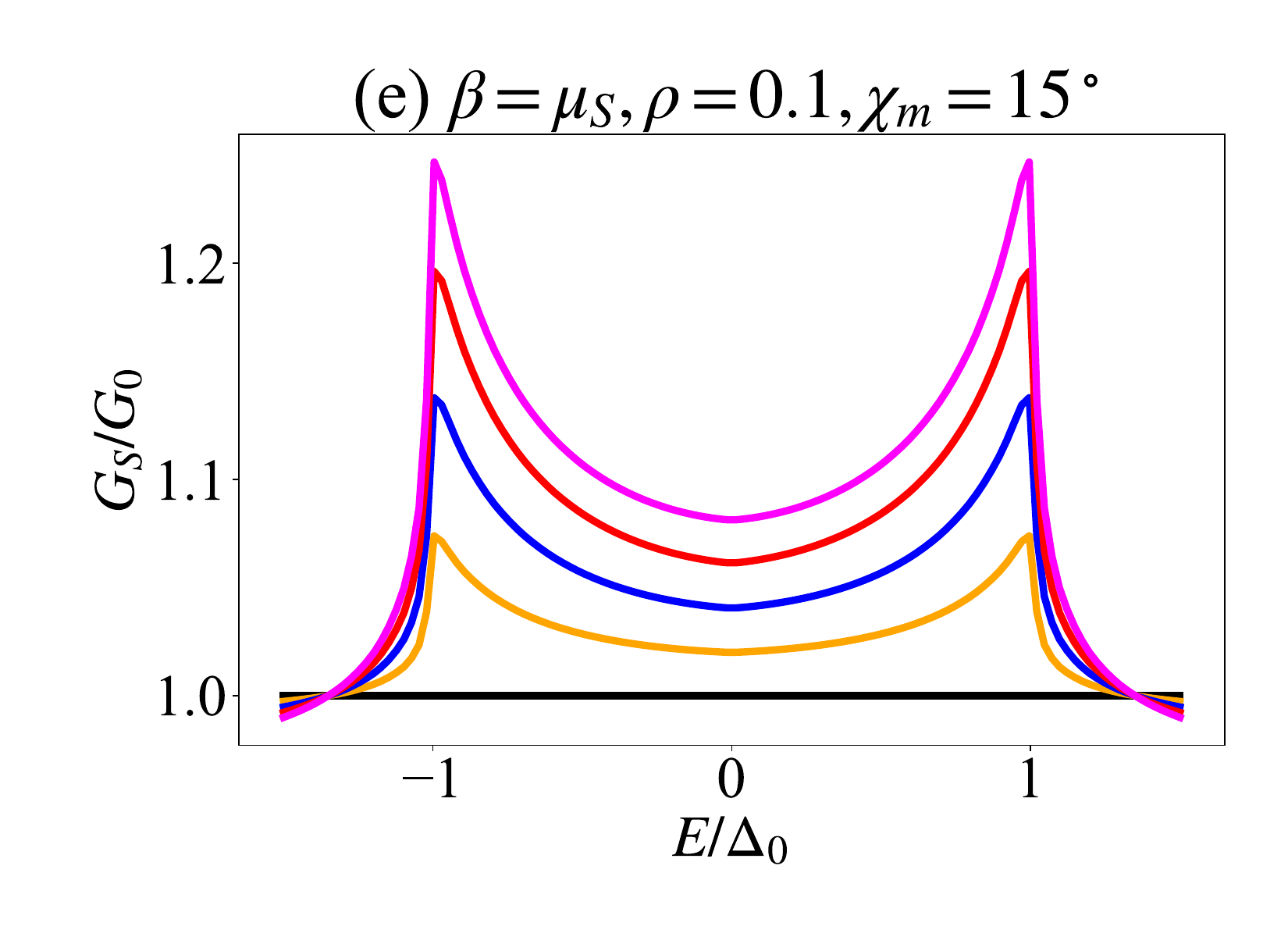}
\hspace{-0.42cm}
\vspace{-0.15cm}
\includegraphics[scale=0.22]{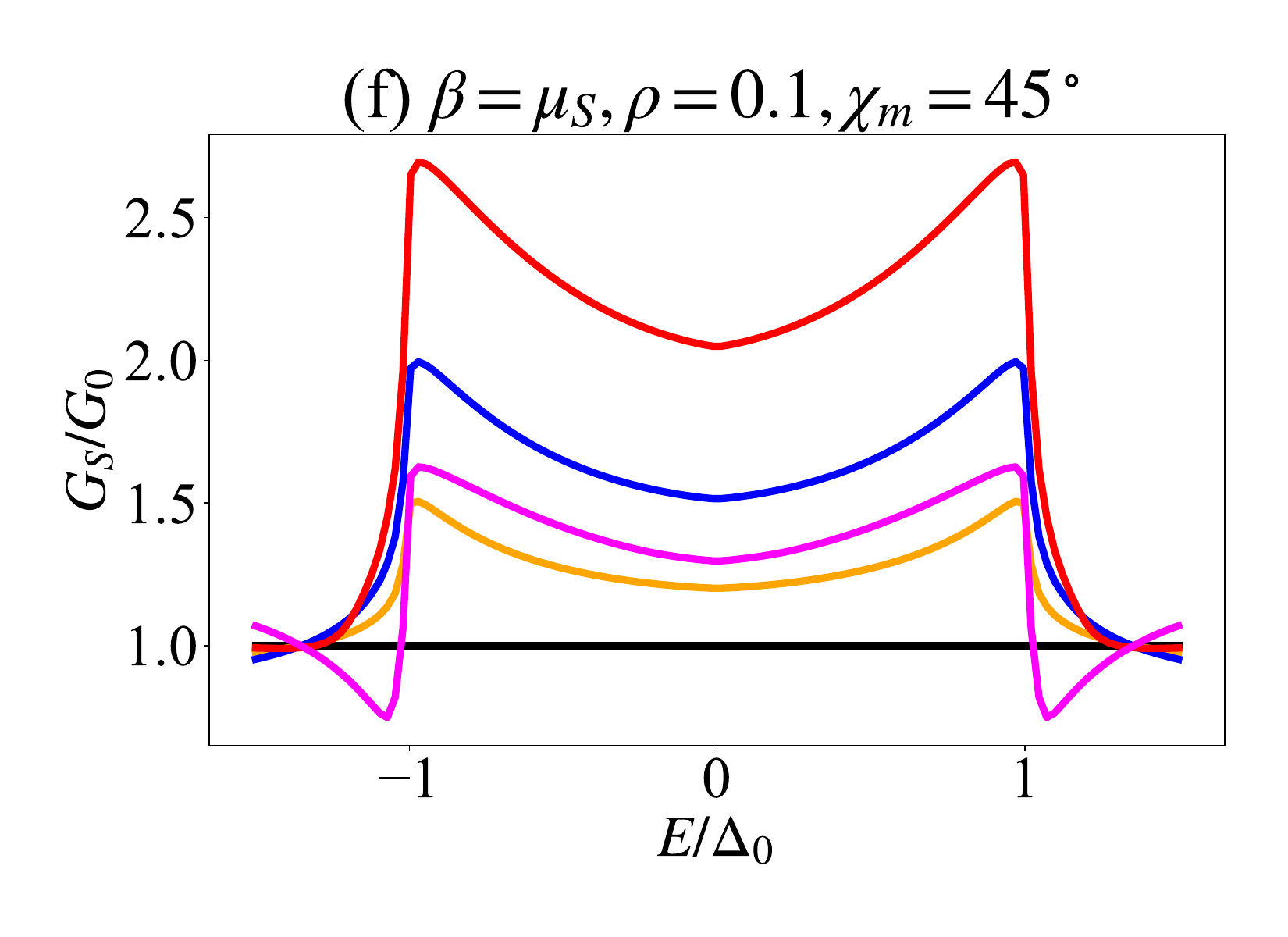}
\hspace{-0.45cm}
\vspace{-0.15cm}}
\centerline{
\includegraphics[scale=0.22]{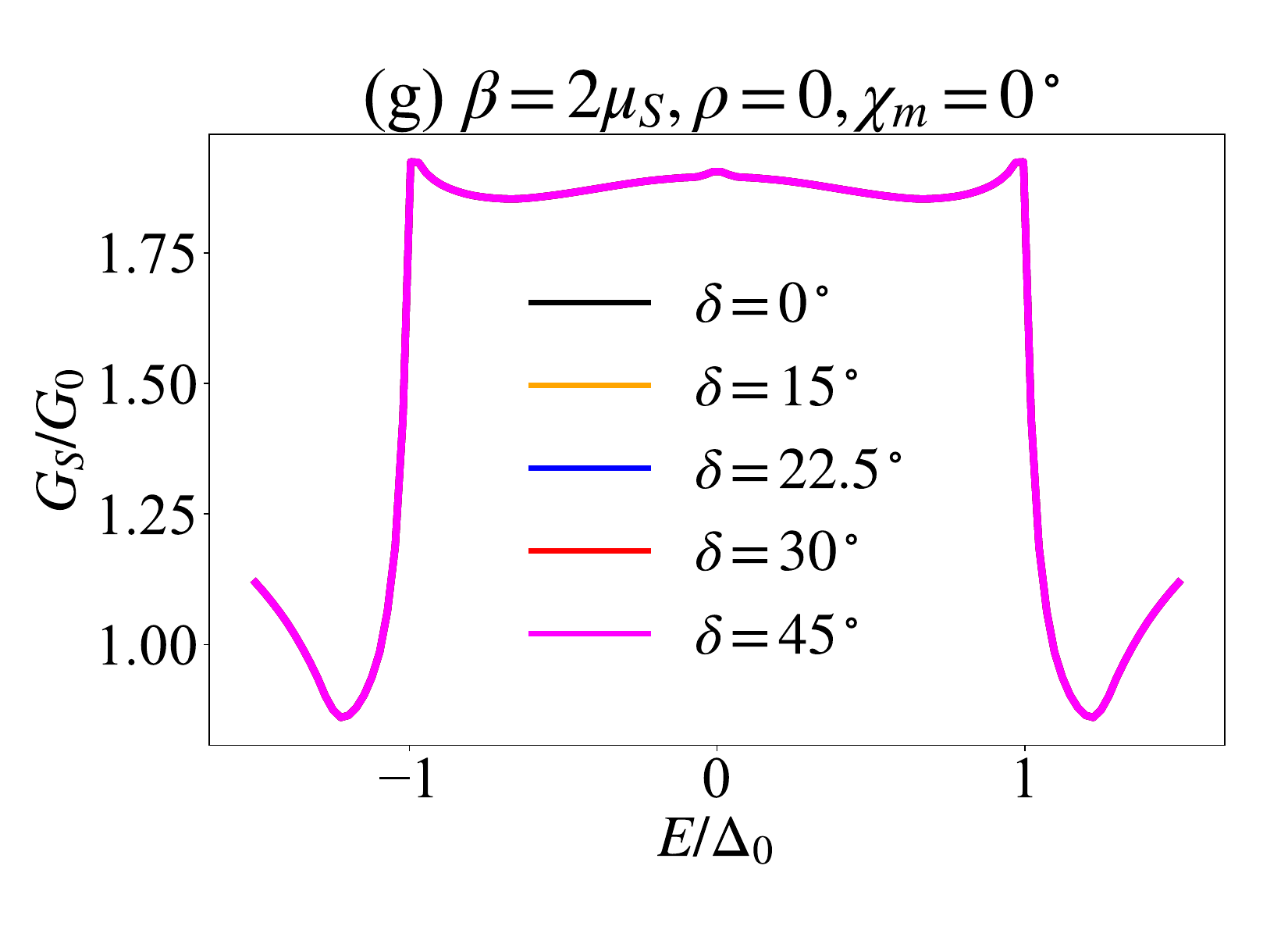}
\hspace{-0.42cm}
\vspace{-0.2cm}
\includegraphics[scale=0.22]{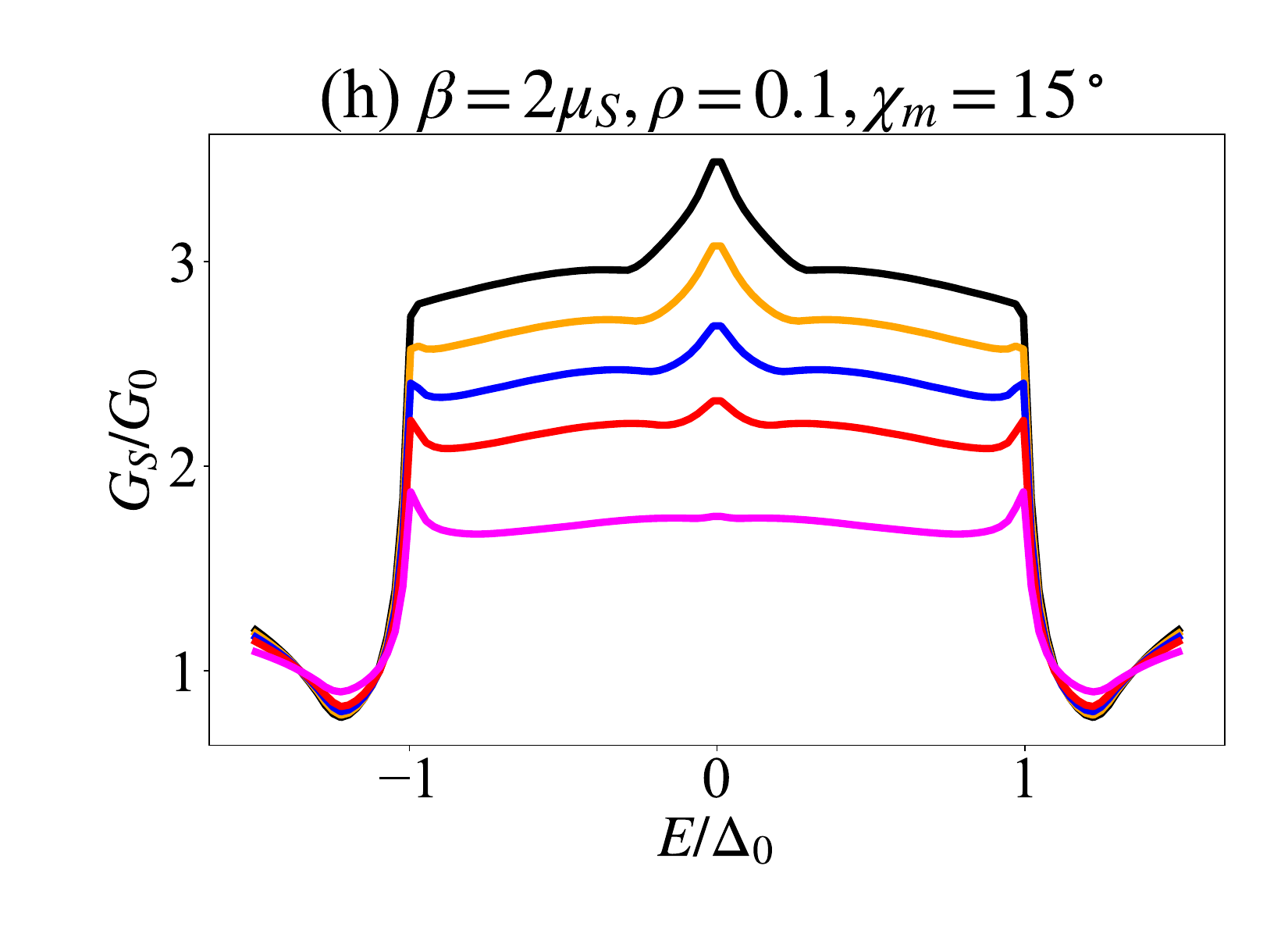}
\hspace{-0.42cm}
\vspace{-0.2cm}
\includegraphics[scale=0.22]{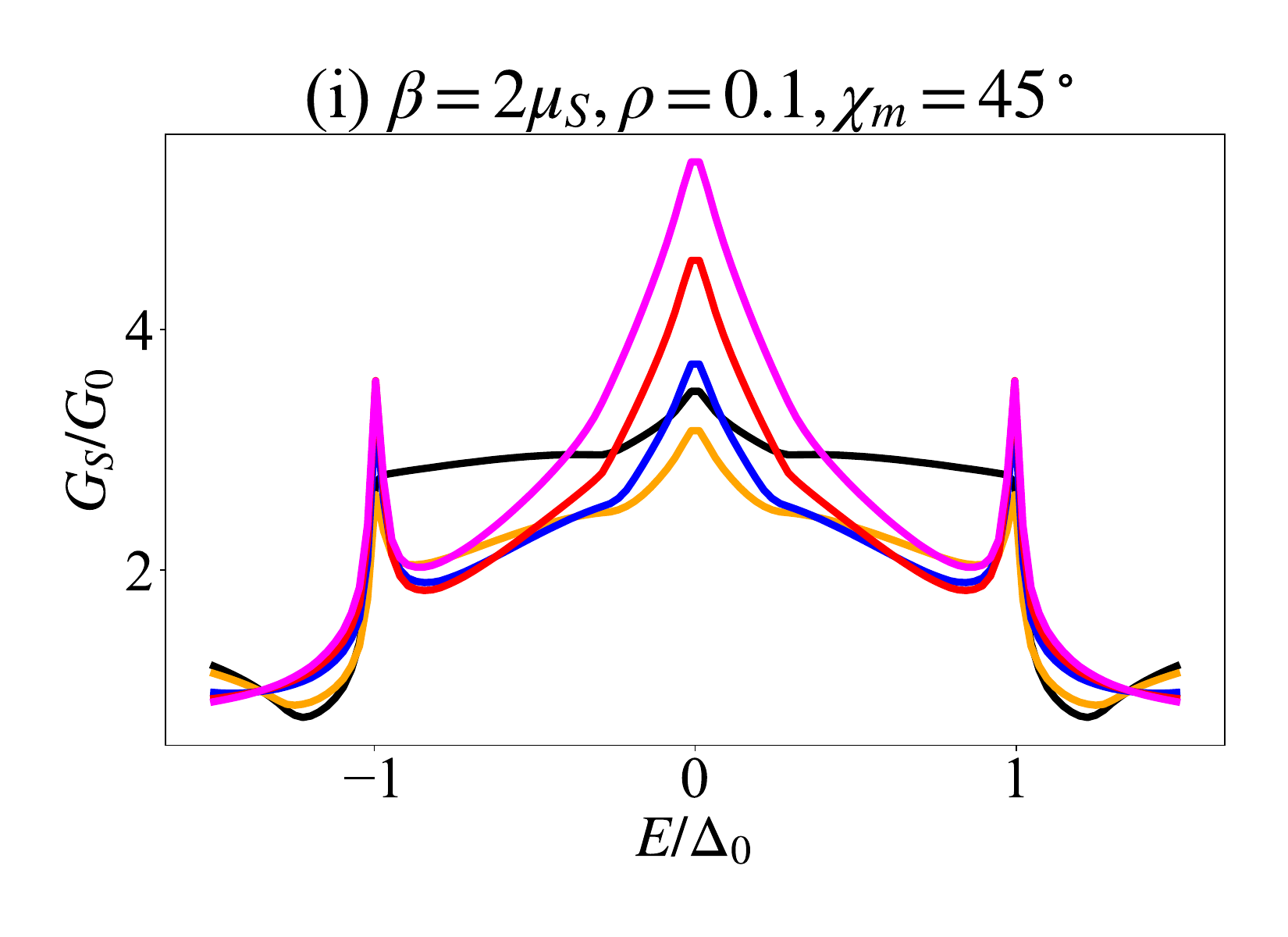}
\hspace{-0.42cm}
\vspace{-0.2cm}}
\caption{Spin conductance spectra for different AM orientation angles ($\delta$) considering $Z_0 = 0.1$ and $\xi_m = 45^\circ$. Panels (a)–(c), (d)–(f), and (g)–(i) correspond to single band ISC ($\beta = 0.5\mu_S$), intermediate ($\beta = \mu_S$), and double band ISC ($\beta = 2\mu_S$) respectively. The interface magnetic moment is $\rho = 0$ and orientation angle $\chi_m = 0^\circ$ in the left column representing non-magnetic barrier while for the middle and right column are for magnetic barrier with $\rho = 0.1$  considering $\chi_m = 15^\circ$ and $45^\circ$ respectively.}
\label{fig3}
\end{figure*}

\section{Results and Discussion}
\subsection{Charge Conductance}

Within the Blonder--Tinkham--Klapwijk (BTK) formalism, the angle-resolved charge conductance in the helicity basis is given by~\cite{Sun2023}
\begin{equation}
G(E,\theta) = \sum_{s,s'=\pm}
\left[
\delta_{ss'} 
- \frac{v^{e}_{s'}}{v^{e}_{s}} |r_{ss'}(E,\theta)|^2 
+ \frac{v^{h}_{s'}}{v^{e}_{s}} |a_{ss'}(E,\theta)|^2
\right]
\label{eq16}
\end{equation}
where $v^{e(h)}_s$ denotes the group velocity of electron-like (hole-like) quasiparticles in helicity channel $s$ and defined as
\begin{equation}
\label{eq17}
v^{e(h)}_s = \frac{1}{\hbar} \frac{\partial E^{e(h)}_s}{\partial k_x}.
\end{equation}

The total normalized conductance is obtained by angular averaging over all incident angles~\cite{Sun2023},
\begin{equation}
\frac{G(E)}{G_N} =
\int_{-\pi/2}^{\pi/2} G(E,\theta)\cos\theta\, d\theta,
\label{eq18}
\end{equation}
where $G_N$ is the normal-state conductance of the junction.

Fig.~\ref{fig2} presents the variation of charge conductance  with quasiparticle energy $E/\Delta_0$ for different
AM orientations $\delta$ and spin-active parameters $\rho$ and $\chi_m$ considering $Z_0 = 0.1$ and $\xi_m = 45^\circ$. It is to be noted that we fix $\xi_m$ and $Z_0$ throughout this work, since variations in $\xi_m$ produce only minor quantitative changes without altering the qualitative transport behavior. A small but finite scalar barrier is chosen to maintain sufficient interface transparency, as larger $Z_0$ suppresses Andreev reflection and weakens the spin-active interfacial effects.
In absence of spin-active interfaces i.e., for $\rho = 0$ and $\chi_m = 0$, the conductance spectra remain nearly independent of $\delta$ for all combinations of $\beta$ and $\mu_S$, as observed from Figs.~\ref{fig2}(a), \ref{fig2}(d), and \ref{fig2}(g). This behavior directly follows from Eq.~(\ref{eq16}), where the reflection amplitudes $r_{ss'}$ and $a_{ss'}$ remain diagonal in helicity space due to the absence of interfacial spin flipping and spin mixing. Consequently, the charge conductance receives independent contributions from each helicity channel. Although the AM hosts a momentum-dependent spin splitting, the field $\mathbf{g}(\mathbf{k},\delta)$ is even under momentum inversion. Consequently, in the absence of spin-active scattering, helicity-resolved contributions from opposite momenta enter symmetrically under angular averaging. This makes the conductance insensitive to the orientation $\delta$. However, a qualitatively distinct behavior emerges for spin-active interfaces, where the interface magnetic moment introduces both spin mixing and spin-flip processes. As a result, different helicity channels coupled, resulting in off-diagonal scattering amplitudes $r_{ss'}$ and $a_{ss'}$ with $s \neq s'$. Consequently, the Andreev reflection probability becomes strongly dependent on the relative orientation between the AM spin texture and the interface magnetic moment. This results in pronounced $\delta$-dependent conductance spectra, observed from Figs.~\ref{fig2}(b), \ref{fig2}(c), \ref{fig2}(e), \ref{fig2}(f), \ref{fig2}(h), and \ref{fig2}(i). Physically, the rotation of the AM spin texture via $\delta$ modifies the projection of the helicity dependent spin onto the interface moment direction. This alters the effective spin-dependent barrier potential experienced by quasiparticles, leading to strong anisotropy in the Andreev reflection probability and the emergence of $\delta$-dependent conductance features. It is observed that for $(\rho,\chi_m) = (0.1, 15^\circ)$, the maximum conductance is found for $\delta = 0^\circ$ across all regimes of $(\beta,\mu_S)$. In contrast, for $(\rho,\chi_m) = (0.1, 45^\circ)$, the angular dependence becomes more pronounced. In this condition, $\delta = 22.5^\circ$ display highest conductance in the single-band regime, while $\delta = 0^\circ$ continues to dominate in the intermediate and double band regimes. This behavior reflects the enhanced sensitivity of spin-selective transport to the relative spin alignment in the single band limit.

The evolution of the conductance spectra with increasing ISOC strength further reveals distinct transport regimes determined by the spin-resolved band structure of the ISC. For single band ISC with $\beta = 0.5\mu_s$, the system exhibits strongly spin-selective Andreev reflection, resulting in pronounced conductance anisotropy together with significant reshaping and asymmetry of the coherence peaks near $E \approx \pm \Delta_0$ as seen from Fig.~\ref{fig2}(a). 
In the intermediate regime i.e. for $\beta = \mu_S$, one spin band touches the Fermi level, resulting in a vanishing Fermi velocity for that channel and a reduced phase space for Andreev processes. This enhances the role of interfacial spin mixing, giving rise to significant peak reshaping and strong $\delta$-dependent anisotropy as observed from Figs.~\ref{fig2}(d)–\ref{fig2}(f).
In contrast, for double band ISC with $\beta  = 2 \mu_S$, both spin-split bands contribute to the transport, resulting in compensation between opposite spin channels giving rise to smoother conductance spectra with comparatively weak anisotropy even in the presence of spin-active scattering as seen from Figs.~\ref{fig2}(g)–\ref{fig2}(i). Overall, these results demonstrate that the charge conductance in AM/ISC junctions is governed by a nontrivial interplay of momentum-dependent spin splitting in the AM, interface-induced spin mixing, and the spin-selective band structure of the ISC.

\begin{figure*}[t]
\centerline{
\includegraphics[scale=0.22]{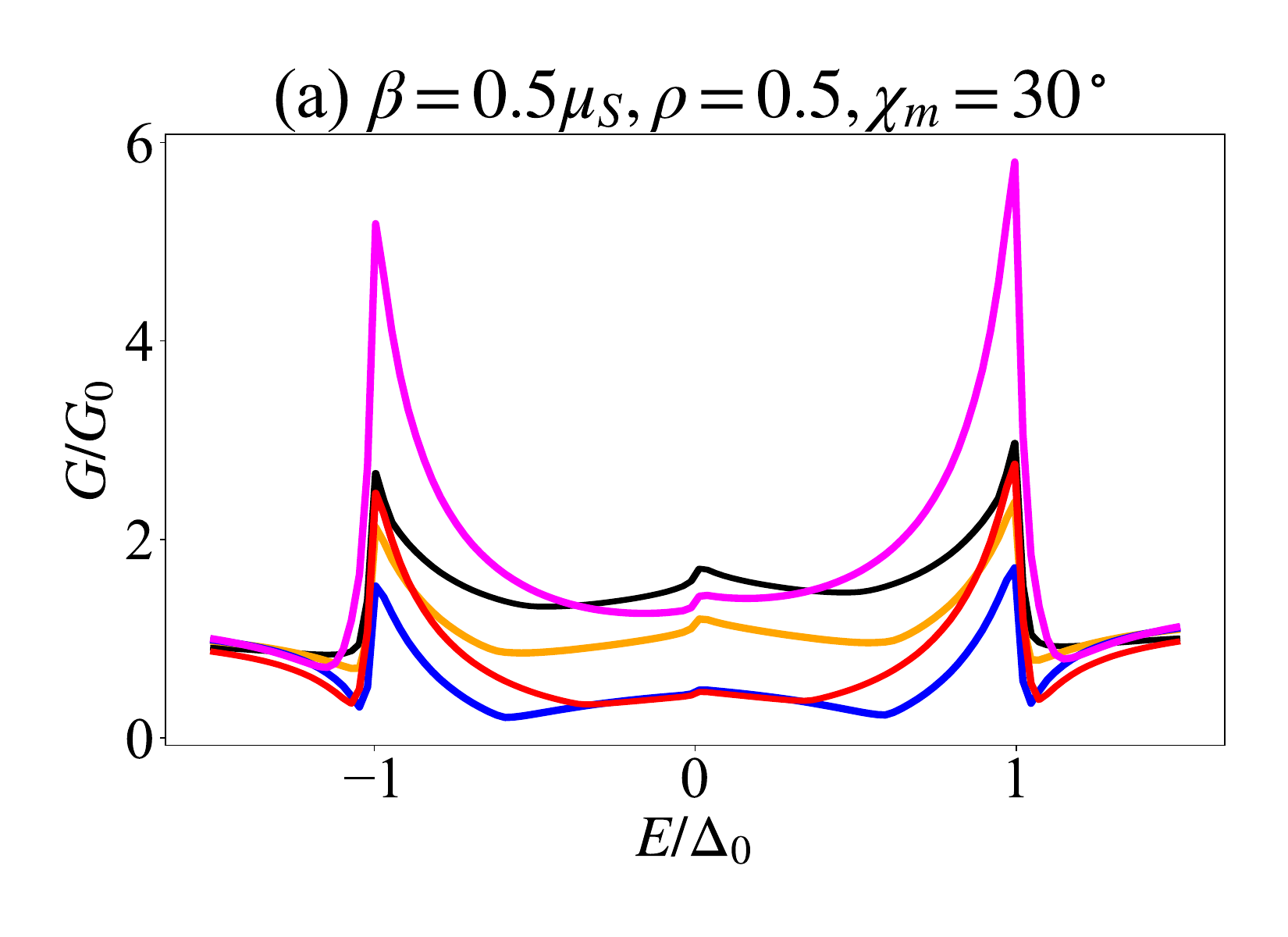}
\hspace{-0.42cm}
\vspace{-0.15cm}
\includegraphics[scale=0.22]{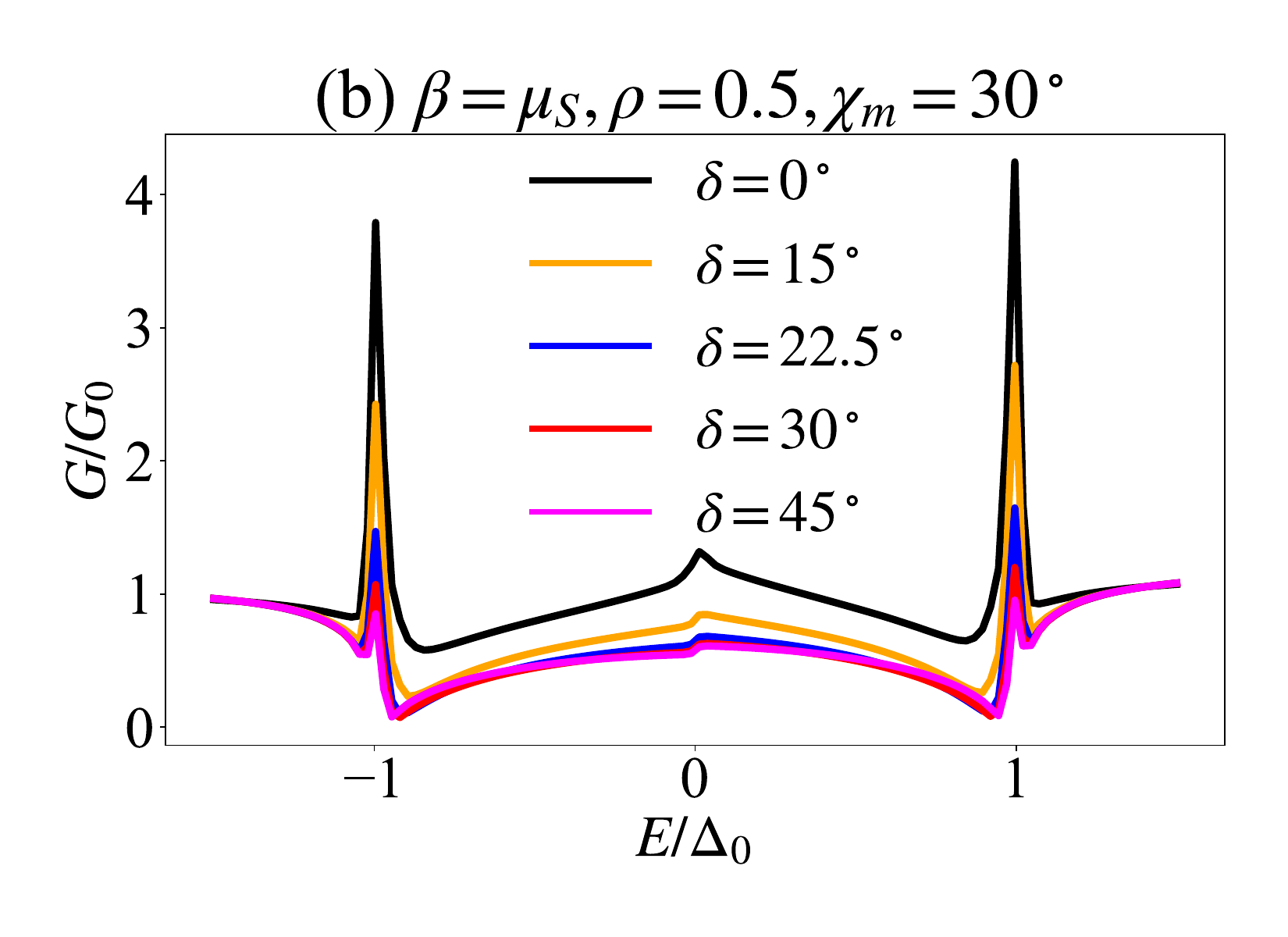}
\hspace{-0.42cm}
\vspace{-0.15cm}
\includegraphics[scale=0.22]{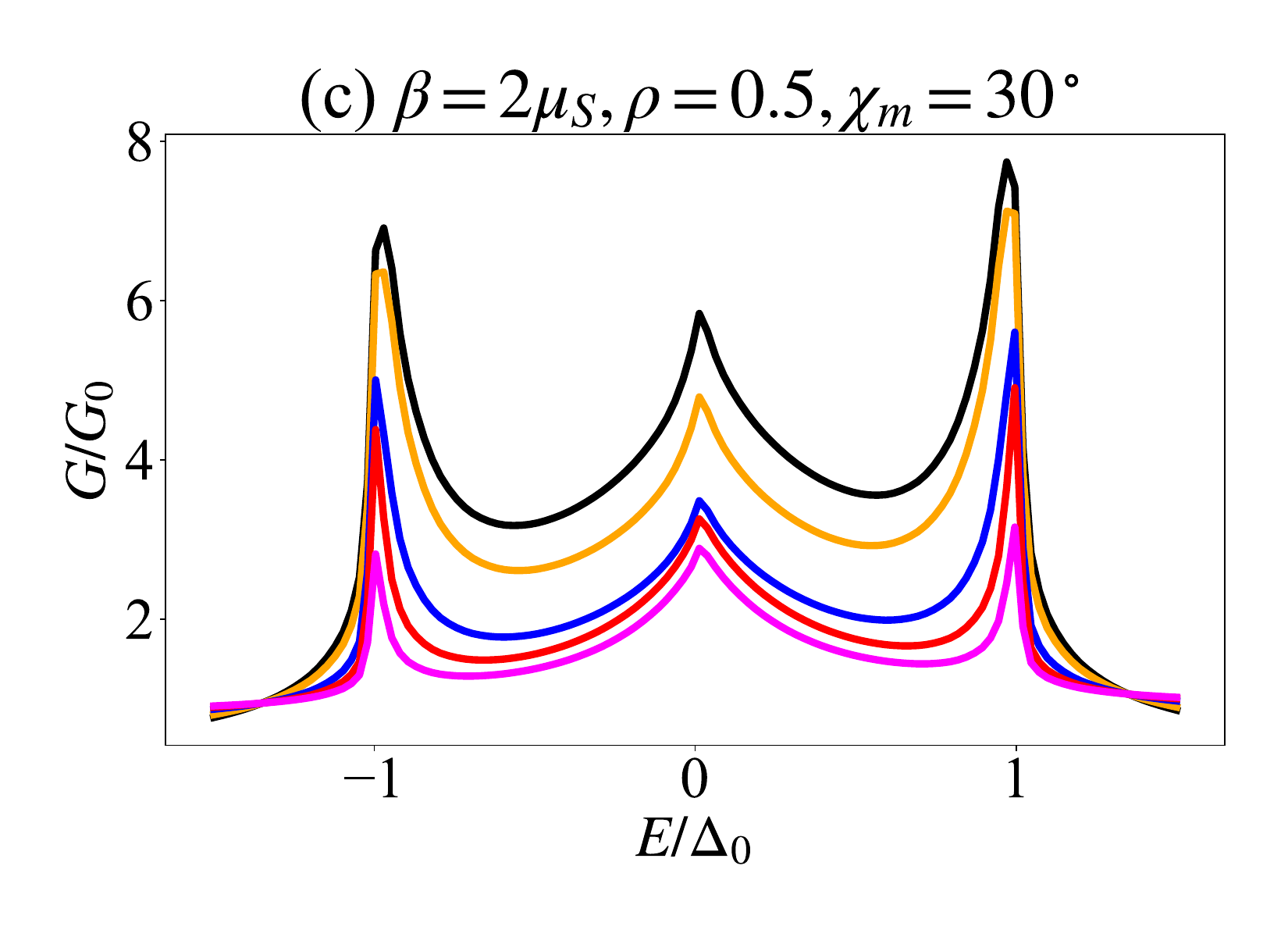}
\hspace{-0.45cm}
\vspace{-0.15cm}}
\centerline{
\includegraphics[scale=0.22]{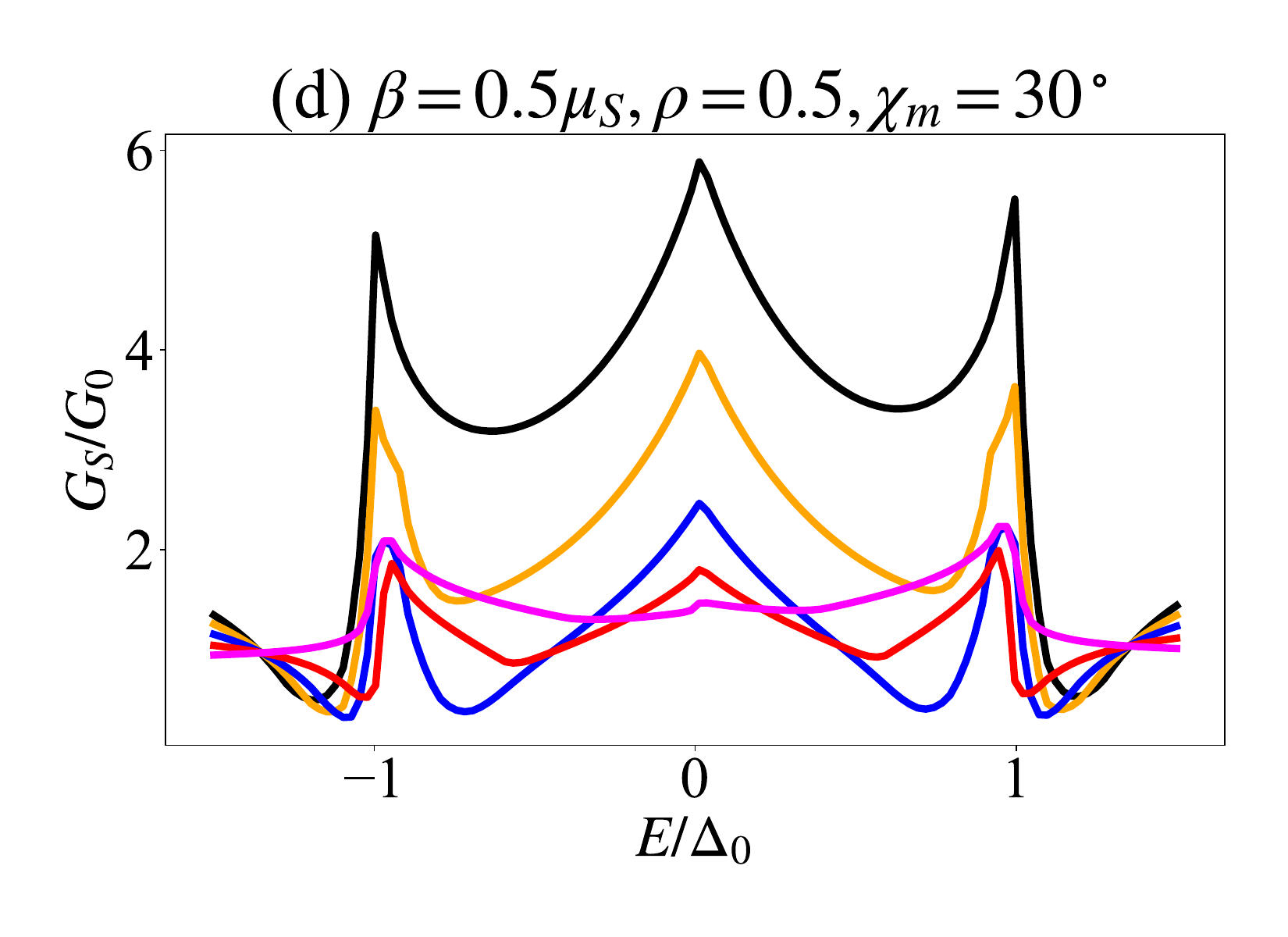}
\hspace{-0.42cm}
\vspace{-0.15cm}
\includegraphics[scale=0.22]{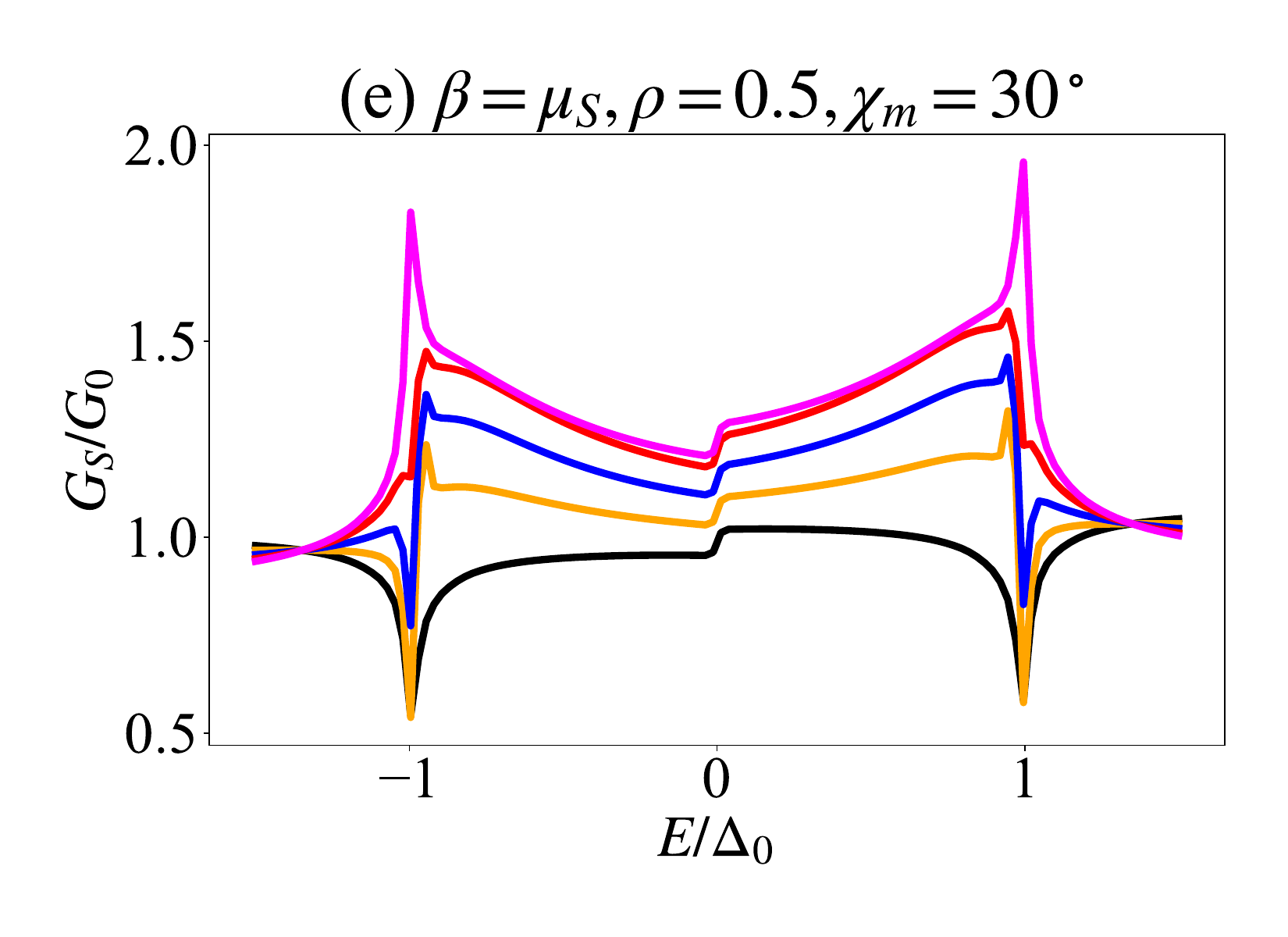}
\hspace{-0.42cm}
\vspace{-0.15cm}
\includegraphics[scale=0.22]{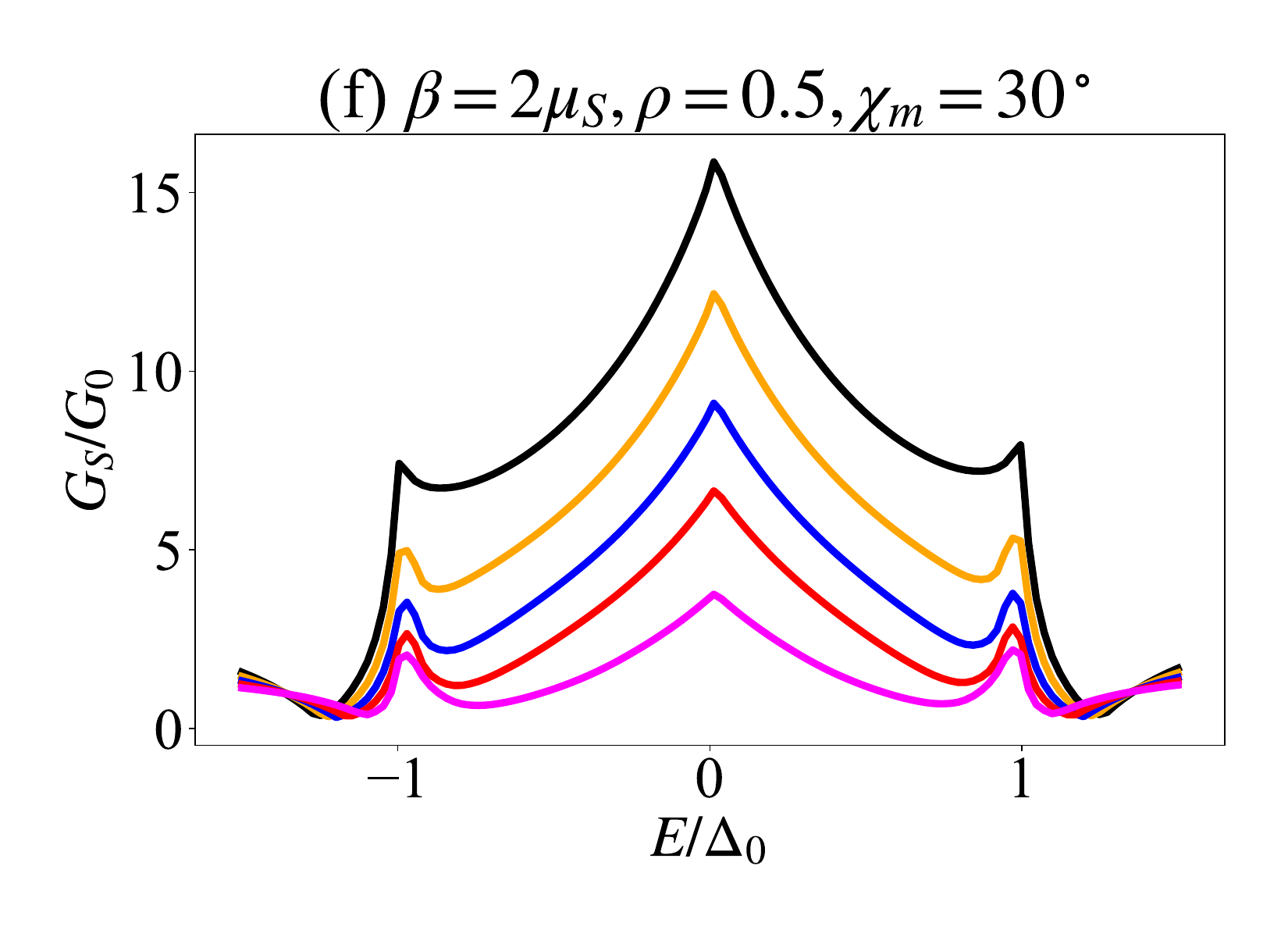}
\hspace{-0.45cm}
\vspace{-0.15cm}}
\caption{(Top panel) Charge conductance spectra, (Bottom panel) Spin conductance spectra for different values of $\delta$ considering $\rho = 0.5$ and $\chi_m = 30^\circ$ with $Z_0 = 0.1$ and $\xi_m = 45^\circ$. Panels (a)–(c) are for single band ISC ($\beta = 0.5\mu_S$) while (d)–(f) are for double band ISC ($\beta = 2\mu_S$) respectively.}
\label{fig4}
\end{figure*} 
\subsection{Spin Conductance}

Since spin is not a good quantum number in the AM region, the spin current must be defined through the expectation value of the spin operator. We focus on the $z$-component of spin, which is physically motivated by the presence of ISOC in the superconducting region, providing a natural spin quantization axis. The angle-resolved spin conductance can be then expressed as 
\begin{align}
G_S(E,\theta) =
\sum_{s,s'}
\Bigg[
- \frac{v^{e}_{s'}}{v^{e}_{s}} 
\langle \Omega^e_{s'} | \sigma_z | \Omega^e_{s'} \rangle
|r_{ss'}(E,\theta)|^2 
\nonumber \\
+ \frac{v^{h}_{s'}}{v^{e}_{s}} 
\langle \Omega^h_{s'} | \sigma_z | \Omega^h_{s'} \rangle
|a_{ss'}(E,\theta)|^2
\Bigg]
\label{eq25}
\end{align}

Although the intrinsic helicity eigenstates of the AM possess vanishing $\sigma_z$ expectation value. However, the presence of spin-active scattering at the interface mixes different helicity channels, generating reflected states with finite spin polarization. Thus a nonzero spin current emerges even in the absence of bulk magnetization. This highlights the important role of the interface in converting helicity imbalance into measurable spin transport. The total spin conductance is obtained by angular averaging~\cite{acharjee1021},
\begin{equation}
G_S(E) =
\int_{-\pi/2}^{\pi/2} G_S(E,\theta)\cos\theta\, d\theta.
\label{eq20}
\end{equation}

The interplay between the momentum-dependent helicity structure of the AM and the spin-active interface leads to strong mixing between different helicity channels. This results in spin-dependent normal and Andreev reflection processes, generating a finite spin current even in the absence of net magnetization. Furthermore, the anisotropic nature of the AM exchange field breaks the symmetry between forward and backward scattering processes when combined with interfacial inversion symmetry breaking. Consequently, quasiparticles incident at opposite angles experience different scattering probabilities, giving rise to nonreciprocal charge transport.

Fig.~\ref{fig3} presents the corresponding the spin conductance spectra for the same set of parameters. In the absence of a spin-active barrier, i.e., $(\rho,\chi_m)=(0,0)$, the spin conductance remains nearly $\delta$-independent and remains almost flat for $\beta \leq \mu_S$, as seen in Figs.~\ref{fig3}(a) and \ref{fig3}(d). This behavior arises because the interface scattering is effectively spin-diagonal, suppressing spin-flip processes. Consequently, Andreev reflection involves nearly equal contributions from opposite spin channels, leading to a cancellation in the net spin current despite the intrinsic momentum-dependent spin splitting  in the AM. In contrast, for $\beta = 2\mu_S$ , the spin conductance becomes strongly energy dependent as observed from Fig.~\ref{fig3}(g). This is due to the strong ISOC in the ISC region, which lifts spin degeneracy and thereby induce a pronounced out of plane spin polarization of quasiparticle states. As a result, the balance between opposite spin contributions in Andreev reflection is broken, leading to spin-selective scattering. The combined effect of spin polarization and energy-dependent velocity mismatch near the gap edges gives rise to a nontrivial spin conductance profile. 

In the presence of a spin-active interface, i.e. for $(\rho,\chi_m)\neq(0,0)$, spin-dependent phase shifts generate finite spin-flip Andreev reflection and enable efficient coupling between the AM spin texture and ISC quasiparticles, producing a pronounced $\delta$-dependent and energy-resolved response. For single band with $\beta = 0.5\mu_S$, the ISC remains nearly spin degenerate, so spin-mixed channels partially cancel each other. This  result in a smooth conductance with moderate anisotropy, maximized near $\delta$ - orientation where the projection of the AM spin texture along the transport direction is largest as seen from Fig.~\ref{fig3}(b) and \ref{fig3}(c). Moreover, for $\chi_m  = 45^\circ$ the orientation angle $\delta = 15^\circ$ support high spin conductance, indicating the tunability of spin conductance via interfacial spin flipping for arbitrary AM orientations.  In the intermediate region i.e., $\beta = \mu_S$, partial spin polarization reduces spin-mixed channels cancellation, leading to more conventional spectra with peaks and dips near $E \sim \Delta_0$, reflecting enhanced spin-flip processes and improved spin matching across the interface as seen from Figs.~\ref{fig3}(e) and \ref{fig3}(f). In the double band regime with $\beta = 2\mu_S$, the quasiparticles become strongly spin polarized, and the interface acts as an efficient spin filter. This results in highly anisotropic and sharply varying conductance with pronounced peaks and sign changes for both $\chi_m =15^\circ$ and $45^\circ$, arising from selective transmission between spin-polarized states and the anisotropic AM bands.
\subsection{Conductance for strong spin mixing}
Figs.~\ref{fig2} and \ref{fig3} demonstrate that the conductance spectra are highly sensitive to the spin-active interfacial parameters and the AM orientation angle. In contrast to the weak spin-mixing regime discussed previously, where scattering is predominantly helicity conserved, the spin-dependent terms of the interface potential now generate strong inter-helicity scattering i.e., $s\neq s'$. Consequently, both charge and spin conductance undergo substantial spectral redistribution accompanied by pronounced angular anisotropy. Fig.~\ref{fig4} presents the conductance spectra in the strong spin-mixing regime with $\rho=0.5$, $\chi_m=30^\circ$.  For $\beta=0.5\mu_S$, the conductance exhibits strongly orientation-dependent coherence peaks and enhanced low-energy response arising from anisotropic spin-selective Andreev reflection as seen from Fig.~\ref{fig4}(a). In the intermediate regime i.e., for $\beta=\mu_S$, significantly enhanced competition between different helicity channels produces strong suppression near the gap edges and significant reshaping of the spectra. For $\beta=2\mu_S$, partial compensation between helicity bands leads to comparatively smoother charge conductance, although finite anisotropy persists as evident from Fig.~\ref{fig4}(c).It is observed that for strong spin mixing, nonreciprocal transport persists throughout all regimes considered. Moreover, the maximum conductance occurs at $\delta=45^\circ$ in the single-band regime, whereas the conductance remains largest for $\delta=0^\circ$ in the intermediate and double-band regimes. The spin conductance displays a much stronger enhancement than in the weak-mixing regime, with pronounced spectral features near $E\approx\pm\Delta_0$ and large magnitude over a broad energy range, indicating efficient spin-selective transport. Here, $\delta = 0^\circ$ exhibits the maximum conductance in the single and double-band regimes, whereas $\delta = 45^\circ$ dominates in the intermediate regime, as observed from Figs.~\ref{fig4}(d)-\ref{fig4}(f).
Overall, strong spin mixing drives the system from predominantly helicity-conserving transport to a regime dominated by inter-helicity scattering, thereby enhancing anisotropic spin filtering and favoring nonreciprocal transport.

\subsection{Mechanism of Nonreciprocity in AMs}
It is important to emphasize that the mechanism of nonreciprocal transport in the present system.  is fundamentally distinct from that in conventional Rashba or ferromagnetic systems. In Rashba systems, nonreciprocity originates from odd-parity spin-orbit coupling of the form $\mathbf{g}(\mathbf{k}) \sim (k_y,-k_x,0)$, which directly breaks inversion symmetry in the bulk. In contrast, the AM exchange field considered here possesses an even-parity $d$-wave form,
$\mathbf{g}(\mathbf{k},\delta) \sim (k_x k_y,\; k_x^2 - k_y^2,\;0),$
which preserves global time-reversal symmetry and does not generate net magnetization. As a consequence, nonreciprocal transport does not arise from bulk symmetry breaking, but instead emerges from the interplay of:
(i) momentum-dependent helicity in the AM,
(ii) inversion symmetry breaking at the interface, and
(iii) interfacial
spin-dependent scattering processes.

\section{Symmetry Analysis and Spin-Dependent Transport}

The BdG Hamiltonian possesses an intrinsic particle-hole symmetry given by
\begin{equation}
\mathcal{C}\,\mathcal{H}_{\text{BdG}}(\mathbf{k})\,\mathcal{C}^{-1}
= -\mathcal{H}_{\text{BdG}}(-\mathbf{k}),
\label{eq21}
\end{equation}
where $\mathcal{C} = \tau_x \mathcal{K}$, with $\tau_x$ acting in particle-hole space and $\mathcal{K}$ denoting complex conjugation. This symmetry ensures that quasiparticle excitations occur in pairs at energies $\pm E$.

In the AM region, the normal-state Hamiltonian contains a momentum-dependent exchange field of the form $\mathbf{g}(\mathbf{k},\delta)\cdot\boldsymbol{\sigma}$. Despite the absence of net magnetization, the system preserves global time-reversal symmetry,
\begin{equation}
\mathcal{T} \mathcal{H}_0(\mathbf{k}) \mathcal{T}^{-1} = \mathcal{H}_0(-\mathbf{k}), 
\quad \mathcal{T} = i\sigma_y \mathcal{K}.
\label{eq22}
\end{equation}
In the absence of phase bias and magnetic ordering, the full BdG Hamiltonian respects an effective time-reversal symmetry. However, the anisotropic structure of $\mathbf{g}(\mathbf{k},\delta)$ lifts spin degeneracy locally in momentum space, resulting in a momentum-dependent spin quantization axis aligned with $\mathbf{g}(\mathbf{k},\delta)$. Consequently, spin no longer remain as a conserved quantum number, and the quasiparticle states are most naturally described in the helicity basis $s=\pm$, characterized by the phase $\phi_{\mathbf{k}}$. This momentum-dependent helicity structure plays a central role in determining the symmetry of scattering processes and ultimately leads to asymmetric transport.

\begin{figure*}[t]
\centerline{
\hspace{-0.5cm}
\includegraphics[scale = 0.2]{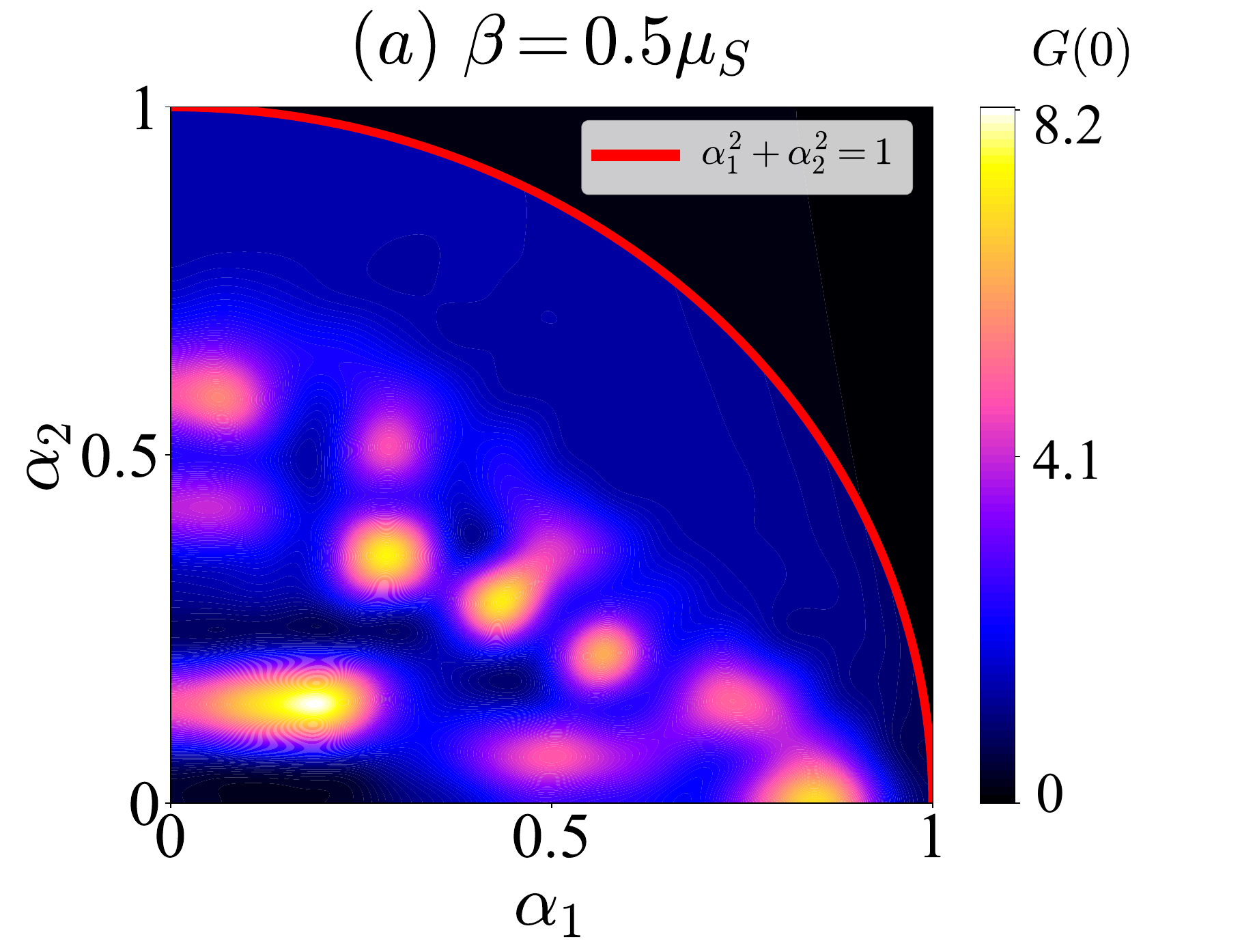}
\hspace{-0.5cm}
\includegraphics[scale = 0.2]{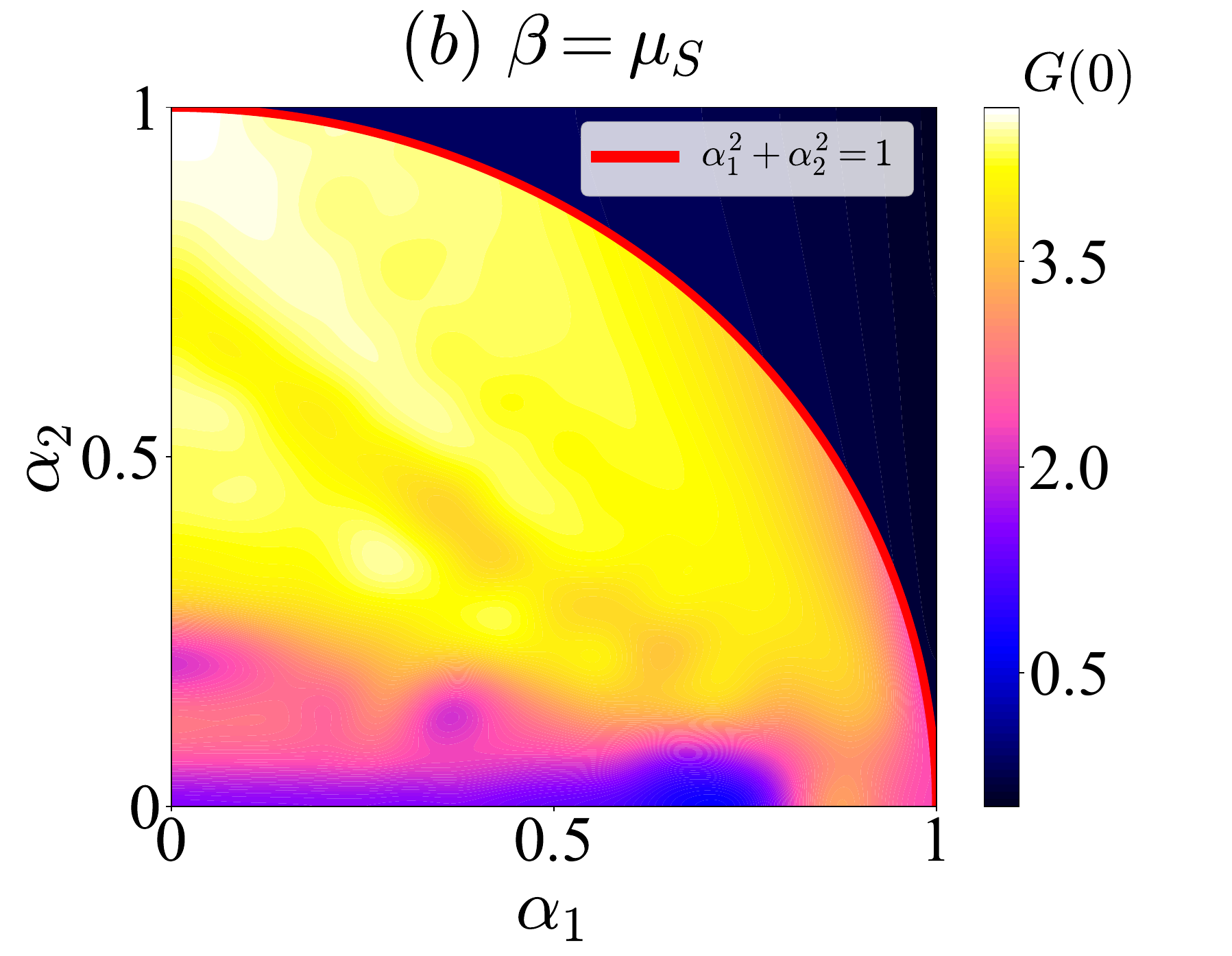}
\hspace{-0.5cm}
\includegraphics[scale = 0.2]{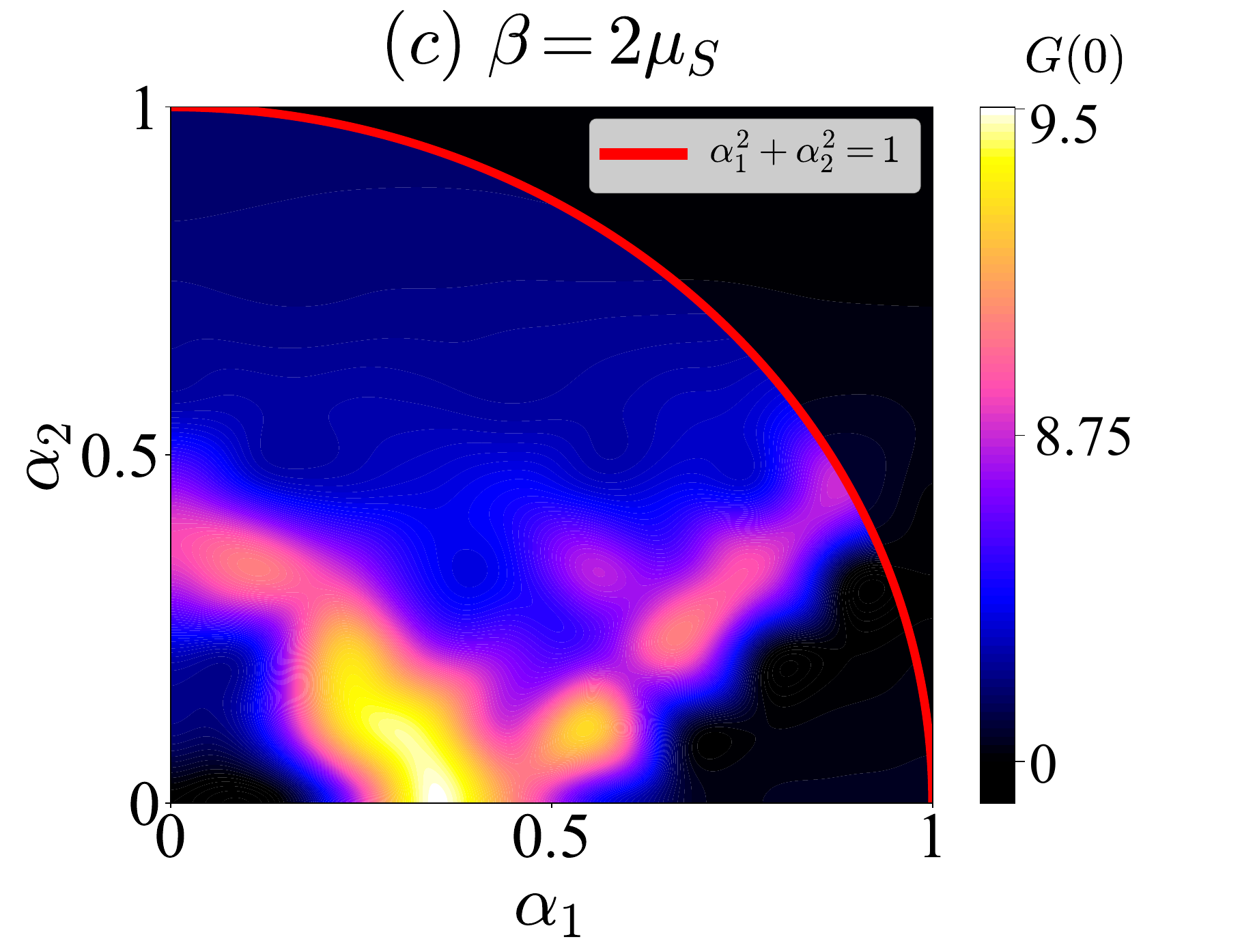}}
\centerline{
\hspace{-0.5cm}
\includegraphics[scale = 0.2]{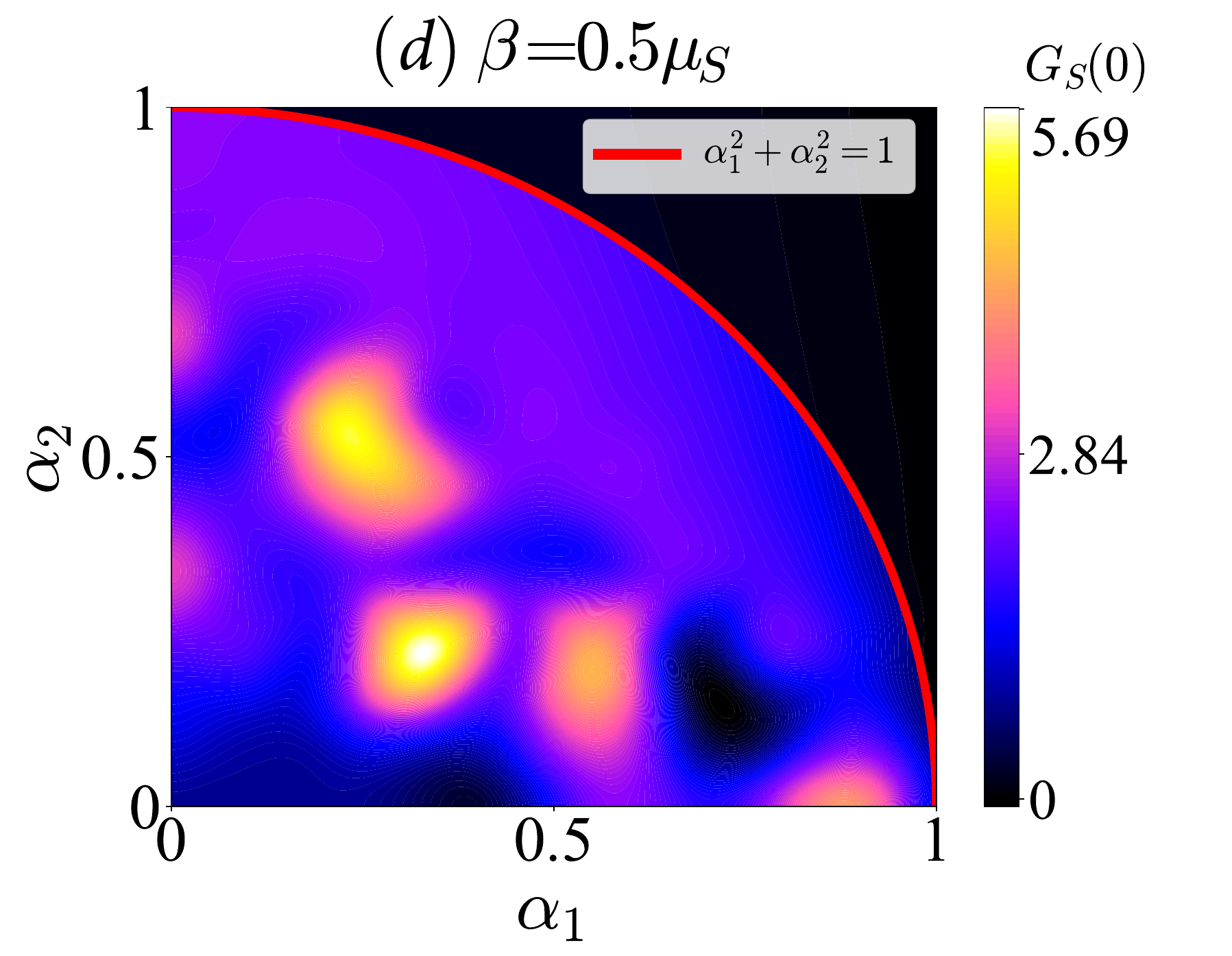}
\hspace{-0.5cm}
\includegraphics[scale = 0.2]{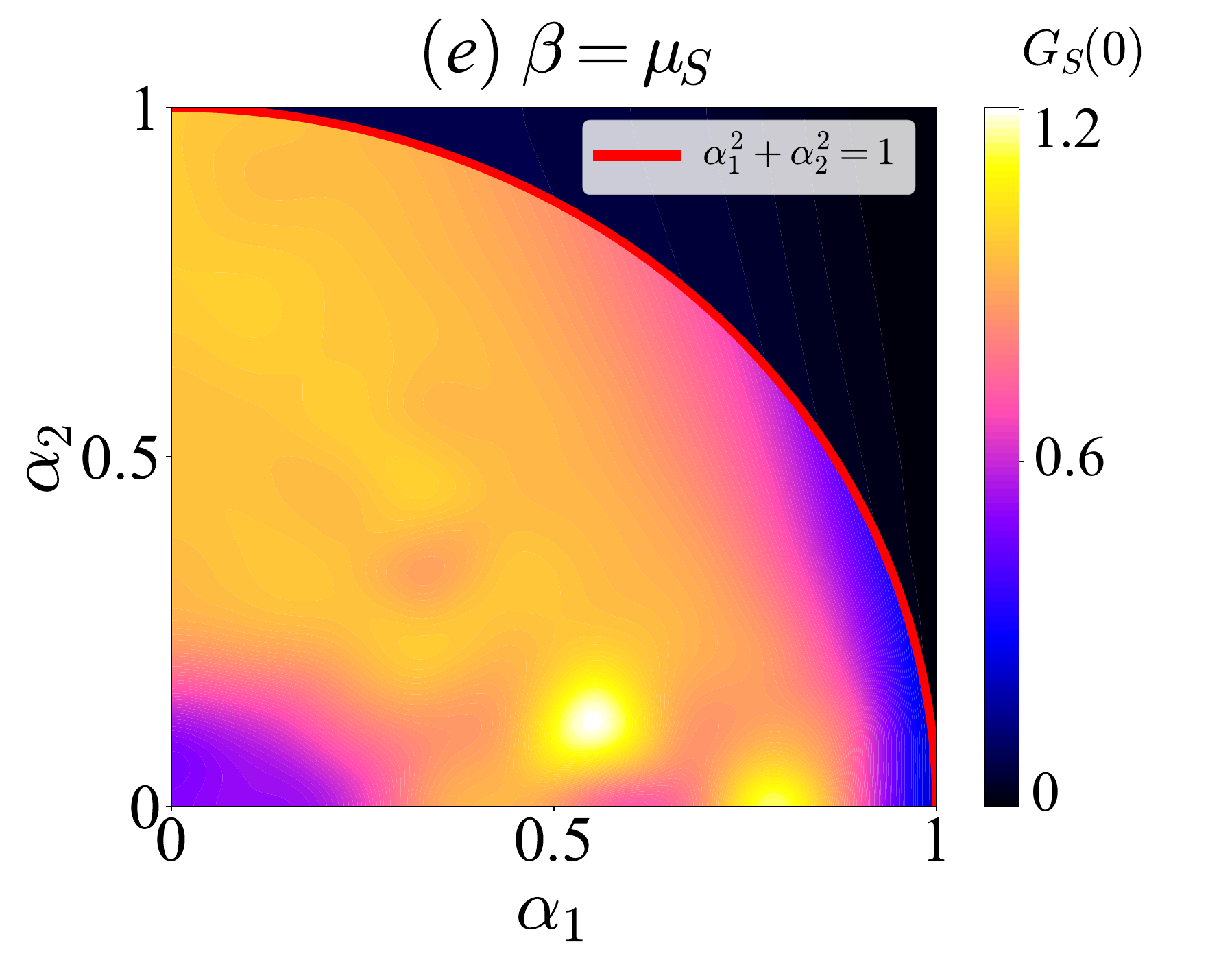}
\hspace{-0.5cm}
\includegraphics[scale = 0.2]{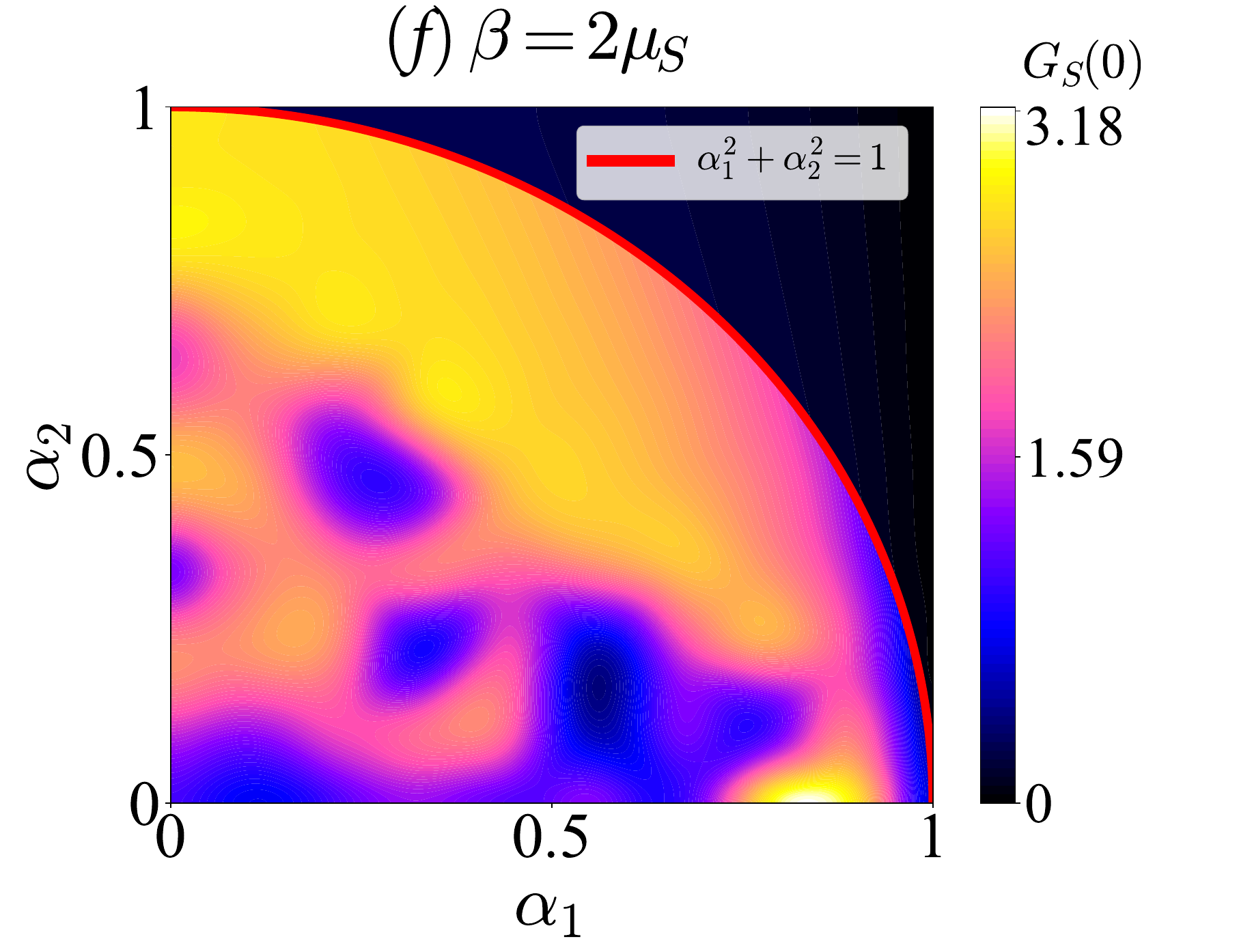}}
\caption{Density plot representing the variation of $G(0)$ (Top Panel) and $G_S(0)$ (Bottom Panel) with $\alpha_1$ and $\alpha_2$ for (a),(d) $\beta = 0.5\mu_S$,  (b),(e) $\beta =  \mu_S$ and (c),(f) $\beta = 2\mu_S$ considering $Z_0 = 0.1$, $\rho = 0.1$ with $(\chi_m, \xi_m) = (45^\circ, 45^\circ)$. The red line corresponds to $\alpha_1^2+\alpha_2^2 = 1$, with $\alpha_c = 1$. The region below this line follows elliptic Fermi surface.}
\label{fig5}
\end{figure*} 

\subsection{Scattering Symmetries}

The transport properties are governed by quasiparticle scattering at the interface. For an incoming quasiparticle in helicity channel $s=\pm$, the normal and Andreev reflection amplitudes are denoted by $r_{ss'}(E,\theta)$ and $a_{ss'}(E,\theta)$, where $s,s'=\pm$ label the helicity indices.

Particle-hole symmetry imposes the constraint
\begin{align}
a_{ss'}(E,\theta) &= [a_{\bar{s}\bar{s}'}(-E,\theta)]^\ast,
\nonumber \\
r_{ss'}(E,\theta) &= [r_{\bar{s}\bar{s}'}(-E,\theta)]^\ast,
\label{eq23}
\end{align}
where $\bar{s}$ denotes the opposite helicity branch.

Time-reversal symmetry further constrains the scattering amplitudes. Under time reversal, $\mathbf{k} \rightarrow -\mathbf{k}$ and the exchange field transforms as $\mathbf{g}(\mathbf{k},\delta) \rightarrow -\mathbf{g}(-\mathbf{k},\delta)$. For the present model, $\mathbf{g}(\mathbf{k},\delta)$ is even in momentum, such that $\mathbf{g}(-\mathbf{k},\delta)=\mathbf{g}(\mathbf{k},\delta)$, and therefore time reversal effectively flips the spin while leaving the parameter $\delta$ unchanged. As a result, the scattering amplitudes satisfy
\begin{equation}
r_{ss'}(E,\theta,\delta) =
r^\ast_{ss'}(E,-\theta,\delta),
\label{eq24}
\end{equation}
with an analogous relation for $a_{ss'}$.

Although the bulk Hamiltonian preserves time-reversal symmetry, the presence of a spin-active interface breaks spatial inversion symmetry. In combination with the momentum-dependent helicity structure of the AM, this leads to an effective asymmetry of the scattering matrix,
\begin{equation}
\mathcal{S}(E,\theta,\delta,\chi_m) \neq \mathcal{S}(E,-\theta,\delta,\chi_m),
\label{eq25}
\end{equation}
which directly results in nonreciprocal transport,
\begin{equation}
G(E,\theta) \neq G(E,-\theta).
\label{eq26}
\end{equation}
This nonreciprocity originates from the interplay between helicity-dependent spin textures and spin-selective interfacial scattering, which lifts the equivalence between forward and backward propagating quasiparticles. Importantly, this mechanism does not rely on breaking time-reversal symmetry, but instead arises from inversion symmetry breaking at the interface together with momentum-dependent spin splitting.

Fig.~\ref{fig5} presents the zero bias charge conductance $G(0)$ (top row) and spin conductance $G_S(0)$ (bottom row) as a function of the AM parameters $(\alpha_1,\alpha_2)$ for different ISOC strengths. The red boundary separates the physically allowed regime with an elliptic Fermi surface from the unphysical region with hyperbolic contours.
In the single band scenario $\beta = 0.5\mu_S$, both $G(0)$ and $G_S(0)$ exhibit highly nonuniform and spin selective maxima across the $(\alpha_1,\alpha_2)$ plane as seen from panels Figs.~\ref{fig5}(a) and ~\ref{fig5}(d) respectively. In this regime, the superconducting quasiparticles remain nearly spin degenerate and transport is primarily governed by the anisotropic projection of the AM spin texture onto the interface. Since $\mathbf{g}(\mathbf{k},\delta)$ has a $d$-wave form, its magnitude and direction vary strongly over the Fermi surface depending on $(\alpha_1,\alpha_2)$. This leads to spatially fluctuating helicity alignment between incident and reflected states which result in selective charge conductance patterns. In case of spin conductance, $G_S(0)$ remains relatively weak and localized, indicating the partial cancellation between opposite spin contributions due to the near degeneracy of quasiparticle states. In the intermediate regime i.e. for $\beta = \mu_S$, the conductance maps become significantly smoother and more uniform as seen from Figs.~\ref{fig5}(b) and ~\ref{fig5}(e) In this scenario the ISOC induces appreciable spin splitting in the ISC region and thereby lifts the degeneracy between opposite spin channels partially. Thus the destructive interference between spin-resolved Andreev processes significantly reduced, thus the overall transmission is significantly enhanced. As a result, $G(0)$ display a broad maxima over a wide region of parameter space indicating efficient Andreev reflection across most $(\alpha_1,\alpha_2)$ regions. Simultaneously, $G_S(0)$ develops a finite and smoothly varying broad regions, reflecting the emergence of a net spin polarization due to the incomplete cancellation between the helicity channels. The reduced sensitivity to $(\alpha_1,\alpha_2)$ in this regime indicates that the transport is now dominated more by the intrinsic spin polarization of the ISC rather than the AM anisotropy. 

\begin{figure*}[t]
\centerline
\centerline{
\includegraphics[scale = 0.22]
{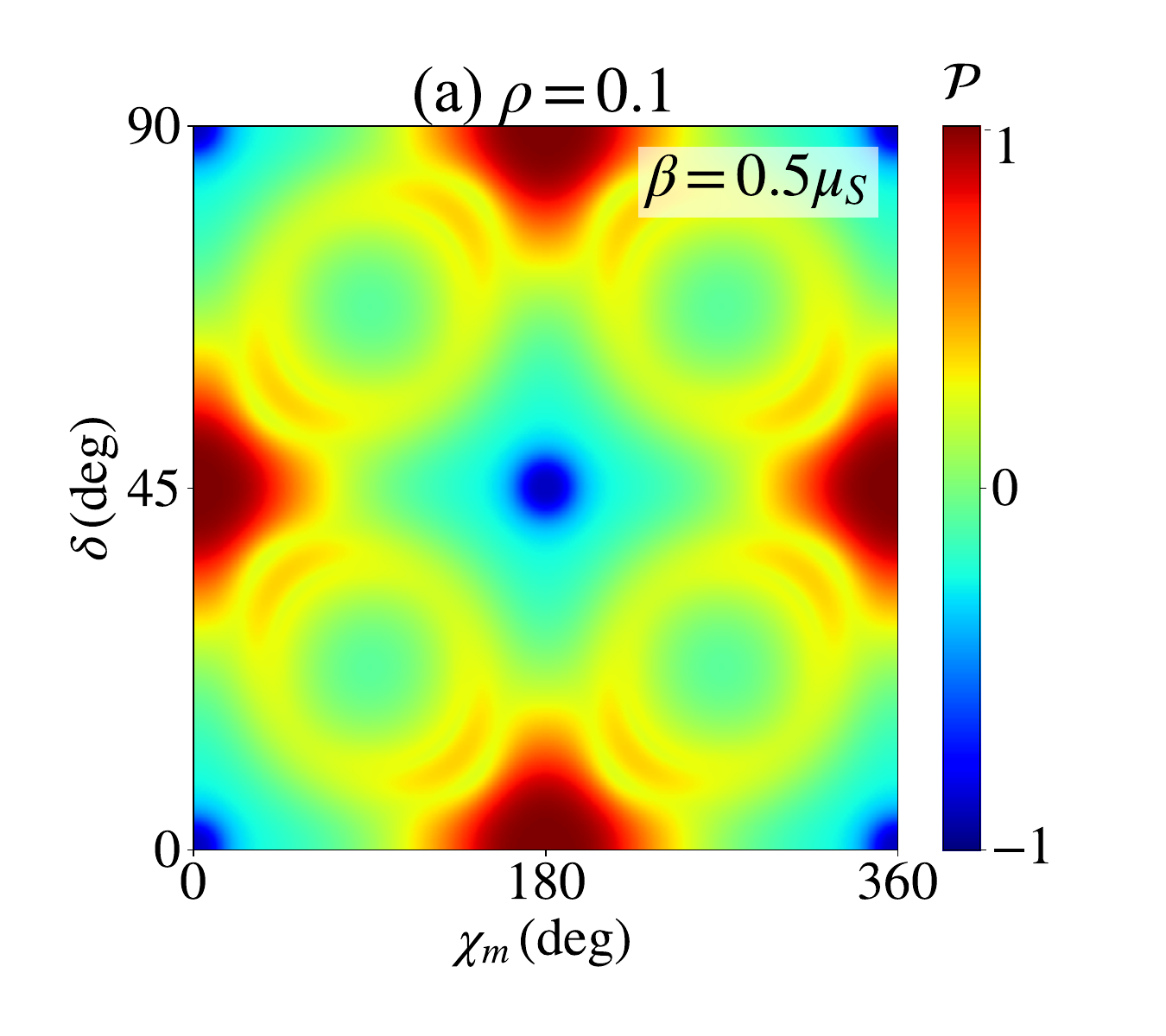}
\hspace{-0.9cm}
\includegraphics[scale = 0.22]{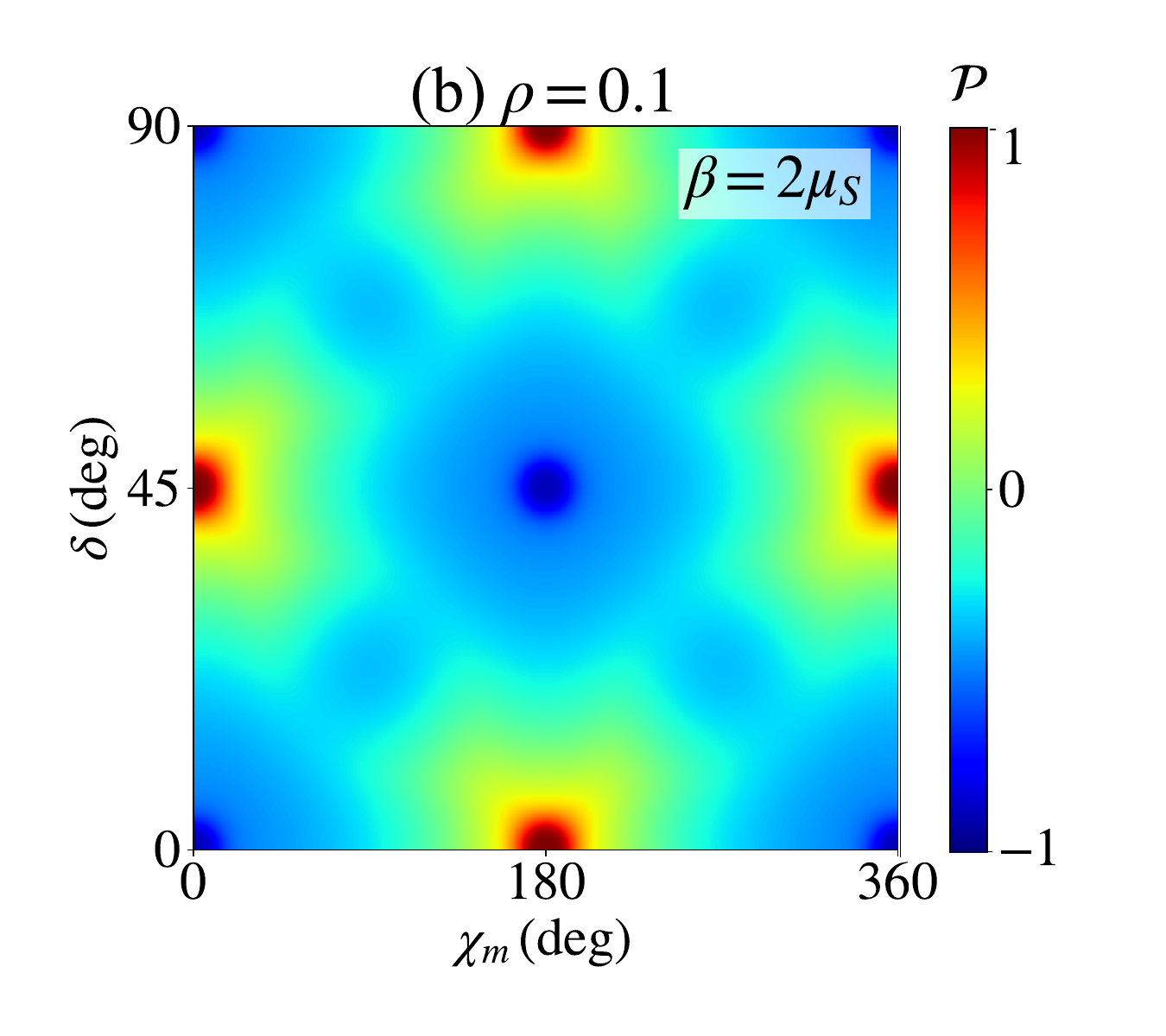}
\hspace{-0.9cm}
\includegraphics[scale = 0.22]{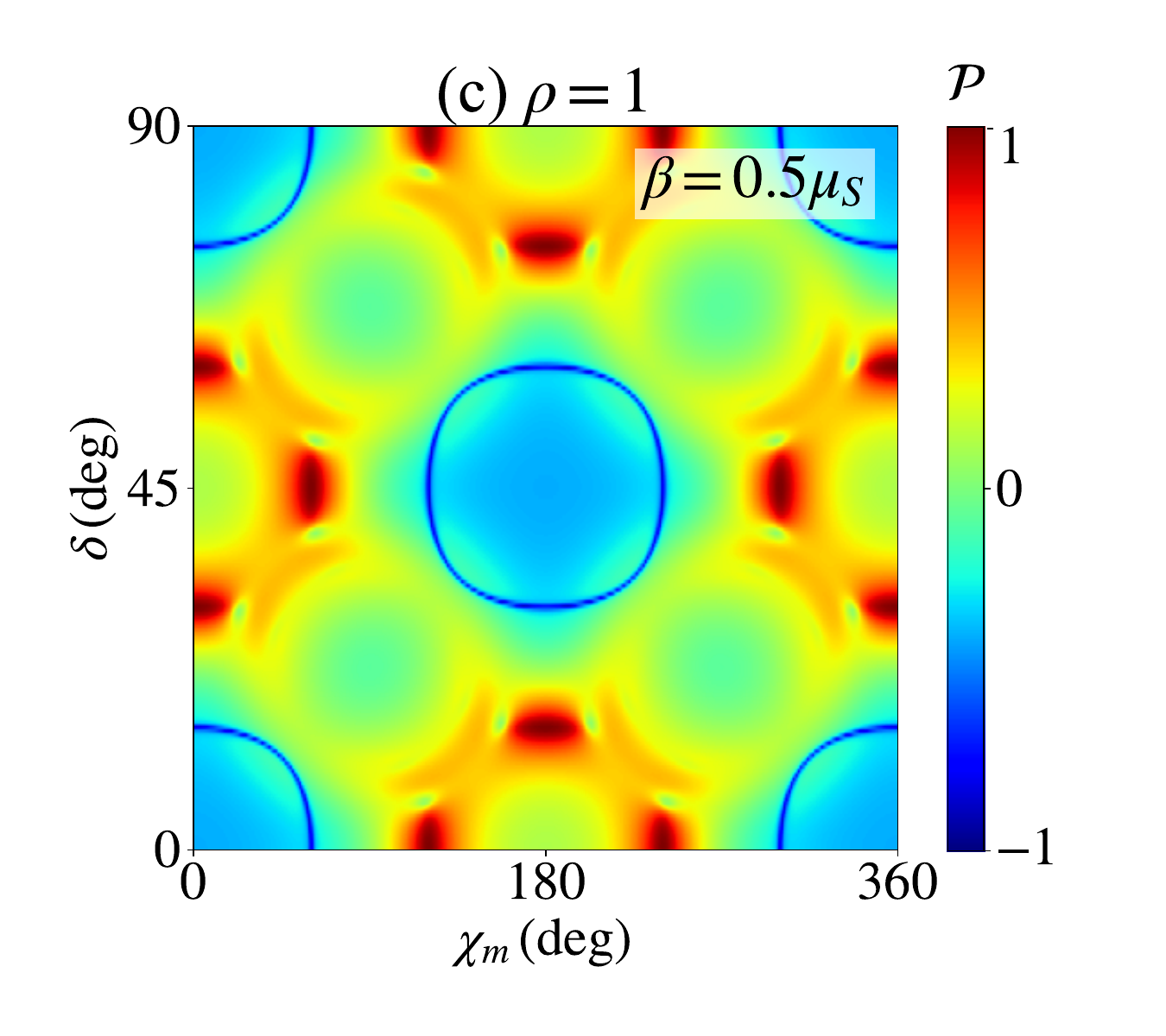}
\hspace{-0.9cm}
\includegraphics[scale = 0.22]{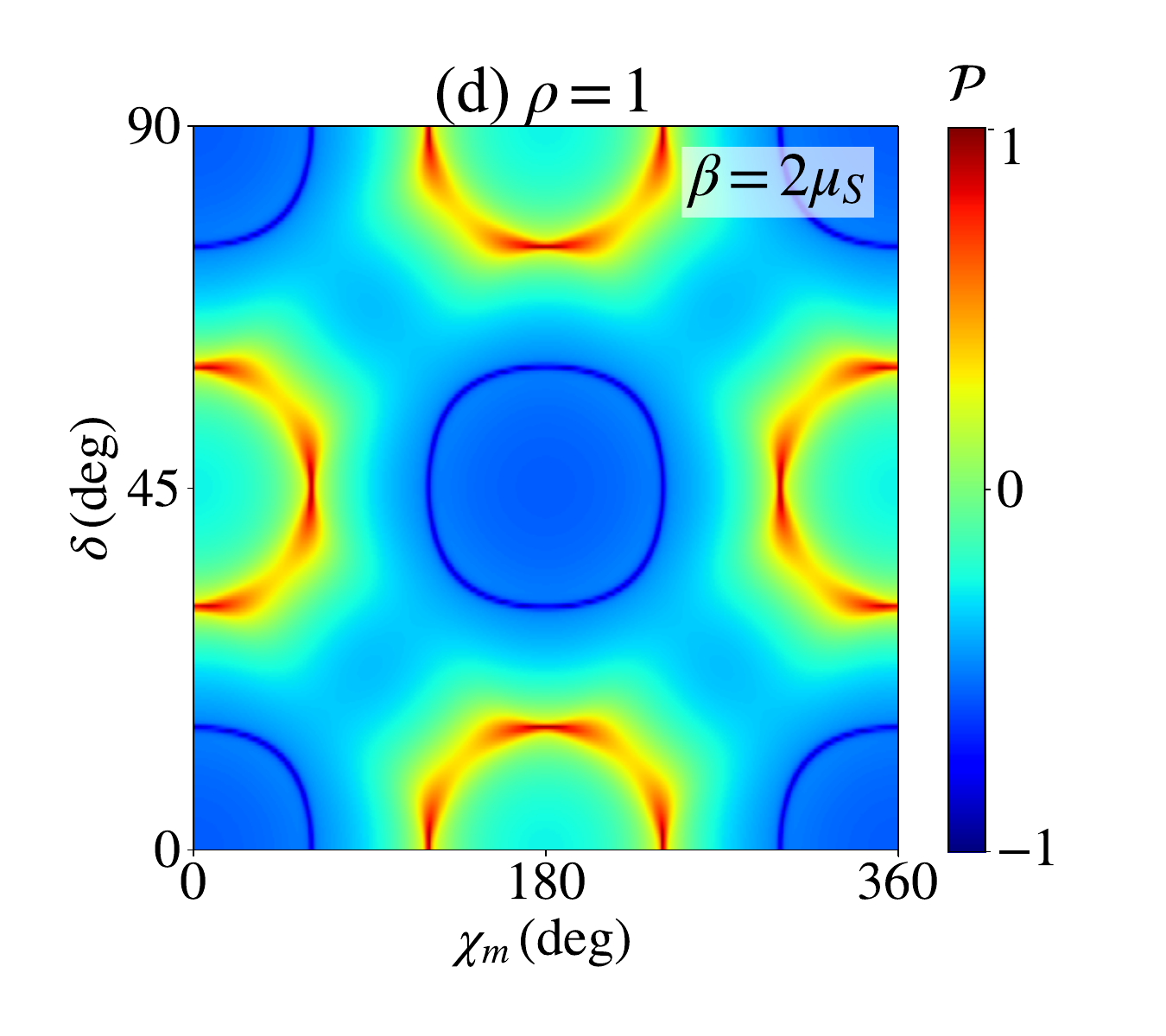}}
\centerline{
\includegraphics[scale = 0.22]
{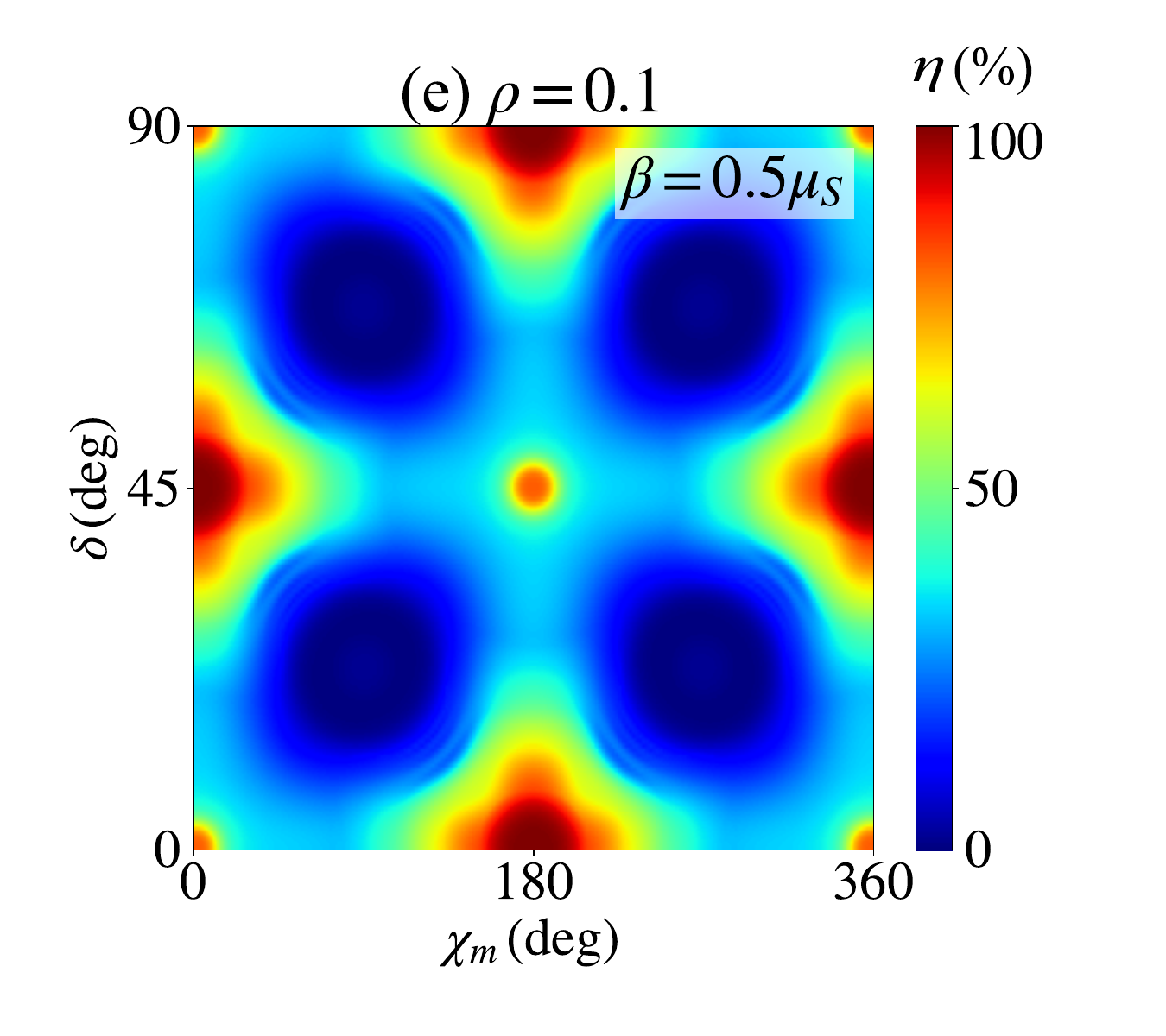}
\hspace{-0.9cm}
\includegraphics[scale = 0.22]{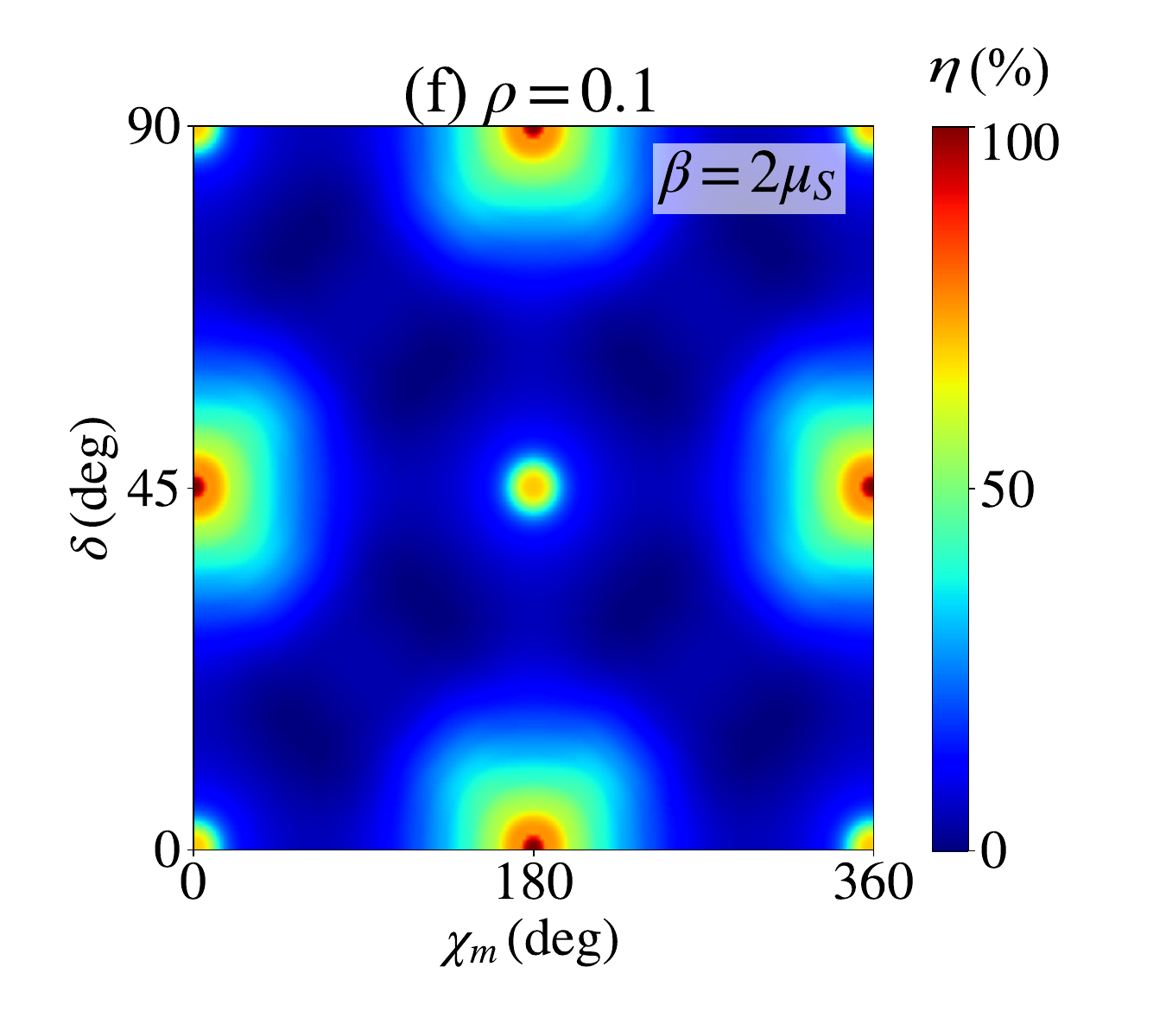}
\hspace{-0.9cm}
\includegraphics[scale = 0.22]{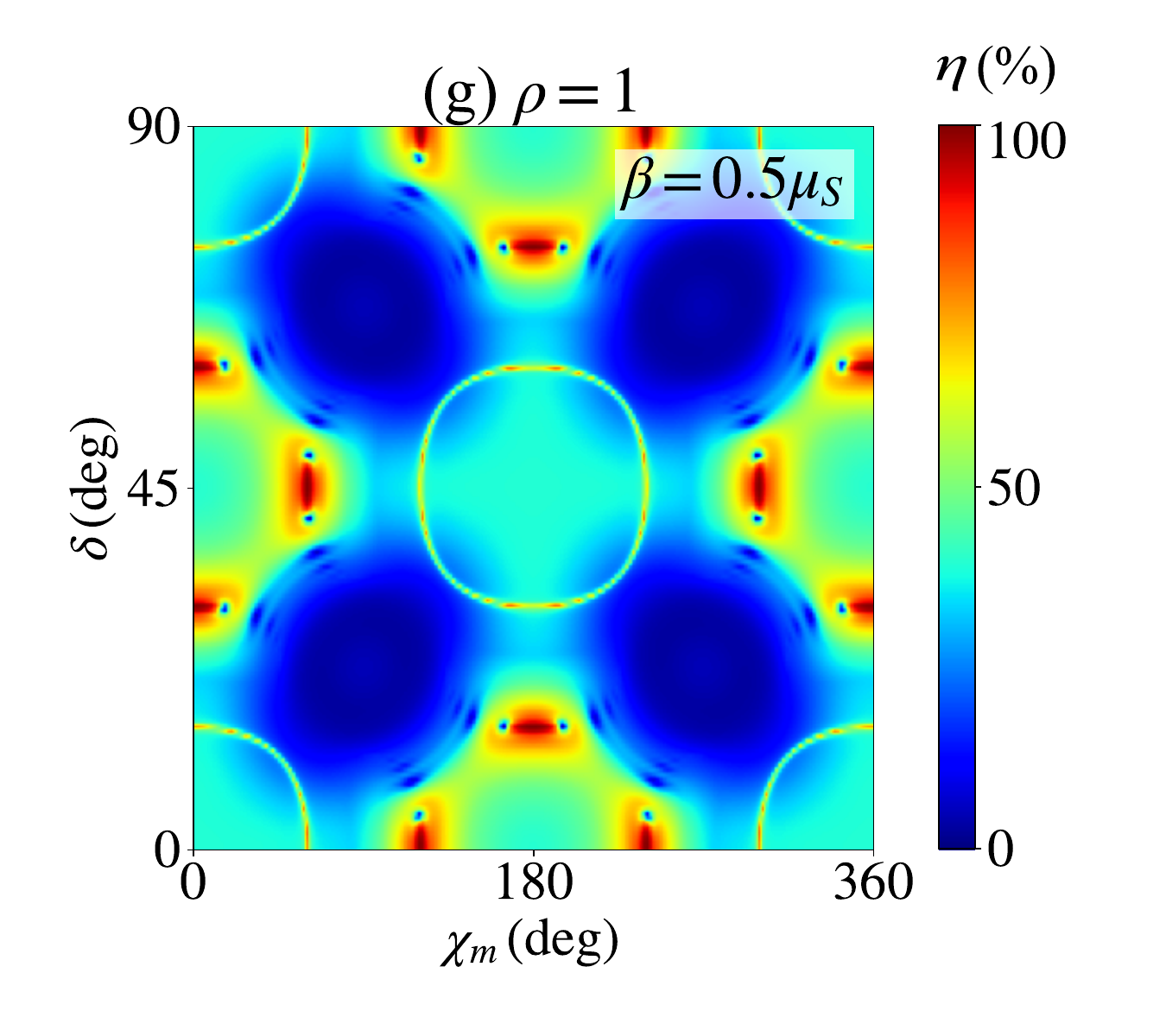}
\hspace{-0.9cm}
\includegraphics[scale = 0.22]{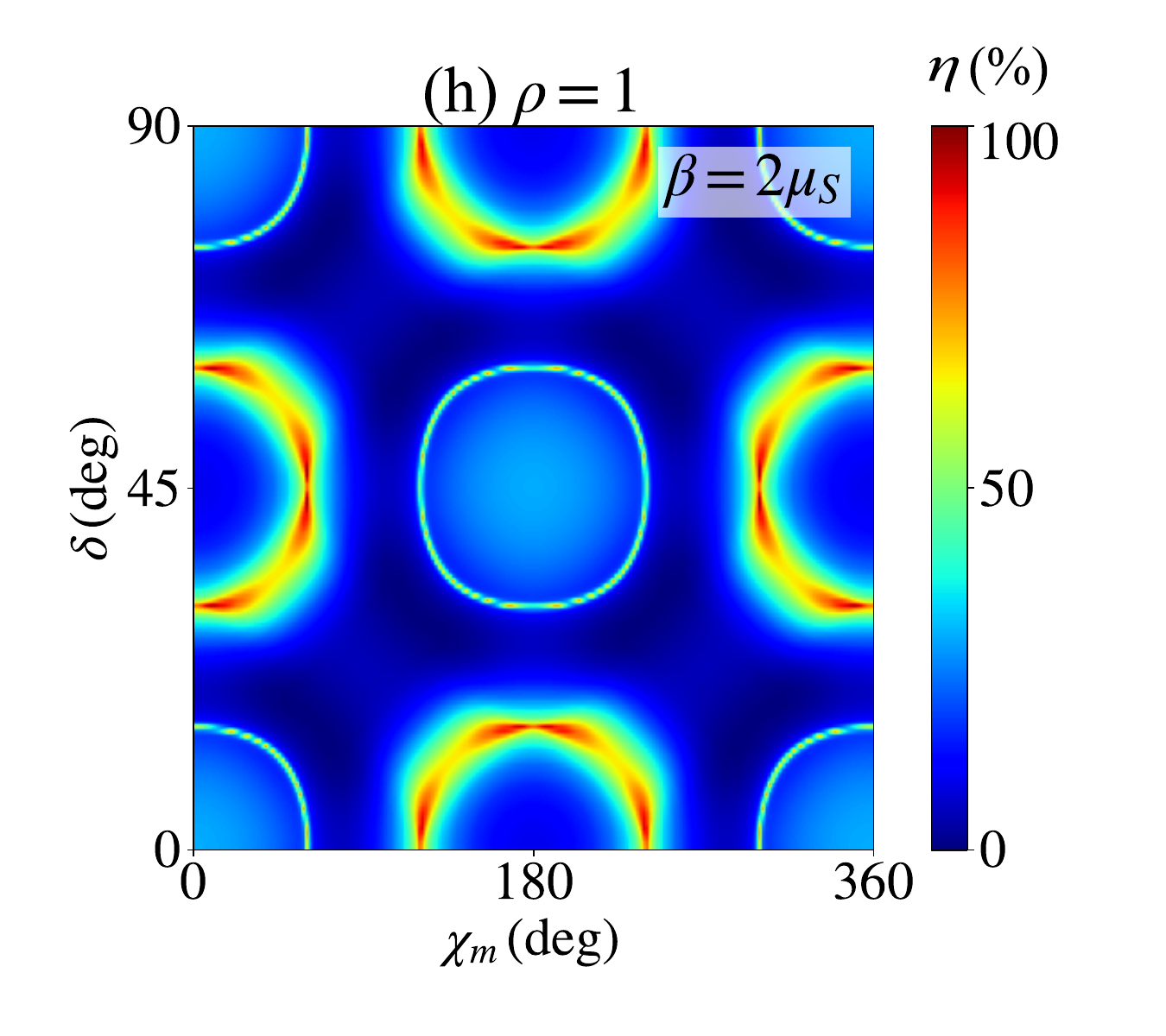}}
\caption{Density plots of spin polarization $\mathcal{P}$ (top panels) and spin-filtering efficiency $\eta$ (bottom panels) as functions of $\delta$ and $\chi_m$. Panels (a),(b),(e) and (f) correspond to weak spin-mixing regime ($\rho=0.1$), while (c),(d),(g) and (h) represent the strong spin-mixing regime ($\rho=1$) considering $Z_0=0.1$ and $\xi_m=45^\circ$.}
\label{fig6}
\end{figure*} 
In the double band regime with $\beta = 2\mu_S$, the conductance profiles again become highly structured but has a qualitatively different origin. The large ISOC of the ISC enforces strong out of plane spin polarization of quasiparticles, which effectively lock their spin orientation. This introduces a strong spin selective Andreev reflection at the interface. In this regime only the regions where the AM helicity texture aligns favorably with the ISC spin polarization contribute to transport. Consequently, $G(0)$ develops pronounced spin selective features, with regions of enhanced conductance corresponding to optimal helicity matching, and suppressed regions where mismatch leads to strong reflection. Contrastingly in this regime, $G_S(0)$ is also enhanced but in a selective way and exhibits sharp spatial variations with the presence of localized maxima and minima, reflecting highly efficient spin filtering. The appearance of both positive and suppressed regions indicates that the balance between spin-conserving and spin-flip Andreev processes is strongly modulated by the anisotropic AM texture. Overall, Fig.~\ref{fig5} demonstrates a crossover from anisotropy dominated transport in the single band regime to nearly isotropic, enhanced transmission at the intermediate regime, followed by strongly spin-selective, helicity filtered transport in the double band regime. This evolution highlights the pivotal role of the $d$-wave AM spin texture in both charge and spin conductance, thereby offering an efficient route for tunable superconducting spin transport.

\section{Spin Polarization and Spin Filtering}
To directly capture the emergence of spin-selective transport inferred from Fig.~\ref{fig5}, we now analyze the spin polarization and spin-filter efficiency of the junction. In AM/ISC hybrid structures, the coexistence of strong ISOC of the ISC and AM order leads to a nontrivial interplay between spin–momentum locking in the ISC and the anisotropic exchange field of the AM. This results in intrinsically spin-dependent scattering amplitudes at the interface. Consequently, the transmission probabilities for spin-up and spin-down quasiparticles become unequal, giving rise to a finite spin imbalance in the transmitted current even in the absence of conventional ferromagnetic polarization.
Thus spin polarization provides a direct measure of the net spin current generated across the junction while the spin-filter efficiency quantifies how effectively the device transmits one spin species over the other, irrespective of the total conductance. Importantly, the dependence of their spin filtering parameter on $E/\Delta_0$ and $\delta$ provides information about the superconducting symmetry and the underlying spin texture. Thus it serves as a sensitive probe of the crossover from weakly spin-mixed to strongly spin-resolved transport regimes. We define the spin polarization of the conductance as
\begin{equation}
\mathcal{P}(E) = \frac{G_\uparrow(E) - G_\downarrow(E)}{G_\uparrow(E) + G_\downarrow(E)},
\label{eq27}
\end{equation}
where $G_{\uparrow(\downarrow)}(E)$ denotes the spin-resolved conductance along the quantization axis. 
In addition, the angle averaged spin-filtering efficiency is defined as
\begin{equation}
\eta
=
\frac{
\displaystyle \int_{-\pi/2}^{\pi/2}
\left|G_\uparrow(E,\theta) - G_\downarrow(E,\theta)\right|
\cos\theta \, d\theta
}{
\displaystyle \int_{-\pi/2}^{\pi/2}
\left[G_\uparrow(E,\theta) + G_\downarrow(E,\theta)\right]
\cos\theta \, d\theta
},
\label{eq28}
\end{equation}

Fig.~\ref{fig6} present the density plots of $\mathcal{P}$ (top panels) and $\eta$ (bottom panels) as functions of $\delta$ and $\chi_m$, for weak ($\rho=0.1$) and strong ($\rho=1$) spin mixing  considering $\beta=0.5\mu_S$ and $\beta=2\mu_S$. It is observed $\mathcal{P} $ and $\eta$ exhibits a smooth angular modulation with characteristic fourfold symmetry in $\delta$, reflecting the underlying $d$-wave structure of the AM exchange field $\mathbf{g}(\mathbf{k},\delta)$. For $\rho=0.1$, the polarization remains moderate and exhibits a smooth angular modulation as seen from  Figs.~\ref{fig6}(a) and \ref{fig6}(b). The maximum $\mathcal{P}$ are observed around the angles $(\chi_m, \delta) \approx (0^\circ, 45^\circ), (360^\circ, 45^\circ), (180^\circ, 0^\circ)$ and $ (180^\circ, 90^\circ)$ for both $\beta = 0.5\mu_S$ and $\beta = 2\mu_S$, These high polarization regions originate from the optimal alignment between the AM spin texture $\mathbf{g}(\mathbf{k},\delta)$ and the interface magnetization axis, which enhances spin-selective scattering and thereby maximizes the imbalance between opposite helicity channels. In contrast, the orientations $(\chi_m, \delta) \approx (0^\circ, 0^\circ),(0^\circ, 90^\circ), (180^\circ, 45^\circ), (360^\circ, 0^\circ) $ and $(360^\circ, 90^\circ)$, exhibit weak polarization due to symmetry-driven cancellation between opposite helicity channels. Increasing the ISOC strength to $\beta=2\mu_S$ reduces the overall magnitude of $\mathcal{P}$ in the weak $\rho$ regime as seen from Fig.~\ref{fig6}(b), as the strong out of plane spin locking in the ISC limits the ability of weak interface scattering to efficiently convert helicity into spin polarization. Moreover, in this case the maxima regions significantly reduced due to the presence of strong ISOC of the ISC region. 

The behavior qualitatively  changes for $\rho=1$ as seen from Figs.~\ref{fig6}(c) and \ref{fig6}(d). In this scenario the polarization develops sharp, spin selective scattered regions with very high value of $\mathcal{P}$. This indicates that in this scenario the interface acts as an efficient spin filter, strongly mixing helicity channels and enabling  selective transmission of specific spin components. The maximum polarization regions now split into three segments and occur at arbitrary angles of $\delta$ and $\chi_m$, where the AM spin texture is optimally aligned perpendicular to the interface magnetization, maximizing spin-flip scattering. Conversely, regions near $\delta \approx 0^\circ$ or $90^\circ$ with $\chi_m \approx 0^\circ$ show reduced polarization due to arbitrary projection of the $d$-wave spin texture along the scattering axis. The spin-filter efficiency $\eta$ follow trends similar to $\mathcal{P}$ but provides a clearer measure of device performance as seen from the corresponding Figs.~\ref{fig6}(g) and \ref{fig6}(h). For $\rho=0.1$, the efficiency remains limited to $\eta \lesssim 50\%$ and broadly distributed, reflecting less spin selectivity as seen earlier from Fig.~\ref{fig6}(c). In contrast, for $\rho=1$, the efficiency increases very sharply over extended regions with arbitrary orientations of $(\chi_m, \delta)$ space. This demonstrates that strong spin-active scattering is essential for achieving high-performance spin filtering. Furthermore, we observe that the region near $(\chi_m,\delta) = (180^\circ, 45^\circ)$ no longer display maximum $\mathcal{P}$ or $\eta$, instead it shows comparatively low values due to strong mixing.
A key observation is that polarization and efficiency, while closely related, encode different physical aspects. The polarization $\mathcal{P}$ is sensitive to the imbalance between spin channels and can change sign depending on the relative alignment of $\mathbf{g}(\mathbf{k},\delta)$ and the interface magnetization. In contrast, $\eta$ measures the magnitude of spin selectivity and is maximized whenever one spin channel dominates, irrespective of its orientation. Thus, regions where $\mathcal{P}$ changes sign correspond to minima in $\eta$, while regions of saturated $|\mathcal{P}|$ directly translate into high efficiency.

The dependence on $\beta$ further highlights the interplay between bulk and interfacial effects. For single band ISC, the spin polarization is primarily governed by the AM helicity texture and interface scattering geometry. However for double band  ISC, it imposes a fixed spin quantization along a particular direction, enhancing spin selectivity when combined with strong spin mixing, but suppressed when the interface is weakly spin active. Overall, these results demonstrate that optimal spin filtering requires a cooperative alignment between the AM orientation angle, the interface magnetic moment, and the ISOC strength of the ISC.

\begin{figure}[t]
\centerline{
\includegraphics[scale = 0.2]{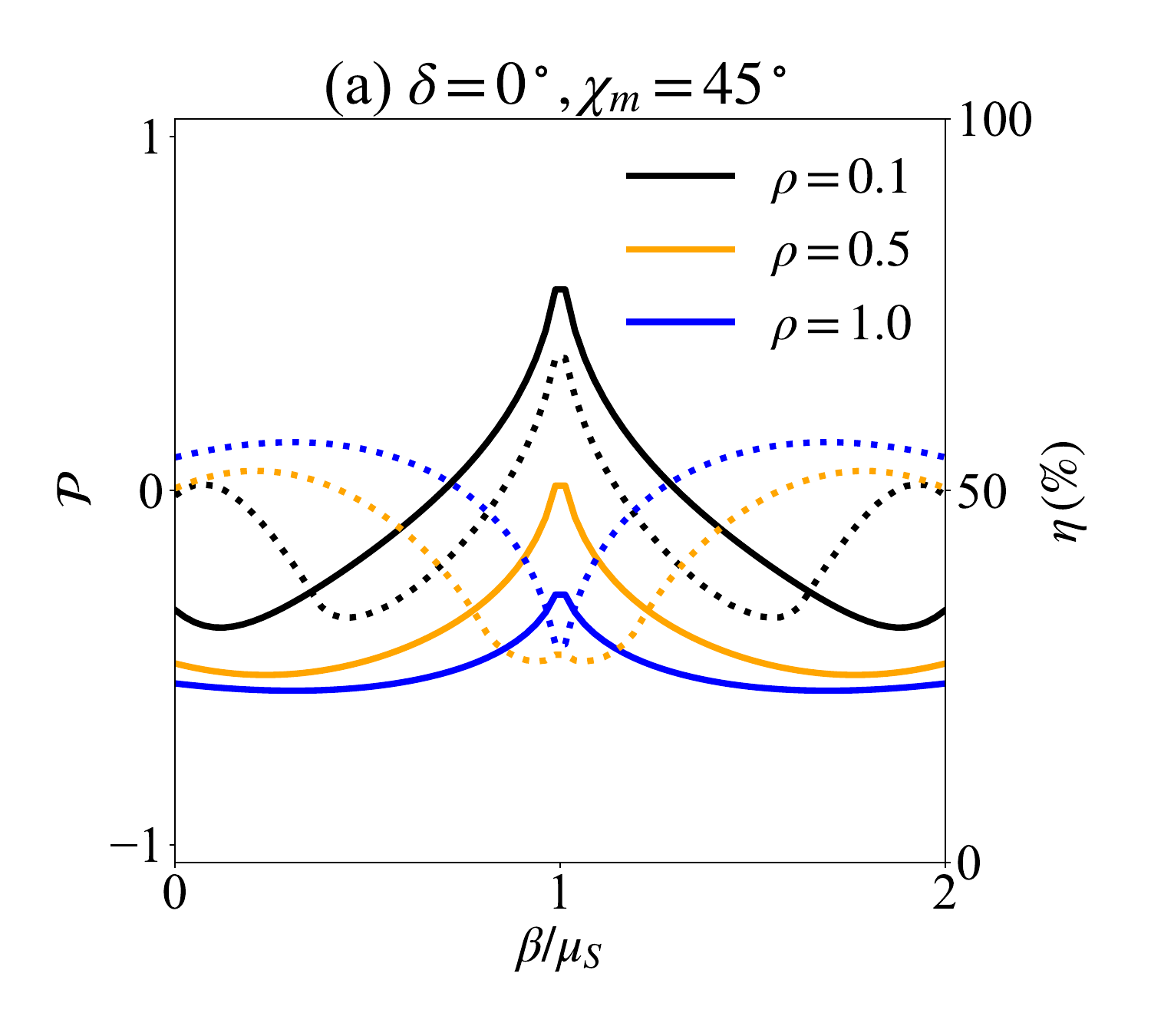}
\hspace{-0.6cm}
\includegraphics[scale = 0.2]{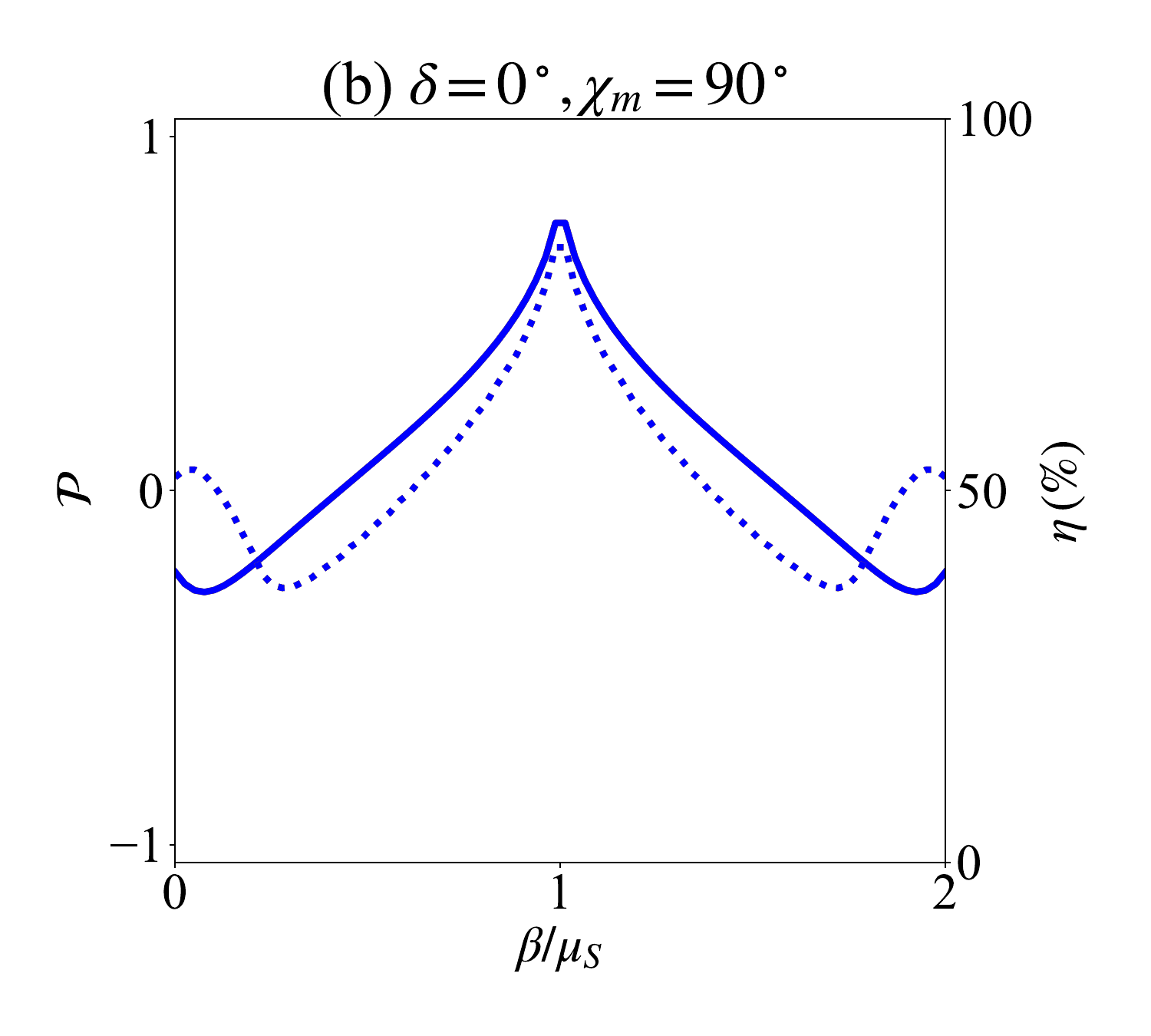} 
}
\centerline{
\includegraphics[scale = 0.2]{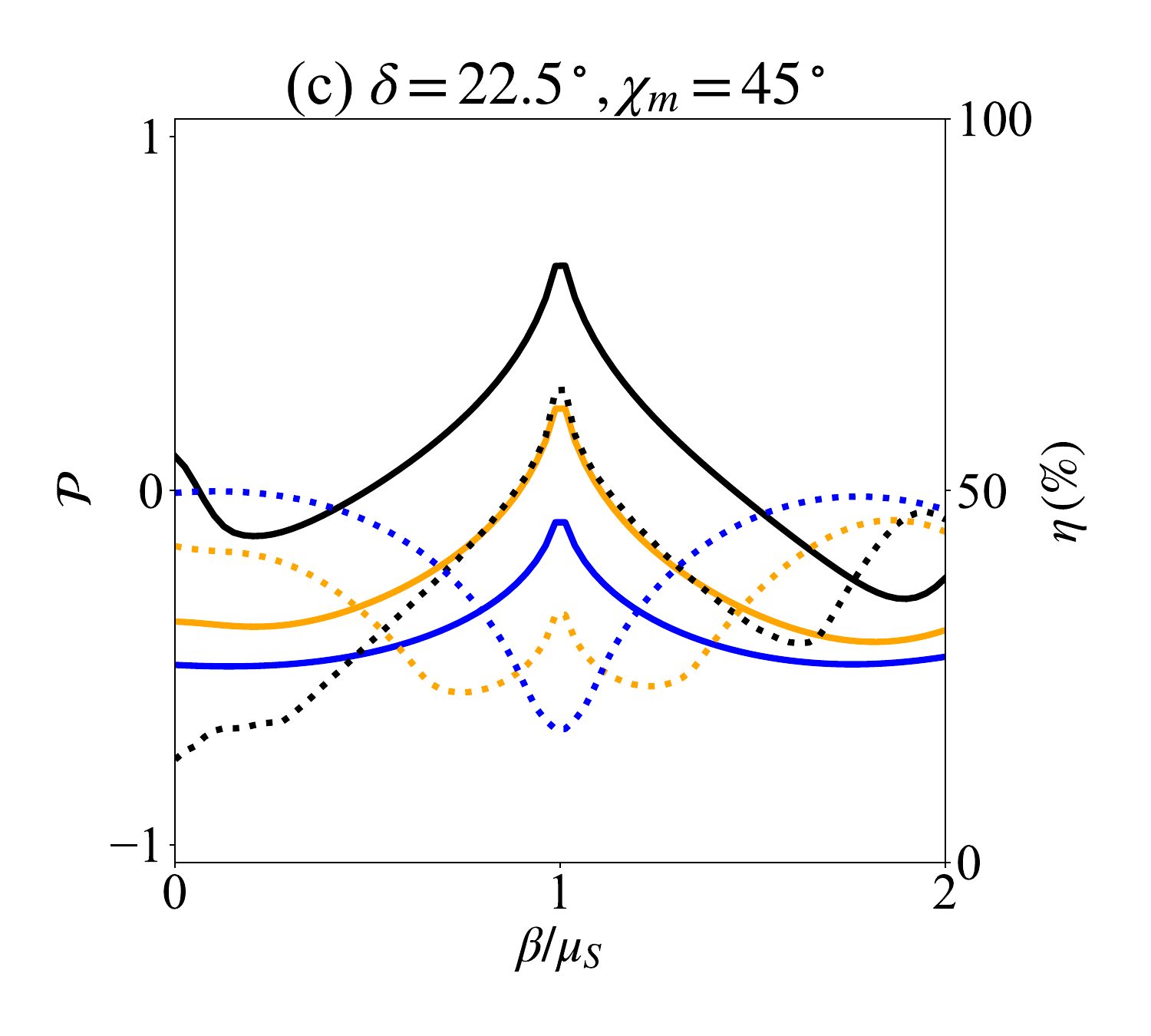}
\hspace{-0.6cm}
\includegraphics[scale = 0.2]{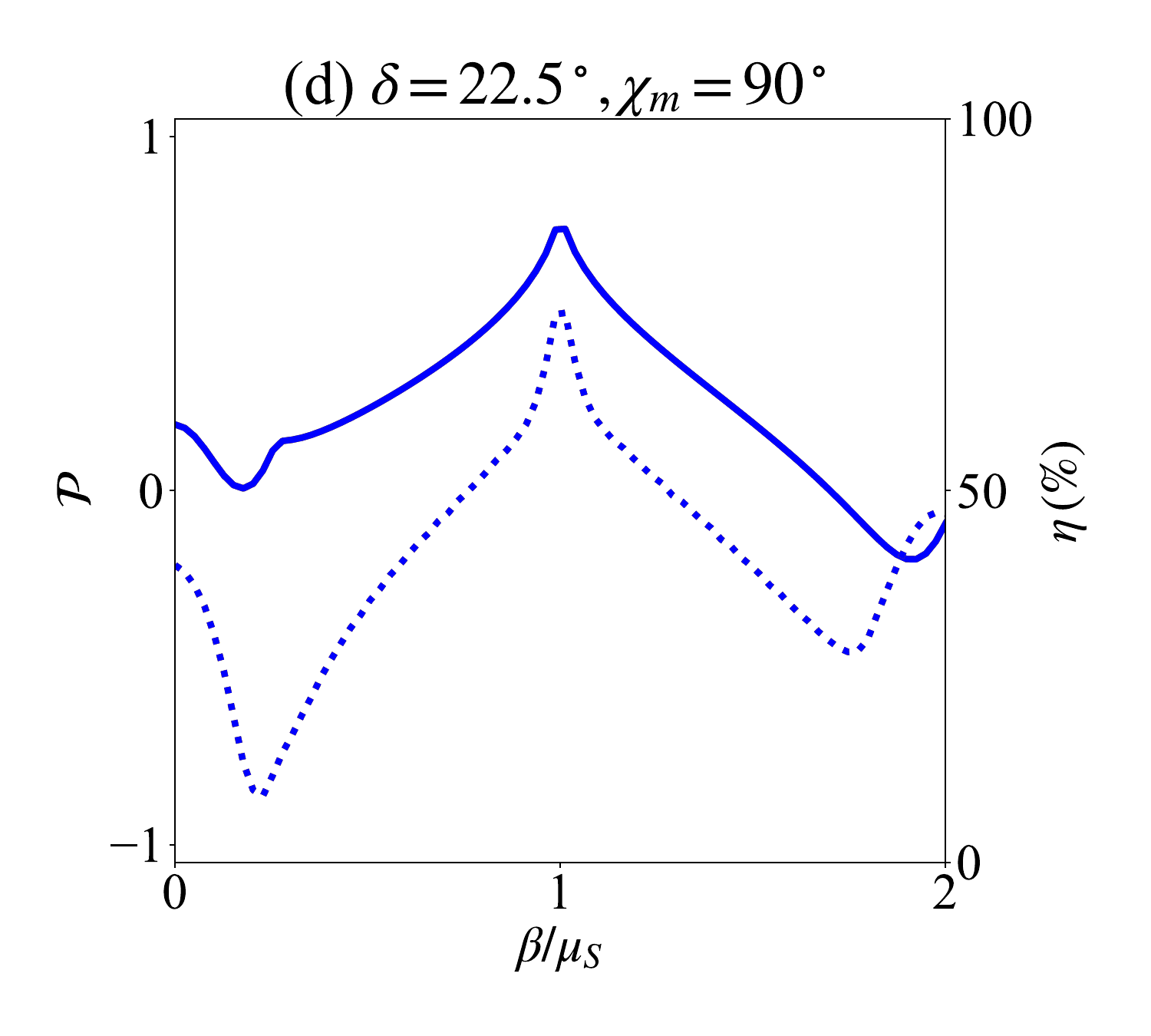} 
}
\caption{Variation of spin polarization $\mathcal{P}$ (solid lines) and spin-filter efficiency $\eta$ (dotted lines) as functions of the normalized ISOC strength $\beta/\mu_S$ for $\delta=0^\circ$ (top panel) and $\delta=22.5^\circ$ (bottom panel) considering different values of $\rho$ with $Z_0 = 0.1$, $E/\Delta_0 = 0.1$ and $\xi_m = 45^\circ$. The left and right panels corresponds to $ \chi_m=45^\circ$ and $90^\circ$ respectively.}
\label{fig7}
\end{figure} 
\begin{figure}[t]
\centerline{
\includegraphics[scale = 0.2]{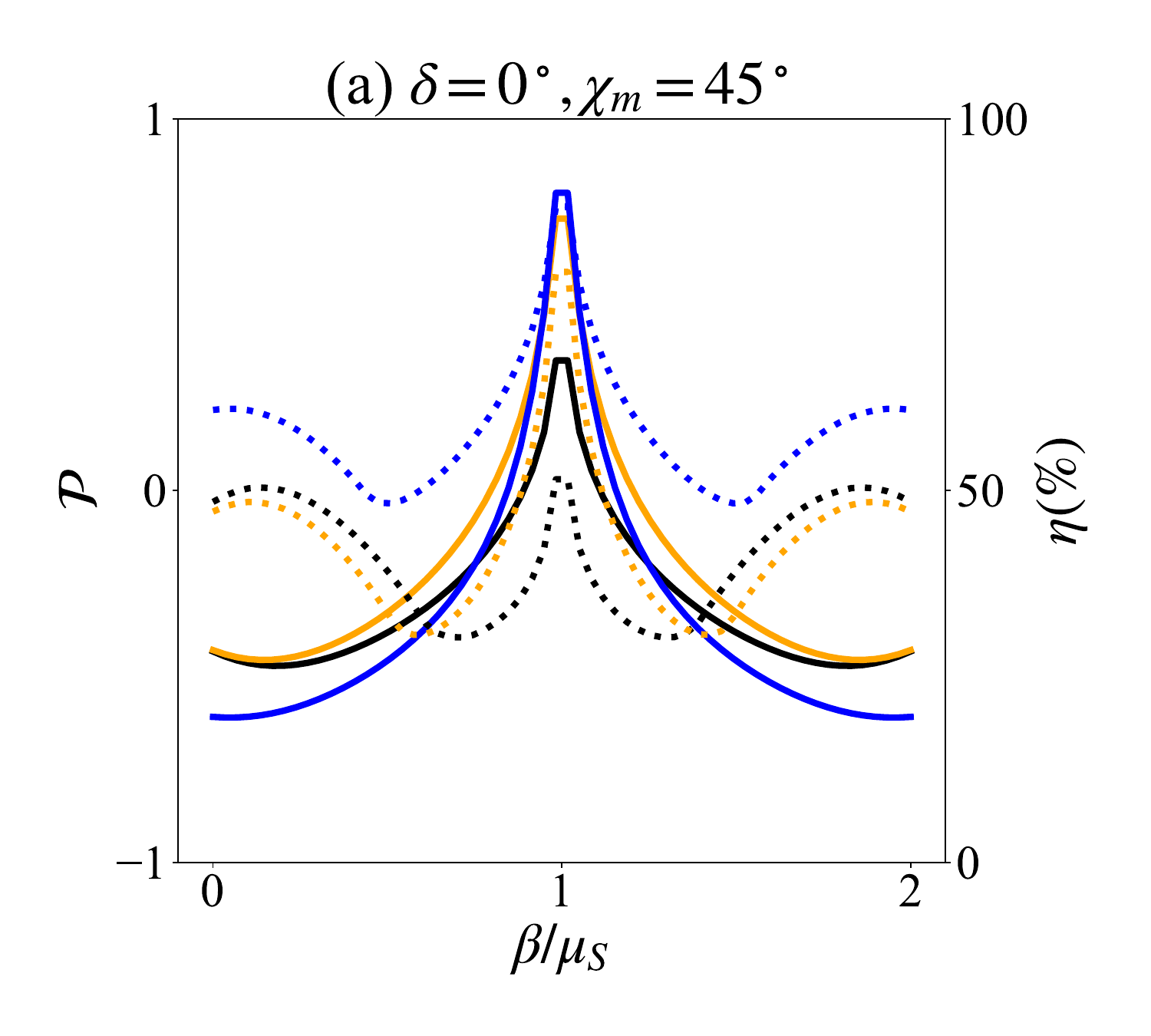}
\hspace{-0.6cm}
\includegraphics[scale = 0.2]{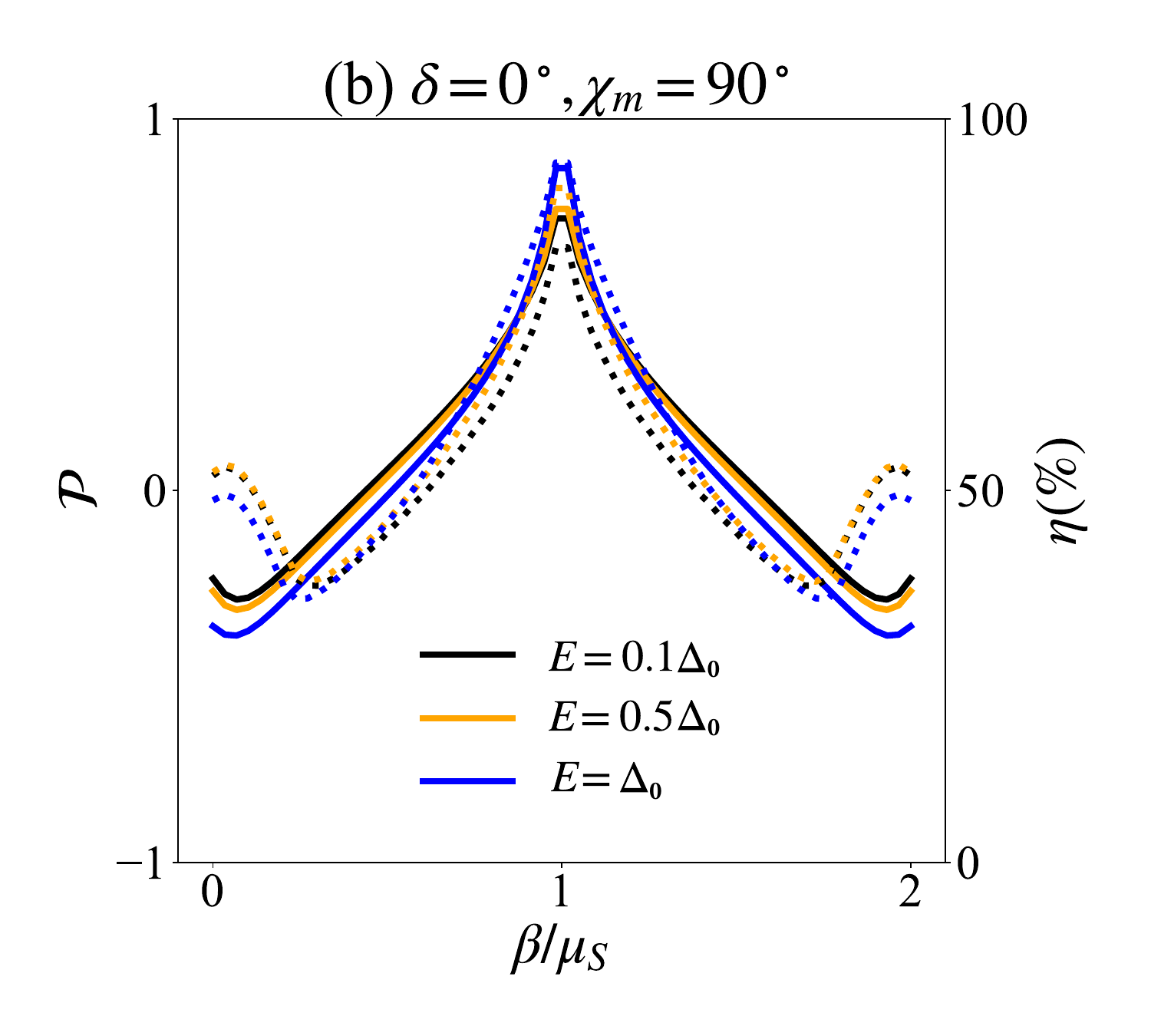} 
}
\centerline{
\includegraphics[scale = 0.2]{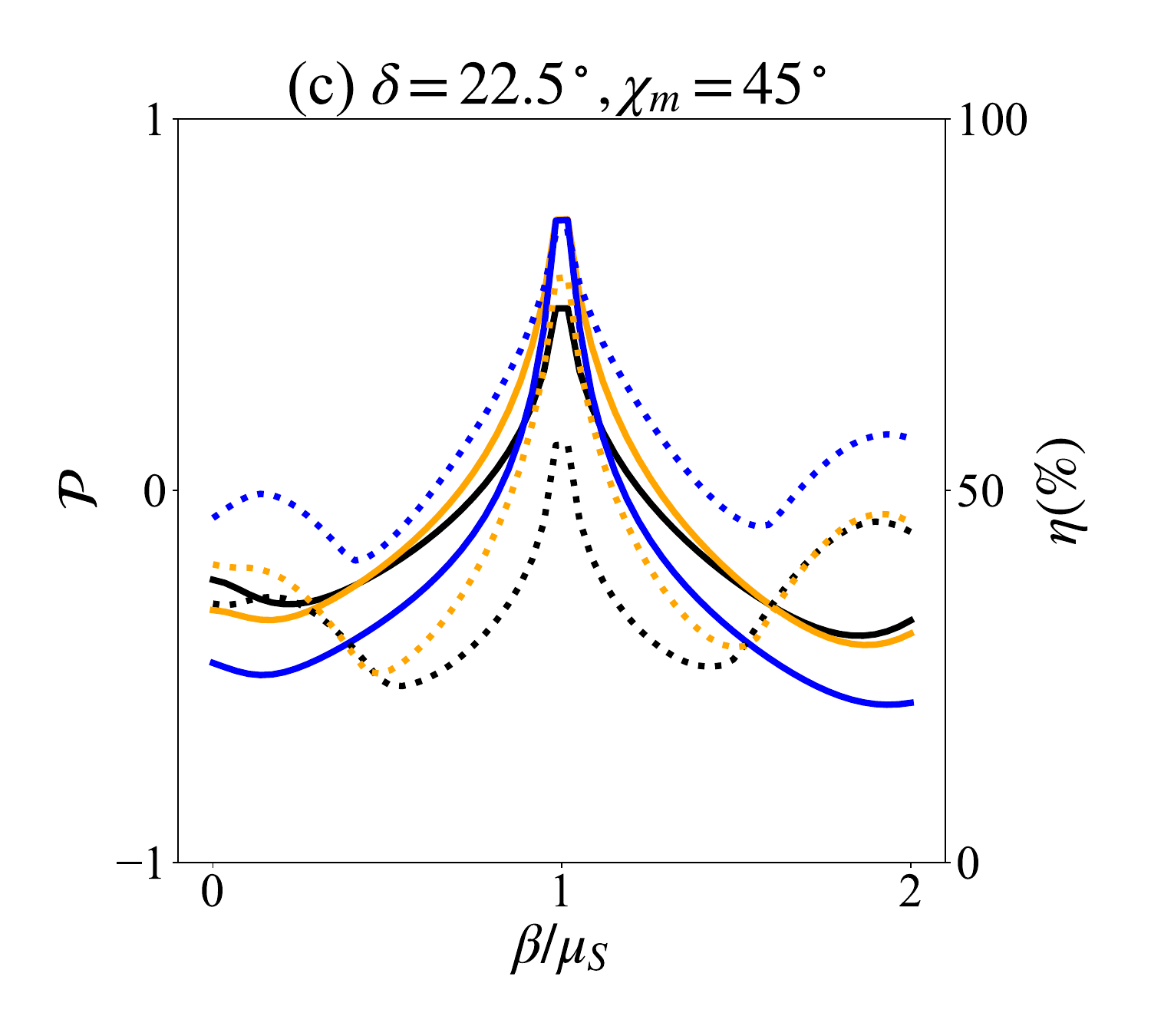}
\hspace{-0.6cm}
\includegraphics[scale = 0.2]{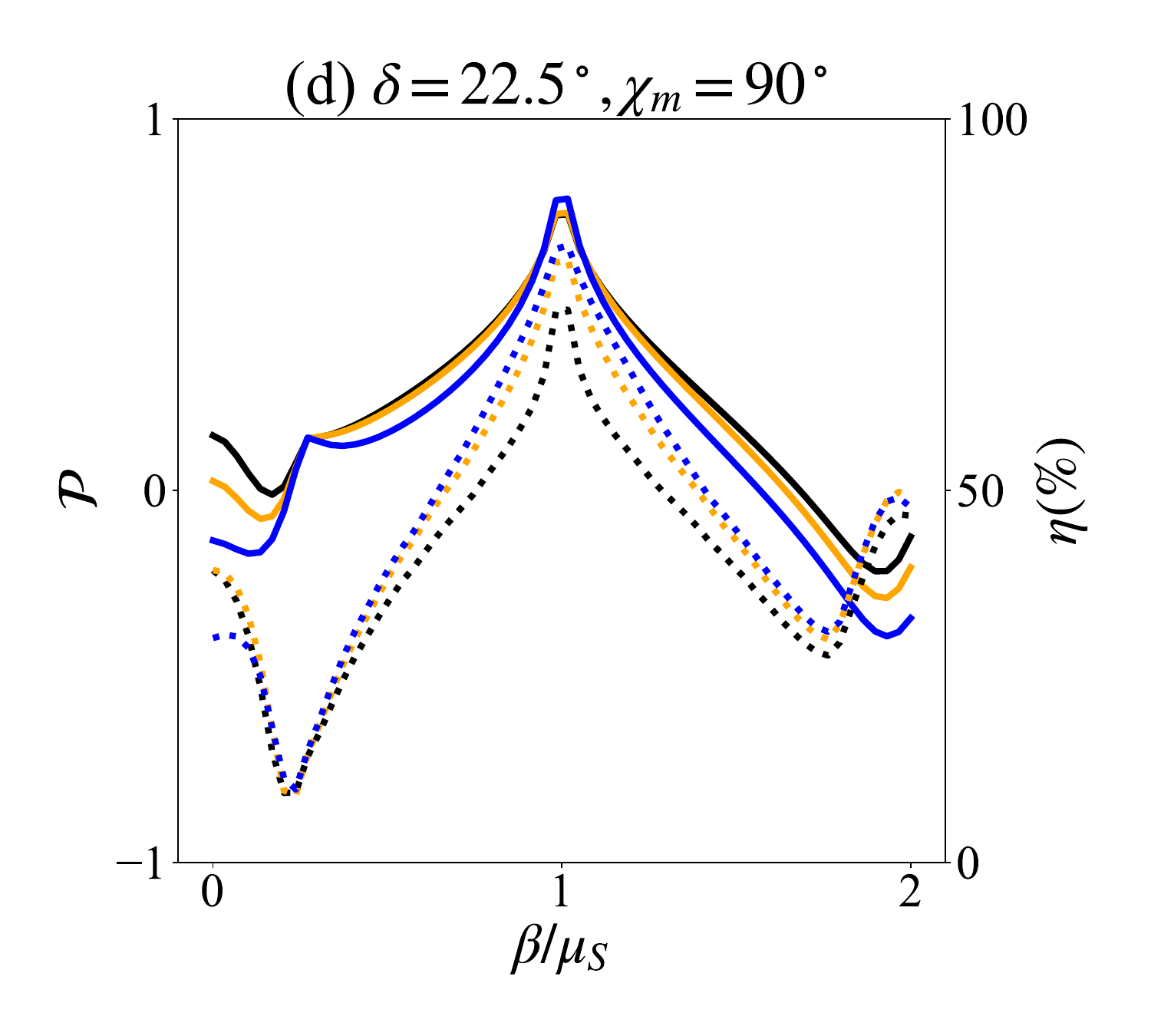} 
}
\caption{Variation of spin polarization $\mathcal{P}$ (solid lines) and spin-filter efficiency $\eta$ (dotted lines) as functions of the normalized ISOC strength $\beta/\mu_S$ for $\delta=0^\circ$ (top panel) and $\delta=22.5^\circ$ (bottom panel) considering different values of $E/\Delta_0$ with $Z_0 = 0.1$, $\rho = 0.3$ and $\xi_m = 45^\circ$. The left and right panels corresponds to $\chi_m=45^\circ$ and $90^\circ$ respectively.}
\label{fig8}
\end{figure} 

\subsection{Dependence on spin active parameters and quasi particle energy}
Fig.~\ref{fig7} and \ref{fig8} illustrate the evolution of the spin polarization $\mathcal{P}$ and spin-filter efficiency $\eta$ as functions of $\beta/\mu_S$ for different values of spin mixing $\rho$ and $E/\Delta_0$ respectively. Fig.~\ref{fig7} reveals a highly nontrivial dependence of $\mathcal{P}$ and $\eta$ on $\beta/\mu_S$, governed by the interplay of spin–valley locking in the ISC, momentum-dependent helicity in the AM, and spin-active interfacial scattering parametrized by $\rho$ and $\chi_m$. For $(\chi_m,\delta)=(45^\circ, 0^\circ)$ , both $\mathcal{P}$ and $\eta$ exhibit an approximate symmetrical pattern around $\beta=\mu_S$ for $\rho=0.1$, with a pronounced extremum at $\beta\sim\mu_S$ as seen from Fig.~\ref{fig7}(a). This reflects a resonance like condition where the ISOC spin splitting in the ISC becomes comparable to the Fermi energy maximizing the mismatch between opposite helicity channels and thereby enhancing spin-selective Andreev reflection processes. However, as $\rho$ increases, strong spin mixing introduces substantial off-diagonal scattering, which redistributes spectral weight between spin channels and suppresses the net polarization. In contrast, the efficiency $\eta$ is maximized away from $\beta\sim\mu_S$ and the maxima are found in the region $\beta\ll\mu_S$ and $\beta\gg\mu_S$, where one spin channel dominates transport more completely. Moreover, the competition between comparable spin channels reduces net selectivity near $\beta=\mu_S$ despite enhanced scattering amplitudes. For $\delta=22.5^\circ$, the spectra becomes asymmetric about $\beta=\mu_S$ in the weak $\rho$ regime as seen from Fig.~\ref{fig7}(b). This asymmetry originates from the intrinsic $d$-wave form of the AM exchange field, which breaks symmetry through anisotropic projection onto the interface magnetization axis. Consequently, the coupling between the AM helicity states and the  spin-polarized quasiparticles of ISC becomes directionally biased, leading to unequal enhancement of scattering amplitudes on either side of $\beta=\mu_S$. The further increase in $\rho$ restores an approximate symmetry, as strong spin mixing averages over helicity-dependent anisotropies and effectively removes the directional selectivity imposed by the AM texture.

A qualitatively different behavior emerges for $\chi_m=90^\circ$, where both $\mathcal{P}$ and $\eta$ become nearly independent of $\rho$ as observed from Figs.~\ref{fig7}(b) and \ref{fig7}(d). In this configuration, the interface magnetization is orthogonal to the dominant spin quantization axis imposed by the ISOC, leading to intrinsically strong spin rotation of incident quasiparticles. Consequently, spin-channel mixing becomes efficient even for weak spin-active scattering making the conductance largely insensitive to further increases in $\rho$. For $\delta=0^\circ$, the spectra remain symmetric about $\beta=\mu_S$, reflecting the residual symmetry of the projected AM spin texture. In contrast, for $\delta=22.5^\circ$, a significant asymmetry is observed from Fig.~\ref{fig7}(d), indicating that the anisotropic helicity the AM continues to imprint itself on the conductance even in the presence of strong spin rotation. An additional key insight is the distinct roles played by $\mathcal{P}$ and $\eta$. While $\mathcal{P}$ measures the imbalance between spin channels and is therefore highly sensitive to interference between helicity-dependent scattering amplitudes, $\eta$ quantifies the absolute spin selectivity and is maximized whenever one channel dominates. This explains why $\eta$ can remain large even when $\mathcal{P}$ is suppressed, particularly in regimes of strong ISOC or strong spin mixing. Overall, the figure demonstrates that the optimal spin filtering does not occur at the resonance condition $\beta\sim\mu_S$, but rather in regimes where either ISOC induced spin locking or interface-induced spin mixing dominates. Moreover, the AM orientation plays a decisive role in controlling symmetry and anisotropy of the response.

\begin{figure*}[t]
\centerline
\centerline{
\includegraphics[scale = 0.25]{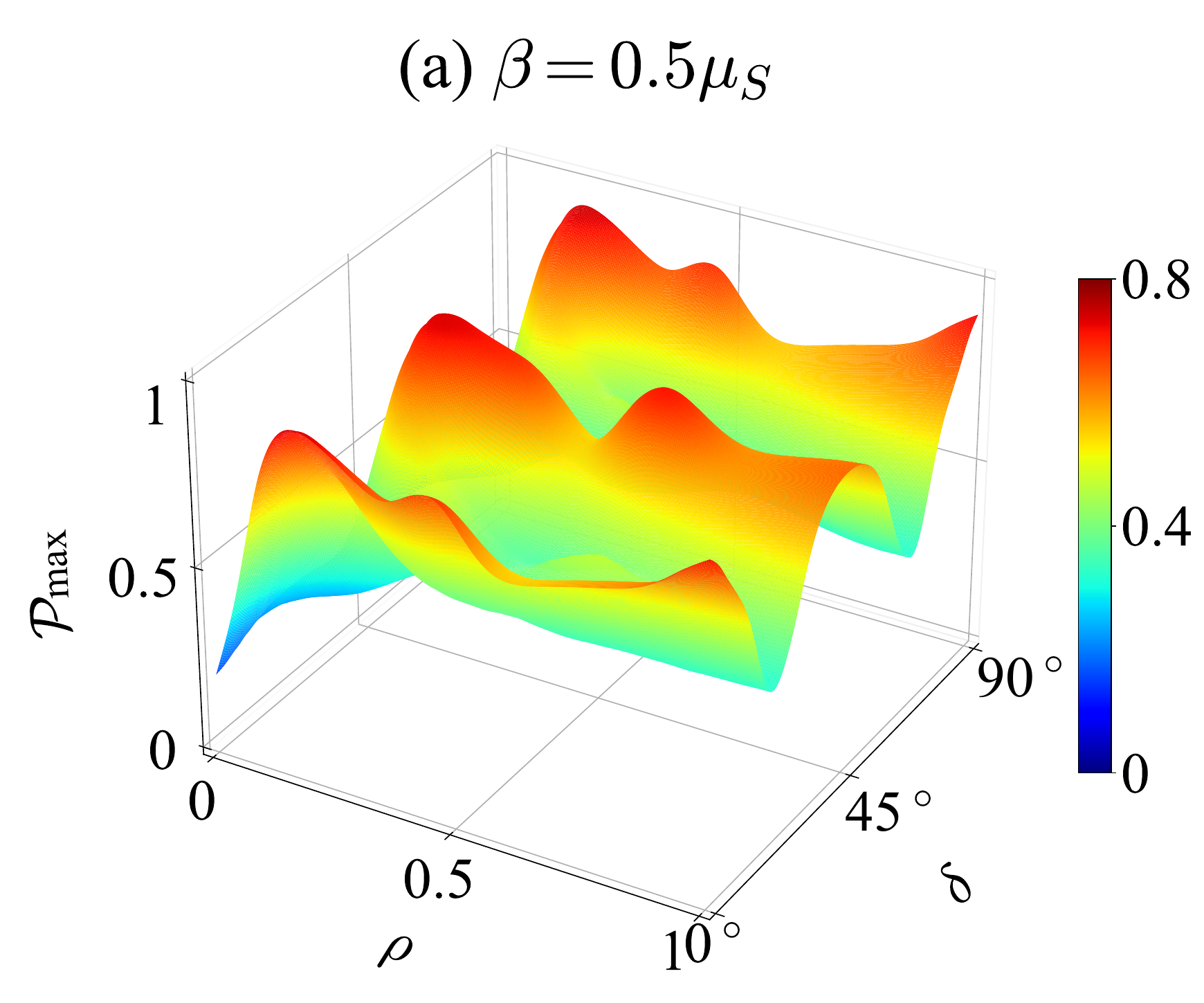}
\includegraphics[scale = 0.25]{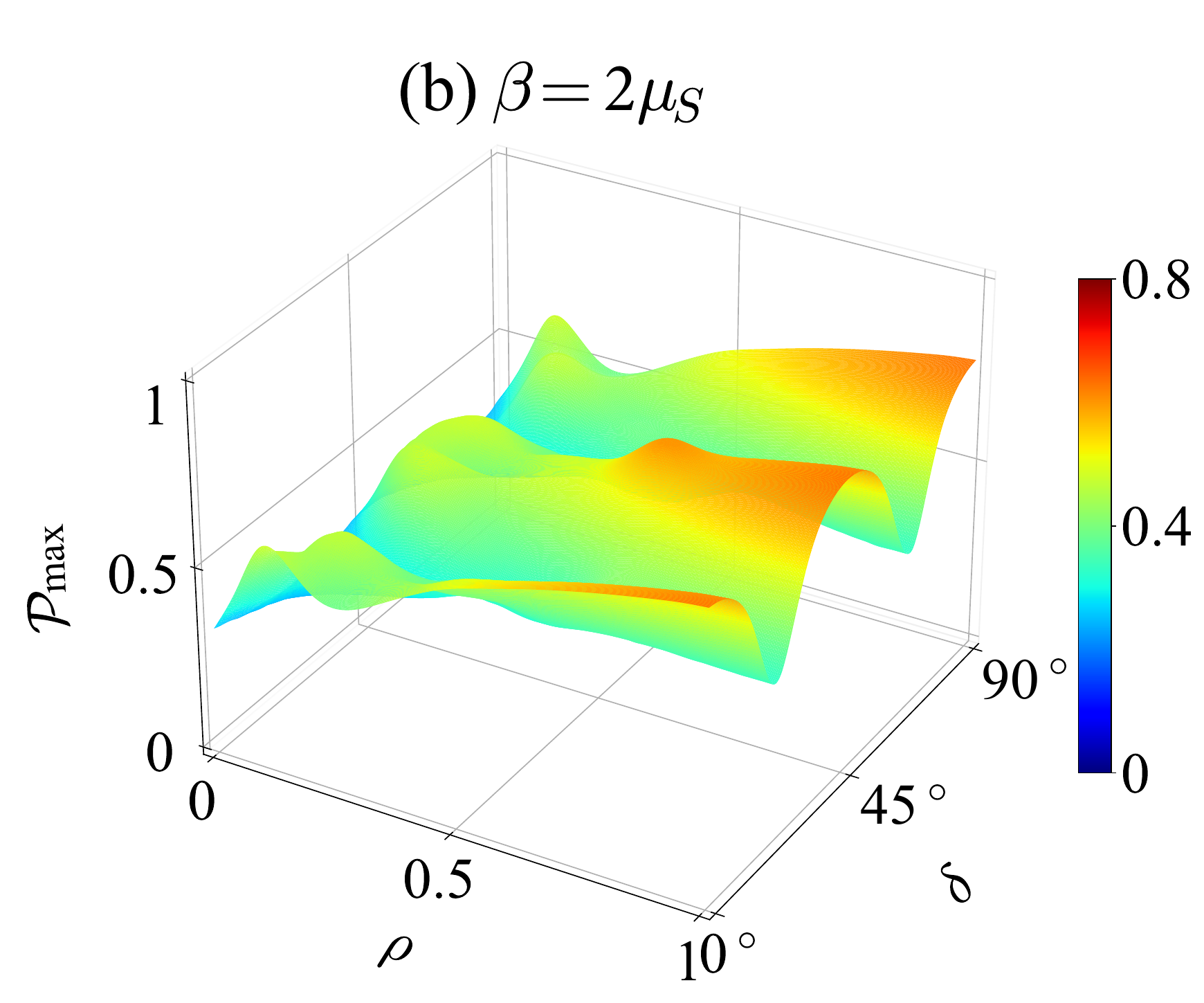}}
\centerline{
\includegraphics[scale = 0.32]{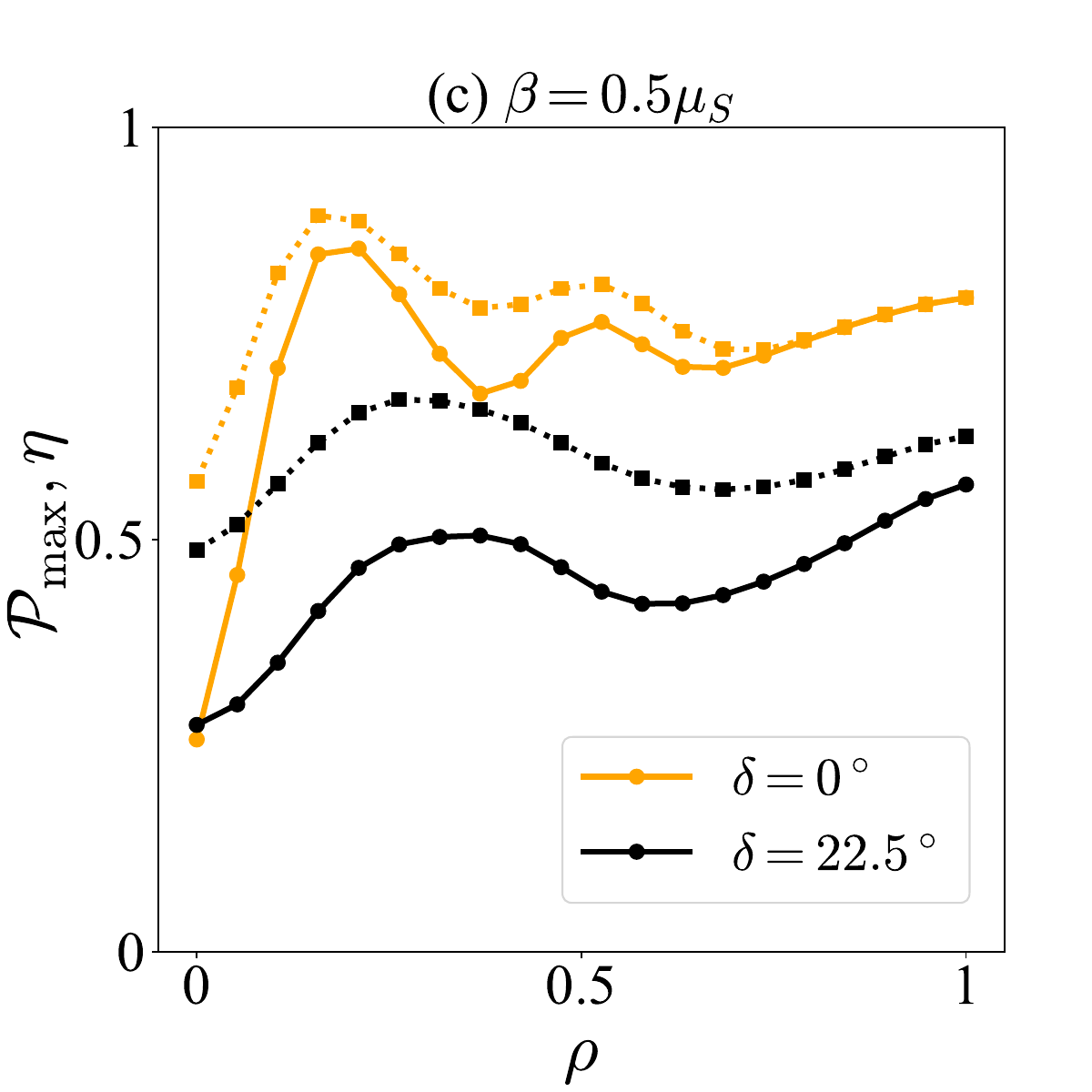}
\hspace{0.1cm}
\includegraphics[scale = 0.32]{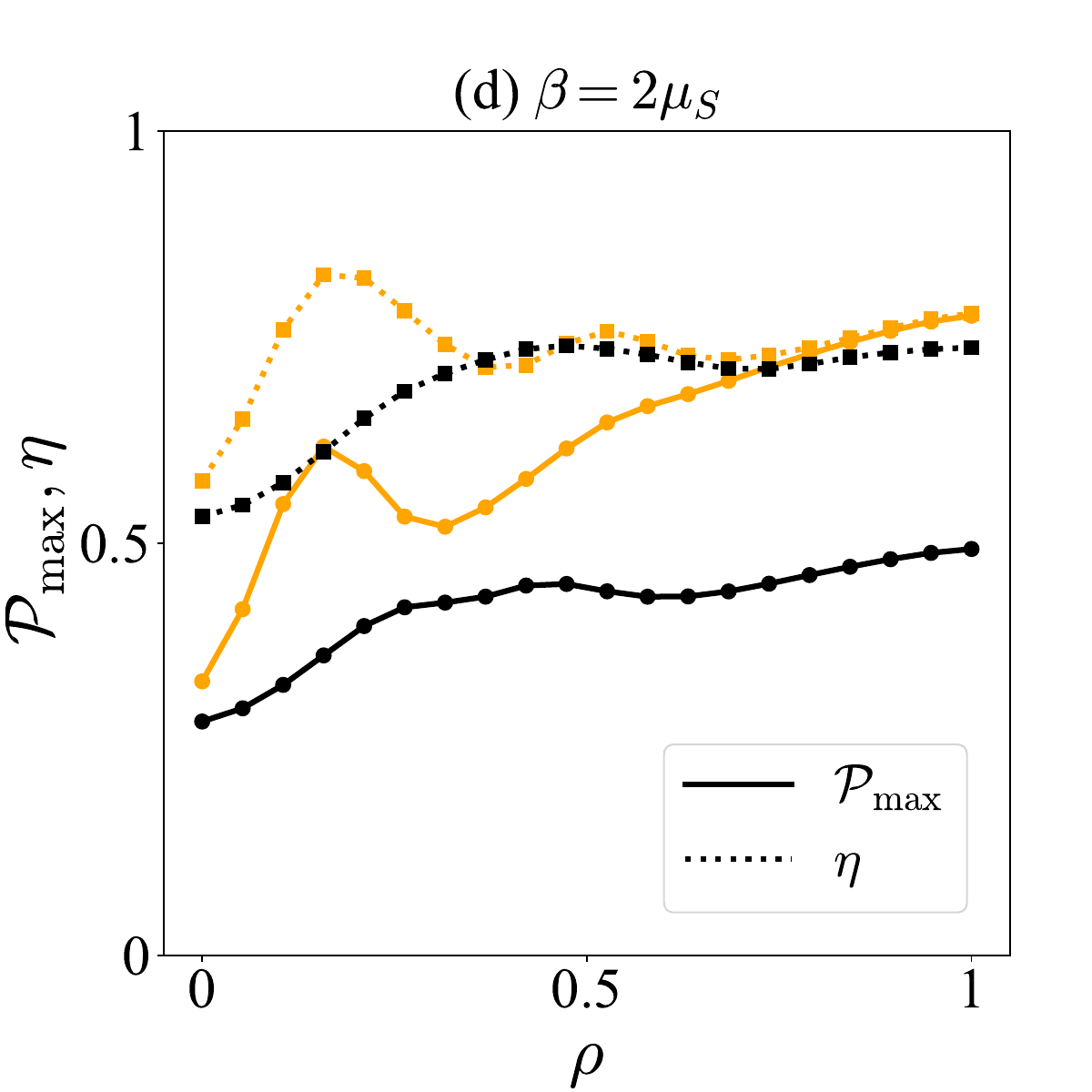}}
\hspace{0.1cm}
\caption{(Top panel) Variation of the $\mathcal{P}_\text{max}$ as a function of $\rho$ and $\delta$ for (a) $\beta = 0.5\mu_S$ and (b) $\beta = 2\mu_S$ considering $Z_0 = 0.1$, $\xi_m = 45^\circ$ and $\chi_m = 45^\circ$. (Bottom panel) Line cuts of $\mathcal{P}_\text{max}$ (solid lines) and $\eta$ (dotted lines) as functions of $\rho$ for (c) $\beta = 0.5\mu_S$ and (d) $\beta = 2\mu_S$ considering $\delta = 0^\circ$ and $22.5^\circ$.}
\label{fig9}
\end{figure*}
Fig.~\ref{fig8} extends the results of Fig.~\ref{fig7} by resolving the spin polarization
and efficiency at finite quasiparticle energies, thereby explaining the interplay between ISOC, helicity mixing and energy-dependent Andreev processes. For $\delta=0^\circ$, both $\mathcal{P}$ and $\eta$ remain symmetric about $\beta=\mu_S$ across all energies as seen from Figs.~\ref{fig8}(a) and \ref{fig8}(b). It indicates that the underlying helicity structure
remains symmetric even away from equilibrium.  The peak near $\beta\sim\mu_S$ arises from optimal matching between the ISOC induced spin polarization in the ISC and the interfacial spin-rotation processes, which maximizes spin-selective Andreev reflection. However, the magnitude and sharpness of this peak strongly dependent on the quasi particle energy. At low energies i.e., for $E\ll\Delta_0$, dominant Andreev reflection enhances spin selectivity. However for $E\sim\Delta_0$ spin selectivity is significantly reduced due to increase in quasiparticle transmission and velocity mismatch, leading to broader and suppressed spectrum. For $\chi_m=90^\circ$, the spectra remain largely insensitive to $\rho$ and retain their symmetry, consistent with the fact that the interface magnetization is orthogonal to the ISOC induced spin
quantization axis, resulting in intrinsically strong spin mixing
even at weak scattering. In contrast, for $\delta=22.5^\circ$, finite-energy transport makes the anisotropic nature of the AM clearly visible, leading to a strong asymmetry around $\beta=\mu_S$ that is much weaker at low bias as observed from Figs.~\ref{fig8}(c) and \ref{fig8}(d). This behavior arises because the $d$-wave helicity field couples differently to the interface magnetization at finite energy, so the two spin channels evolve unequally. The effect is most pronounced for $\chi_m=45^\circ$, where partial alignment between spins allows strong interference between spin-conserving and spin-flip Andreev processes, making both $\mathcal{P}$ and $\eta$ highly energy dependent. In contrast, for $\chi_m=90^\circ$, spin mixing is already maximized, so increasing energy mainly redistributes spectral weight without significantly changing spin selectivity. Overall, Fig.~\ref{fig8} demonstrates that finite-energy transport does not merely renormalize the low bias behavior, but qualitatively enhances the role of anisotropic helicity, energy-resolved Andreev processes, and velocity mismatch, thereby provide efficient spin filtering in AM/ISC junctions.

\subsection{Maximum Polarization}
To further quantify the optimal regimes of spin-selective transport beyond the trends discussed above, it is necessary to analyze the global maxima of spin polarization and spin-filter efficiency across the $(\rho, \delta)$ parameter space. 
Fig.~\ref{fig9} provides a global view of the evolution of $\mathcal{P}_\text{max}$ and the corresponding $\eta$ as functions of $\rho$ and $\delta$, consolidating the key trends identified in Fig.~\ref{fig7} and Fig.~\ref{fig8} into a unified phase-space picture.
The upper panels show that $\mathcal{P}_\text{max}$ is not inherently a monotonic function of either $\rho$ or $\delta$, but instead emerges from the competition between three mechanism: spin-selective transmission at the interface, helicity driven spin precession due to ISOC, and momentum dependent spin splitting of the AM. For $\delta=0^\circ$, the band structure remains symmetric at the dispersion level, implying no intrinsic preference for a given spin channel. Nevertheless, finite polarization emerges from the combined effect of ISOC induced spin - momentum locking, directional constraints imposed by transport, and spin-dependent scattering, which together prevent complete cancellation between opposite helicity contributions. As $\rho$ increases, spin filtering is initially enhanced; however, beyond an optimal value, excessive spin mixing redistributes spectral weight between channels and suppresses $\mathcal{P}_\text{max}$, leading to the ridge-like structure observed in Figs.~\ref{fig9}(a) and \ref{fig9}(b). In contrast, for $\delta\neq 0$, the system becomes anisotropic in momentum space. This breaks the balance between opposite spin channels and redistributes spectral weight unevenly and hence  reduce the value of $\mathcal{P}_\text{max}$ and shifts its optimal value with respect to $\rho$. Notably, pronounced maxima occur corresponding to high symmetry configurations near $\delta \approx 0^\circ, 45^\circ$ and $90^\circ$. It is due to the reason that at $\delta=0^\circ$ although the band structure is symmetric at the dispersion level, so it does not intrinsically favor any spin channel. The finite polarization instead arises from the combined effects of ISOC induced spin -  momentum locking, directional transport, and spin selective scattering, which prevent exact cancellation between opposite helicity contributions. For $\delta=45^\circ$, the band structure maximizes the overlap between transport momentum, ISOC helicity and interface spin selectivity, leading to a robust and uniform preference for one spin channel thereby having maximum value of $\mathcal{P}$. At $\delta=90^\circ$, the system recovers a symmetry related configuration with reversed spin texture, which again supports strong polarization. 

The lower panels highlight the role of ISOC strength. 
For $\beta=0.5\mu_S$, the system lies away from the resonance condition $\beta \sim \mu_S$, resulting in incomplete spin rotation.  Hence $\mathcal{P}_\text{max}$ is found to be moderate in this scenario as seen from Fig.~\ref{fig9}(c). However, $\eta$ still remains relatively large as partial filtering still survives without strong cancellation between channels. Furthermore, both $\mathcal{P}_\text{max}$ and $\eta$ exhibit non-monotonic dependence on $\rho$, reflecting the crossover from weak to strong spin mixing. However, at $\beta=2\mu_S$, the system enters a regime of strong helicity-induced spin locking, where spins are already significantly rotated before scattering.  Consequently, further increase in $\rho$ has a diminished effect, and both $\mathcal{P}_\text{max}$ and $\eta$ tend to saturate as seen from Fig.~\ref{fig9}(d). Importantly, across all panels, $\eta$ does not necessarily follow the trend of $\mathcal{P}_\text{max}$. It is due to the reason that high polarization requires imbalance between spin channels, whereas high efficiency requires both strong transmission and selectivity so the two quantities peak in different regions of parameter space. Overall, Fig.~\ref{fig9} demonstrates that optimal spin filtering is achieved not by maximizing spin mixing or ISOC of the ISC independently but by tuning the system to a regime where ISOC induced spin rotation and interface induced spin selectivity cooperate without inducing excessive spin mixing, with this balance strongly modulated by the AM anisotropy.

\section{Conclusions}
In summary, we have theoretically investigated the spin-resolved transport, spin filtering and nonreciprocal effects in AM/ISC junctions with a spin-active interface. We observe that both charge and spin conductance remain nearly independent of the AM orientation in the absence of spin-dependent interfacial scattering, exhibiting isotropic and helicity decoupled transport despite momentum-dependent spin splitting. The inclusion of spin-active scattering induces strong helicity mixing, leading to pronounced orientation dependent conductance. We found significantly different behavior for different ISOC strength of the ISC. For single band ISC, the conductance spectra exhibit strong anisotropy and spin-selective features with modified coherence peaks.
The intermediate regime shows enhanced sensitivity to interfacial scattering, resulting in significant spectral redistribution and angular dependence. In the double-band regime, compensation between spin channels produces smoother spectra with reduced anisotropy. In the strong spin-mixing limit, spectral weight redistribution suppresses conventional Andreev reflection and generates highly anisotropic characteristics.
The spin conductance is found to be suppressed and orientation independent  in the absence of spin-active scattering for weak and intermediate ISOC strengths, while it show strongly energy dependent characteristics in the strong ISOC regime. Spin-active interfaces generate finite, highly anisotropic spin conductance with sharp spectral features, indicating efficient spin-selective transport.
The spin polarization and spin-filter efficiency exhibit nonmonotonic dependence on system parameters and distinct angular modulation. Maximum polarization occurs at AM orientations $\delta = 0^\circ, 45^\circ,$ and $90^\circ$. Although the spin-filter efficiency reaches $\sim 86\%$ in the single-band regime and $\sim 77\%$ in the double-band regime for $\delta=0^\circ$ orientation but it can be tuned through suitable choice of spin active parameters. Finite-energy analysis shows enhanced selectivity at low energies and reduced response near the superconducting gap. The emergence of angle asymmetric conductance under spin-active scattering demonstrates nonreciprocal transport. Overall, our results establish AM/ISC junctions as a promising and highly tunable platform for superconducting spintronics. The demonstrated control over spin polarization, efficiency, and nonreciprocity provides a clear pathway for designing next-generation devices based on directional spin transport, superconducting spin filtering and anisotropic quantum functionalities.

\appendix

\section{Derivation of Spin-Active Interface Boundary Conditions}
In this appendix, we provide a detailed derivation of the boundary conditions at the spin-active interface located at $x=0$. The starting point is the Bogoliubov--de Gennes (BdG) equation,
\begin{equation}
\check{\mathcal{H}}_{\text{BdG}}(x)\Psi(x) = E \Psi(x),
\end{equation}
where the total Hamiltonian includes a delta-function interface potential describing both spin-independent and spin-dependent scattering processes. The interface Hamiltonian is modeled as
\begin{equation}
H_{\text{int}} = \mathcal{V}\,\delta(x),
\end{equation}
where, $\mathcal{V} = \Phi_0 + \mathbf{\Phi}_m \cdot \boldsymbol{\sigma} $. We parameter $\Phi_0$ corresponds to the scalar barrier potential while $\mathbf{\Phi}_m$ represents the magnetic component of the interface potential.

In the Nambu basis, the interface contribution takes the form
\begin{equation}
\check{H}_{\text{int}} =
\begin{pmatrix}
\mathcal{V} & 0 \\
0 & -\mathcal{V}^\ast
\end{pmatrix}\delta(x).
\end{equation}
To obtain the boundary conditions, we integrate the BdG equation over an infinitesimal region across the interface, i.e., over the interval $[-\epsilon, \epsilon]$ and take the limit $\epsilon \to 0$. This yields
\begin{equation}
\int_{-\epsilon}^{\epsilon} dx \left[ \check{\mathcal{H}}_{\text{BdG}} - E \right]\Psi(x) = 0.
\end{equation}
The dominant contribution arises from the kinetic energy term involving second-order spatial derivatives. Explicitly, we evaluate
\begin{equation}
-\frac{\hbar^2}{2m}\int_{-\epsilon}^{\epsilon} dx \, \partial_x^2 \Psi(x)
= -\frac{\hbar^2}{2m} \left[ \partial_x \Psi(0^+) - \partial_x \Psi(0^-) \right].
\end{equation}
The delta-function term contributes as
\begin{equation}
\int_{-\epsilon}^{\epsilon} dx \, \check{H}_{\text{int}} \Psi(x)
= 
\begin{pmatrix}
\mathcal{V} & 0 \\
0 & -\mathcal{V}^\ast
\end{pmatrix}
\Psi(0).
\end{equation}
Combining these contributions, we obtain the discontinuity condition for the derivative of the wave function,
\begin{equation}
\partial_x \Psi(0^+) - \partial_x \Psi(0^-)
=
\frac{2m}{\hbar^2}
\begin{pmatrix}
\mathcal{V}& 0 \\
0 & -\mathcal{V}^\ast
\end{pmatrix}
\Psi(0).
\end{equation}
The wave function itself remains continuous across the interface, i.e.,
\begin{equation}
\Psi(0^+) = \Psi(0^-).
\end{equation}
To express the boundary condition in a convenient form, we decompose the magnetic interface potential as
\begin{equation}
\mathbf{\Phi}_m \cdot \boldsymbol{\sigma}
=
\Phi_x \sigma_x + \Phi_y \sigma_y + \Phi_z \sigma_z .
\end{equation}
Defining the transverse combinations
\begin{equation}
\Phi_\pm = \Phi_x \pm i\Phi_y,
\end{equation}
the spin-dependent interface potential can be written as
\begin{equation}
\mathbf{\Phi}_m \cdot \boldsymbol{\sigma}
=
\begin{pmatrix}
\Phi_z & \Phi_- \\
\Phi_+ & -\Phi_z
\end{pmatrix}.
\end{equation}
Here, the longitudinal component $\Phi_z$ gives rise to spin-dependent phase shifts and therefore produces spin mixing, whereas the transverse components $\Phi_\pm$ induce spin-flip scattering between opposite spin channels. Comparing this form with the interface matrix $\hat{\Lambda}$ introduced in Eq.~(\ref{eq15}), one obtains
\begin{align}
\Omega_1 &= 2\rho \Phi_0 \cos\chi_m,\\
\Omega_2 &= 2\rho \Phi_0 \sin\chi_m\, e^{-i\xi_m},
\end{align}
where $\Omega_1$ characterizes the spin-mixing contribution and $\Omega_2$ represents the spin-flip scattering amplitude at the interface. Consequently, the interface matrix $\hat{\Lambda}$ incorporates both spin-conserving and spin-flip scattering processes, thereby coupling different spin and helicity channels across the AM/ISC junction.

\section{Helicity Eigenstates of the Altermagnetic Hamiltonian}

We now derive the helicity eigenstates of the altermagnetic Hamiltonian. The single-particle Hamiltonian is given by
\begin{equation}
\hat{\mathcal{H}}_0(\mathbf{k}) = \xi_{\mathbf{k}} \hat{I} + \lambda \mathbf{g}(\mathbf{k}) \cdot \boldsymbol{\sigma},
\end{equation}
where $\mathbf{g}(\mathbf{k}) = (g_x, g_y, 0)$. Writing the spin-dependent term explicitly, we obtain
\begin{equation}
\mathbf{g} \cdot \boldsymbol{\sigma} =
\begin{pmatrix}
0 & g_x - i g_y \\
g_x + i g_y & 0
\end{pmatrix}.
\end{equation}
Defining the complex phase $\phi_{\mathbf{k}}$ through
\begin{equation}
g_x + i g_y = |\mathbf{g}| e^{i\phi_{\mathbf{k}}},
\end{equation}
the Hamiltonian takes the form
\begin{equation}
\hat{\mathcal{H}}_0 =
\begin{pmatrix}
\xi_{\mathbf{k}} & \lambda |\mathbf{g}| e^{-i\phi_{\mathbf{k}}} \\
\lambda |\mathbf{g}| e^{i\phi_{\mathbf{k}}} & \xi_{\mathbf{k}}
\end{pmatrix}.
\end{equation}
The eigenvalues follow from the secular equation,
\begin{equation}
\det(\hat{\mathcal{H}}_0 - E \hat{I}) = 0,
\end{equation}
which yields
\begin{equation}
E_{\pm} = \xi_{\mathbf{k}} \pm \lambda |\mathbf{g}|.
\end{equation}
The corresponding normalized eigenvectors are obtained by solving the linear system explicitly. For the positive helicity branch we can write
\begin{equation}
\psi_+ = \frac{1}{\sqrt{2}}
\begin{pmatrix}
1 \\
e^{i\phi_{\mathbf{k}}}
\end{pmatrix},
\end{equation}
while for the negative branch we have
\begin{equation}
\psi_- = \frac{1}{\sqrt{2}}
\begin{pmatrix}
1 \\
- e^{i\phi_{\mathbf{k}}}
\end{pmatrix}.
\end{equation}
These eigenstates define the helicity basis used to construct the quasiparticle wave functions in the altermagnetic region.

\section{Spin Expectation Values in the Helicity Basis}

We next evaluate the spin expectation values in the helicity eigenstates. Considering the positive helicity state $\psi_+$, the expectation value of $\sigma_z$ is given by
\begin{equation}
\langle \sigma_z \rangle_+ = \psi_+^\dagger \sigma_z \psi_+.
\end{equation}
Substituting the explicit form of $\psi_+$ and carrying out the matrix multiplication yields
\begin{equation}
\langle \sigma_z \rangle_+ = \frac{1}{2}(1 - 1) = 0.
\end{equation}
Similarly, evaluating the expectation values of $\sigma_x$ and $\sigma_y$, one finds
\begin{equation}
\langle \sigma_x \rangle_+ = \cos \phi_{\mathbf{k}}, \quad
\langle \sigma_y \rangle_+ = \sin \phi_{\mathbf{k}}.
\end{equation}
An identical calculation for the negative helicity state leads to
\begin{equation}
\langle \boldsymbol{\sigma} \rangle_{\pm} = \pm (\cos \phi_{\mathbf{k}}, \sin \phi_{\mathbf{k}}, 0).
\end{equation}
This explicitly demonstrates that the spin expectation value is locked to the momentum-dependent direction of the altermagnetic exchange field, confirming that spin is not a conserved quantum number in this system.

\section{Derivation of the Conductance Formula}

Finally, we derive the expression for the charge conductance within the BTK formalism. The current carried by a quasiparticle state is given by
\begin{equation}
J_x = \Psi^\dagger \hat{v}_x \Psi,
\end{equation}
where the velocity operator is defined as $\hat{v}_x = \partial \hat{\mathcal{H}}_0 / \partial k_x$. For an incident quasiparticle in helicity channel $s$, the incoming current is proportional to the group velocity $v^e_s$. The reflected current consists of contributions from all helicity channels,
\begin{equation}
J_r = \sum_{s'} v^e_{s'} |r_{ss'}|^2,
\end{equation}
while the Andreev-reflected holes contribute
\begin{equation}
J_A = \sum_{s'} v^h_{s'} |a_{ss'}|^2.
\end{equation}
Taking into account that holes carry opposite charge, the normalized conductance is obtained as
\begin{equation}
G(E,\theta) =
\sum_{s,s'}
\left[
\delta_{ss'} - \frac{v^e_{s'}}{v^e_s}|r_{ss'}|^2
+ \frac{v^h_{s'}}{v^e_s}|a_{ss'}|^2
\right].
\end{equation}

\end{document}